\documentclass[twocolumn]{aastex631}

\usepackage{latexsym,amsmath,amssymb}
\usepackage{graphics,graphicx}
\usepackage{natbib}
\usepackage{rotating}
\newcommand{\degmark}{$^\circ$}
\newcommand{\nh}{N${\rm _H}$}
\def \rchisq {$\chi_{\nu} ^{2}$}
\newcommand{\rxte}{{\it RXTE}}

\newcommand{\eins}{{\it EINSTEIN}}

\newcommand{\suz}{{\it Suzaku}}
\newcommand{\xmm}{{\it XMM-Newton}}

\newcommand{\swi}{{\it Swift}}
\newcommand{\cha}{{\it Chandra}}
\newcommand{\nustar}{\textit{NuSTAR}}
\newcommand{\nicer}{\textit{NICER}}

\newcommand{\msun}{$\rm M_{\odot}$}

\newcommand{\fluxcgs}{erg~s$^{-1}$~cm$^{-2}$}
\newcommand{\lumcgs}{erg~s$^{-1}$}
\newcommand{\chisq}{$\chi ^{2}$}

\shorttitle{X-ray Spectroscopy of Z Cha with \xmm}
\shortauthors{Balman, \c{S}. }

\begin{document}

\title{X-ray Spectroscopy of the Dwarf Nova Z Chamaeleontis in Quiescence and Outburst Using the \xmm\ Observatory}

\correspondingauthor{\c{S}\"olen Balman}
\email{solen.balman@gmail.com, solen.balman@istanbul.edu.tr}

\author[0000-0001-6135-1144]{\c{S}\"olen Balman}
\affiliation{Department of Astronomy and Space Sciences, Faculty of Science,  Istanbul University, Beyazit, 34119, Istanbul, Turkey}
\affiliation{Kadir Has University, Faculty of Engineering and Natural Sciences, Cibali 34083, Istanbul, Turkey}

\author[0000-0002-4162-8190]{Eric M. Schlegel}
\affiliation{Department of Physics and Astronomy, University of Texas-San Antonio, San Antonio, TX 78249, USA }

\author[0000-0002-4806-5319]{Patrick Godon}
\affiliation{Department of Astrophysics \& Planetary Science, Villanova University, 800 Lancaster Avenue, Villanova, PA 19085, USA}
\affiliation{Henry A. Rowland Department of Physics \& Astronomy, Johns Hopkins University, Baltimore, MD 21218, USA}

\author[0000-0002-0210-2276]{Jeremy J. Drake}
\affiliation{Lockheed Martin Solar and Astrophysics Laboratory, 3251 Hanover St, Palo Alto, CA 94304, USA}




\begin{abstract}

We present  X-ray spectroscopy of the SU UMa-type dwarf nova (DN) Z Cha using the EPIC and RGS instruments onboard the \xmm\ Observatory. The quiescent system can be modeled by collisional equilibrium or nonequilibrium plasma models, yielding a kT of 8.2-13.0 keV at a luminosity of (5.0-6.0)$\times$10$^{30}$\lumcgs. The spectra yield better \rchisq\ using partial covering absorbers of cold and photoionized nature. The ionized absorber has an equivalent \nh=(3.4-5.9)$\times$10$^{22}$ cm$^{-2}$ and a log($\xi$)=3.5-3.7 with (50-60)\% covering fraction when VNEI model (XSPEC) is used. The line diagnosis in quiescence shows no resonance lines with only detected forbidden lines of  Ne, Mg, Si. The H-like C, O, Ne, and Mg are detected. The strongest line is O VIII with (2.7-4.6)$\times$10$^{-14}$\fluxcgs. The quiescent X-ray emitting plasma is not collisional and not in ionization equilibrium which is consistent with hot ADAF-like accretion flows. The  line diagnosis in outburst shows He-like O, and Ne with intercombination lines being the strongest along with  weaker resonance lines. This indicates the plasma is more collisional and denser, but yet not in a collisional equilibrium, revealing ionization timescales of (0.97-1.4)$\times$10$^{11}$ s\ cm$^{-3}$.  The R-ratios in outburst yield electron densities of (7-90)$\times$10$^{11}$ cm$^{-3}$ and the G-ratios yield electron temperatures of (2-3)$\times$10$^{6}$ K. The outburst luminosity is (1.4-2.5)$\times$10$^{30}$\lumcgs. The flow is inhomogeneous in density. All detected lines are narrow with widths limited by the resolution of RGS yielding Keplerian rotational velocities $<$1000 km s$^{-1}$. This is too low for boundary layers, consistent with the nature of ADAF-like  hot flows.

\end{abstract}



\section{Introduction} \label{sec:intro}

Cataclysmic Variables (CVs) are close binary systems where a white dwarf (WD) accretes matter from a late-type Roche-Lobe-filling main sequence star \citep{1995Warner}.  
In nonmagnetic (or weakly magnetic) CVs, an accretion disk is assumed to reach all the way to the WD surface.
Dwarf novae (DNe) are a class of nonmagnetic CVs where
ongoing accretion at a low rate (quiescence) is interrupted every few weeks to months or sometimes longer durations by
intense accretion (outburst) of days to weeks where $\dot{\rm M}$ increases 
\citep[see][for a review]{2020Balman,2020Hameury}. There are also a few magnetic CVs (i.e., intermediate polars) that show DN-type outbursts. Standard accretion disk theory  \citep{1973Shakura} predicts half of the accretion luminosity to emerge from the disk and the other half from the boundary layer (BL) very close to the WD  \citep{1974Lynden-Bell}. 
The theoretically-expected BL emission is such that during low-mass accretion states ($\dot M_{acc}$$<$10$^{-(9-9.5)}$M$_{\odot}$), it is optically thin in the hard X-rays  \citep{1993Narayan,1999Popham}, and for high accretion rate
states ($\dot M_{acc}$$>$10$^{-(9-9.5)}$M$_{\odot}$), it is optically thick in the soft X-rays and EUV; kT$\sim$10$^{5-5.6}$ K \citep{1995Popham,1995Godon,2015Hertfelder,2017Hertfelder}. It has been shown that the optically thin BLs can be radially extended, advecting part of the energy to the WD as a result of their inability to cool  \citep{1993Narayan}. The standard disk is often found inadequate to model high-state CVs (Nova-likes; NLs), as well as some eclipsing quiescent dwarf nova in the UV and/or optical, generating a spectrum that is bluer than the observed UV spectra indicating that a hot optically thick inner disk is considerably colder than predicted by theory \citep{2007Puebla,2010Linnell}. As a result, recent standard disk models have used a truncated inner optically-thick disk \citep[see,][]{2017Godon} that models the UV data adequately. Moreover, the Balmer jump (absorption due to H-Balmer series) is missing in CVs \citep{1989laDous,1998Knigge}. These observations in the UV and the optical indicate that the radial temperature profile in standard disk flows need to be modified, decreasing it in the inner disk and increasing it in the outer disk.  A recent model alternative to the basic standard accretion disk flow has used large magnetically-controlled zones that transfer mass and angular momentum outwards which flattens the spectral continuum in the optical  and UV to produce this observed effect \citep{2019Nixon}.  A more recent model \citep{2021Hubeny} produces similar effects by changing the temperature structure in the standard disk where a vertical temperature structure is not maintained and  most of the dissipation occurs from the heated surface layers (instead of mid-plane as in a standard disk flow)  without a need for inner truncation. However, these models can not accommodate and explain the extent and/or emission characteristics of the X-ray region (e.g., hard X-rays). The accretion flows in high-state CVs, NLs and DNe are well-explained in the context of radiatively-inefficient advective hot flows as opposed to standard optically-thick accretion flows, that form in the inner disk, which explain most of the complexities in the X-rays and other wavelengths \citep{2022Balman,2020Balman}. Shock formation in ADAF flows (advection dominated accretion flows) around accreting WDs has been calculated \citep{2021Datta} and a shock occurs around 1.3$\times$10$^9$ cm from the WD surface (about 1.5-2.5 R$_{wd}$ depending on WD mass) which may explain the hard X-ray emission.

\subsection{X-ray Characteristics of DNe}\label{DNX}

One of the earliest comprehensive studies on the X-ray emission  of accreting nonmagnetic CVs using the \eins\ IPC (0.2-4 keV) 
showed that they mostly emit hard X-ray emission in this range with 
luminosities $\le$ a few$\times$10$^{32}$ \lumcgs \citep{1985Patterson}.  
The DN in quiescence (low $\dot{\rm M}$ systems) were observed with
 subsequent X-ray telescopes (e.g., \rxte, \xmm, \cha, \suz, \swi, \nustar, and \nicer) and the results seem in accordance with one-dimensional
  numerical simulations of the optically-thin BLs in standard steady-state disks in quiescence with
low accretion rates \citep[e.g.,]{2002Szkody,2005Pandel,2006Rana,2009Ishida,2011Nucita,2011Balman,2017Wada,2023Kimura,2023Dutta,2023Dobrotka} 

On the other hand, studies of X-ray variability and inner disk structure of DN in quiescence (and outburst) 
reveal optically-thick  disk truncation and plausible formation of hot (coronal) flows
in the inner parts of the quiescent DN accretion disks \citep{2012Balman,2019Balman,2020Balman,2016Dobrotka,2023Dobrotka}.
Moreover, the optically thick disk truncation, as in X-ray binaries, is supported with simulations and light curve modeling of some DN \citep{2021Hameury}.
The aperiodic timing variability is used  to derive broadband noise power spectra
where the break frequencies in the characteristic power law red noise structure show the change  in and the diminishing Keplerian flow of a standard Keplerian disk into a sub-Keplerian advective hot flow.  The range of break frequencies is 1-6 mHz for quiescent dwarf novae, translating to a transition radius of  (3-10)$\times$10$^{9}$cm \citep{2012Balman,2020Balman}.  Detailed cross correlation analyses of DN light curves show 90-180 min lags of X-rays with respect to UV, indicating propagation delays and change of flow structure \citep{2012Balman} into advective hot flows within DN disks. SS Cyg shows 0.26-3.11 s optical lags (the X-rays) indicating X-ray reprocessing  and  irradiation in the disk \citep{2022Nishino}.  \citet{2018Aranzana} find 5 s soft-lags in the optical (red lags the blue) revealing reprocessing and thermal-timescale effects.  

The optically thin hard X-ray emission with virial temperatures in the X-ray emitting region plus the lack of soft X-ray emission at high states have been interpreted as the existence of ADAF-like radiatively inefficient advective accretion flows near the WD \citep[][and refrences therein]{2020Balman,2021Kimura,2022Balman}. From an analysis of Kepler data of MV Lyr (an NL; high state CV), it was proposed  that, in the very inner region (e.g., BL), the geometrically thin, optically thick disk is surrounded by a geometrically thick and optically thin disk --- the “sandwich” model \citep{2014Scaringi}. This sandwich model can explain the connection between X-rays and optical together with the detection of the same frequency (flickering) in both bands in other DN systems \citep[][and refrences therein]{2023Dobrotka}.

DN outbursts are brightenings of the accretion disks as a result of thermal-viscous instabilities summarized
in the DIM model \citep[Disk Instability Model;][]{2001Lasota,2004Lasota,2008Lasota,2020Hameury}. Other supporting calculations are
tidal thermal instability in DN outbursts \citep{1996Osaki} and enhanced mass transfer aiding the outbursts \citep{2000Hameury}.
During the outburst stage, X-ray spectra of DN differ from the quiescence since the
accretion rates are higher \citep[10$^{-10}$-10$^{-8}$ M$_{\odot}$ yr$^{-1}$;][]{2011Knigge}, 
the BL is expected to be optically thick, emitting EUV/soft X-rays \citep[see the X-ray reviews in][]{2006Kuulkers,2017Mukai}.
Such soft X-ray/EUV components with temperatures in a range 5-30 eV are detected only from  about six systems as compared to all DN
\citep[e.g.,][]{1995Mauche,2000Mauche,1996Long,2004Mauche,2009Byckling,2023Kimura}. On the other hand, as a more
frequently detected emission component (all DN), DNe show hard X-ray emission during the outburst stage at  lower flux levels and X-ray
temperature compared with the quiescence \citep[e.g.,][]{2010Collins,2011Fertig,2003Wheatley,2004McGowan,2009Ishida,2015Balman,2023Dutta,2023Dobrotka}.
Few DN show increased levels of hard X-ray emission \citep[GW Lib \& U Gem:][]{2009Byckling,2006Guver,2021Takeo}.
The total X-ray luminosity during outburst is 10$^{30}$ -10$^{34}$ erg s$^{-1}$ where the upper level is observed when the soft component is detected.
Grating spectroscopy of the outburst data (SS Cyg \& U Gem) indicate  large widths for lines with velocities in excess of 1000 km s$^{-1}$
mostly of H and He-like emission lines (C, N, O, Ne, Mg, Si, Fe, ect.) \citep{2004Mauche,2006Rana,2008Okada}.
A characteristic of some DN outburst light curves are the
UV and X-ray delays of several hrs up to 2 d, in rise to outburst (with respect to optical) indicating possible disk truncation \citep{1994Meyer,1999Stehle,2003Schreiber}.
During outburst no X-ray  eclipses are detected in the eclipsing systems (particularly of soft X-ray emission)
or no distinct orbital variations are seen \citep[e.g.,][]{1999Pratt,2009Byckling}.

\subsection{Z Chamaeleontis (Z Cha)}

Z Cha is a dwarf nova of SU UMa sub-class showing normal and superoutbursts 
\citep{1978Bateson,1979Smak}. The importance of Z Cha amoung the DN lies in the complexity of its eclipse structure which is
a superposition of total eclipses of both the WD and the hotspot \citep{1974Warner,1979Bailey}. Other similar DNe 
are OY Car, HT Cas, and V2051 Oph \citep{1990Woodhorne}.
The detailed eclipse structure reveals the system parameters. The binary is highly inclined with i=81.6-81.9 degrees 
\citep{1986Wood,2007Smak}, thus deep eclipses
affect the optical light curves. Z Cha has an orbital period of 1.78 hrs with a WD mass of 0.54-0.84 M$_{\odot}$ and
a secondary mass of 0.081  M$_{\odot}$ with a ratio of secondary to primary mass, q=0.2  \citep{1986Wood,1988Wade,2002Baptista,2007Smak}.
A distance of 97 pc has been long accepted for Z Cha \citep{1986Wood}. The GAIA archive\footnote{https://gea.esac.esa.int/archive}\  yields a distance of 120$\pm$33 pc using the parallax measurement  which is in agreement with the accepted value. The X-ray properties observed with $ROSAT$ PSPC are a temperature kT=2.6-19 keV and an N$_H$=(1.3-2.3)$\times$10$^{20}$ cm$^{-2}$
translating to a luminosity of 2.5$\times$10$^{30}$ erg s$^{-1}$ (bolometric) at an 
accretion rate of $\sim$10$^{-12}$  M$_{\odot}$\ yr$^{-1}$ \citep[see][]{1997Teeseling}.
The quiescent phase \xmm\ data of Z Cha 
has been previously presented in \citet{2011Nucita} and is reanalyzed in this paper and the differences in results will be discussed in the
analysis and results section. The archival \xmm\ data of Z Cha in the outburst stage is analyzed for the first time and discussed in this paper.

\section{The Observation and data}\label{sec:obs}

The \xmm\ Observatory \citep{2001Jansen} has three 1500 cm$^2$
X-ray telescopes each  equipped with an European Photon Imaging Camera (EPIC)
at the focus where two of them have metal oxide semiconductor  (MOS) CCDs
\citep{2001Turner} and the last one uses pn CCDs \citep{2001Struder}
for data recording. About half the X-rays
are diverted, by reflection grating arrays (RGA), to the reflection
grating spectrometers \citep[RGS;][]{2001denHerder} which
provide high resolution ($\lambda$/$\Delta$$\lambda$ $\sim$100-800) X-ray spectroscopy
in the 0.33-2.5 keV range. The EPIC cameras acquire data in
the 0.1-15 keV range with a field of view (FOV) $\sim$ 30 arc minutes
diameter. The optical monitor (OM), a photon-counting instrument,
is a co-aligned 30-cm optical/UV telescope, providing a
possibility of observing  simultaneously in the X-ray and optical/UV wavelengths
\citep{2001Mason} with an imaging capability in three broad-band ultraviolet
filters and three optical filters, spanning 1800-6000 \AA.

Z Cha was observed (pointed observation) by \xmm\  on two different occasions: in quiescence (OBS ID=0205770101), and in outburst (OBS ID=0306560301). 
The pointed observation during quiescence was obtained between 2003 December 19 UT 20:46:36
and 2003 December 21 UT 00:52:53 with a duration of $\sim$ 100 ks.
A thin optical blocking filter was used with all the EPIC cameras. The
pn instrument was operated  in the full frame imaging mode. 
The MOS1 and MOS2 CCDs were used in the large window imaging mode. 
Z Cha was detected with a count rate of 1.1$\pm$0.004 c s$^{-1}$, 0.36$\pm$0.002 c s$^{-1}$
and 0.34$\pm$0.002 c s$^{-1}$ using EPIC pn, MOS1 and MOS2 instruments, respectively. The combined
RGS (1$+$2)  spectrum has a count rate of 0.024$\pm$0.0006 c s$^{-1}$.
The pointed observation during outburst was obtained between 2005 September 30 UT 01:45:43
and 2005 October 01 UT 01:56:20 with a duration of $\sim$ 90 ks. 
A medium  optical blocking filter was used with all the EPIC cameras
where the pn data was obtained in the full frame imaging mode and the MOS CCDs were operated with the large window imaging mode.
Z Cha was detected with a count rate of 0.31$\pm$0.002 c s$^{-1}$, 0.081$\pm$0.001 c s$^{-1}$
and 0.076$\pm$0.001 c s$^{-1}$ using EPIC pn, MOS1 and MOS2 instruments, respectively. The combined
RGS (1$+$2)  spectrum has a count rate of 0.0093$\pm$0.0006 c s$^{-1}$.

We analyzed the pipeline-processed data using Science Analysis Software (SAS)
version 17.0.0-21.0.0\ . The pipeline-processed events were also checked using the SAS task {\it epproc} 
and {\it emproc} to make sure there are no differences relating to software or processing versions. The RGS spectra and response files were created with the SAS task {\it rgsproc}.
Data (single- and double-pixel events, i.e., patterns 0--4 with Flag=0 option for pn and patterns $\le$12 for MOS) 
were extracted from
a circular region of radius 35$^{\prime\prime}$-40$^{\prime\prime}$ for pn, MOS1 and MOS2
in order to perform spectral analysis. The background events were extracted from a
source-free zone close to the source and normalized to the same extraction area. The EPIC spectra were created using
the task {\it especget}. The RGS spectra were created with the {\it rgsproc} task and co-added utilizing  the
SAS task  {\it rgscombine}. {\it especget}  and  {\it rgsproc} tasks also create the necessary response and the ancillary files
for the analysis. The task descriptions and documentations are available on-line at the
ESA website\footnote{http://xmm2.esac.esa.int/sas/}.

For our timing analysis purposes, we utilized the data collected with the
EPIC cameras in the large window imaging mode, separetely, using the pn and MOS 1 and 2,
and the OM data using the fast imaging mode ($\ge$0.5 s time
resolution)  with the B filter (3980-4334 \AA). 
The time resolution of
imaging modes of pn CCDs are 70 ms which is more than adequate for our timing analysis.
We checked/cleaned the pipeline-processed/produced event files from existing flaring
episodes for both of the Z Cha observations. 
The OM data were analyzed using  {\it omfchain} with 0.5 sec time resolution (bin-time).
EPIC light curves were created with the  task {\it evselect}. Identical region files used in the spectral analysis were used to extract source and background light curves. 
Further reductions and analysis of spectra and light curves were performed using  HEASoft\footnote{https://heasarc.gsfc.nasa.gov/docs/software/heasoft/}\ (version 6.24-6.32) using XRONOS or XSPEC software within the  HEASoft distributions.

\section{Temporal analysis and optical eclipse times in the X-ray data}\label{sec:temp}

We extracted light curves from the quiescent and outburst \xmm\ EPIC data sets between 0.2-10 keV using the
SAS task {\it evselect} at 0.01 sec resolution for quiescence and outburst.  Circular regions with a range of radius
35$^{\prime\prime}$-40$^{\prime\prime}$ 
were used for the photon extraction to calculate both the source and the background light curves
(devoid of other contaminating sources). Finally, background-subtracted light curves were attained using the XRONOS 
task  {\it lcmath}.  Barycentric correction was applied  on both of the event files (outburst and quiescence). 
Next, we folded the EPIC pn, MOS1-2 light curves in outburst and quiescence over the orbital period using the ephemerides
by \citet{2002Baptista}; T$_0$=(BJD)2440264.68070($\pm$4)$+$0.0744993048($\pm$5)$\times$E days. 
We note here that \citet{2011Nucita} presents
a quiescent optical B-band light curve analysis for Z Cha in detail together with the eclipses and the EPIC temporal analysis (quiescence) which we omit here. 

\begin{figure*}[ht]
\begin{center}
\vspace{-0.2cm}
\includegraphics[height=1.8in,width=3.9in,angle=0]{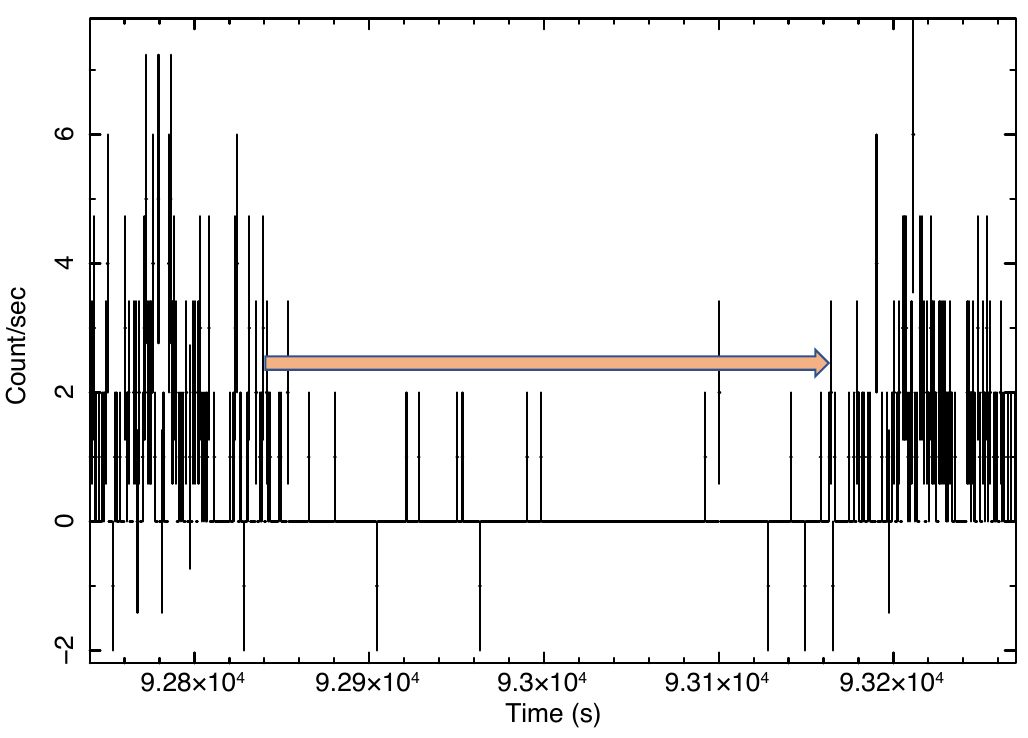}\\
\vspace{-0.1cm}
\hspace{0.2cm}
\includegraphics[height=1.8in,width=4.1in,angle=0]{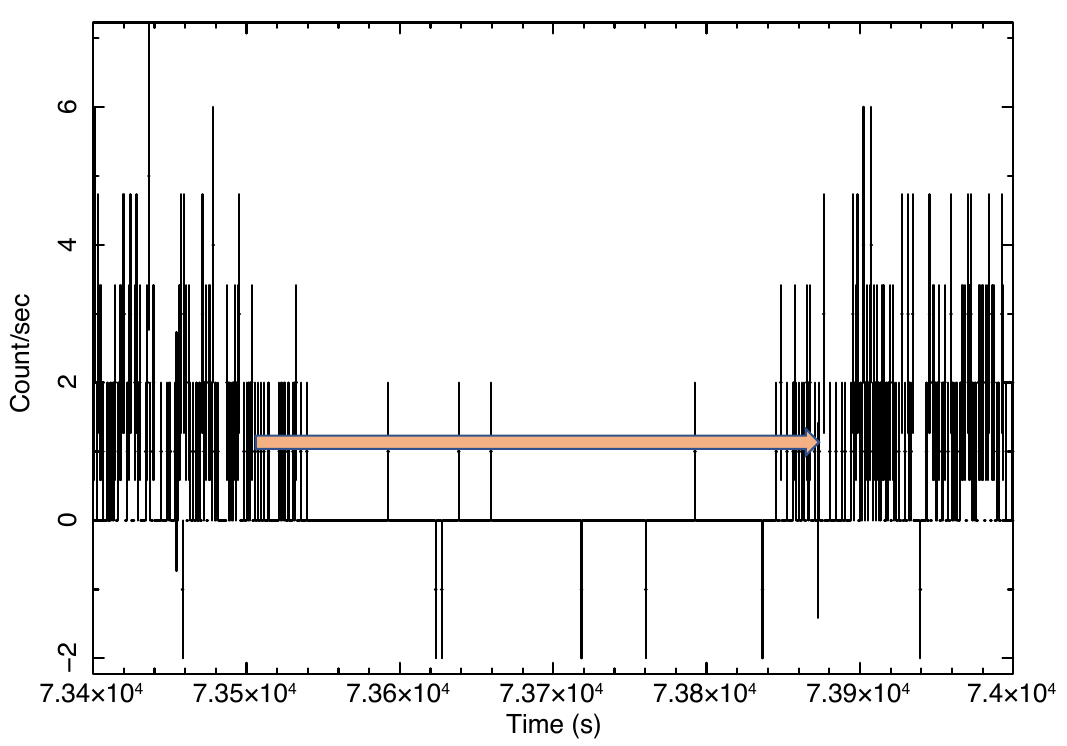}\\ 
\vspace{-0.1cm}
\includegraphics[height=1.75in,width=3.9in,angle=0]{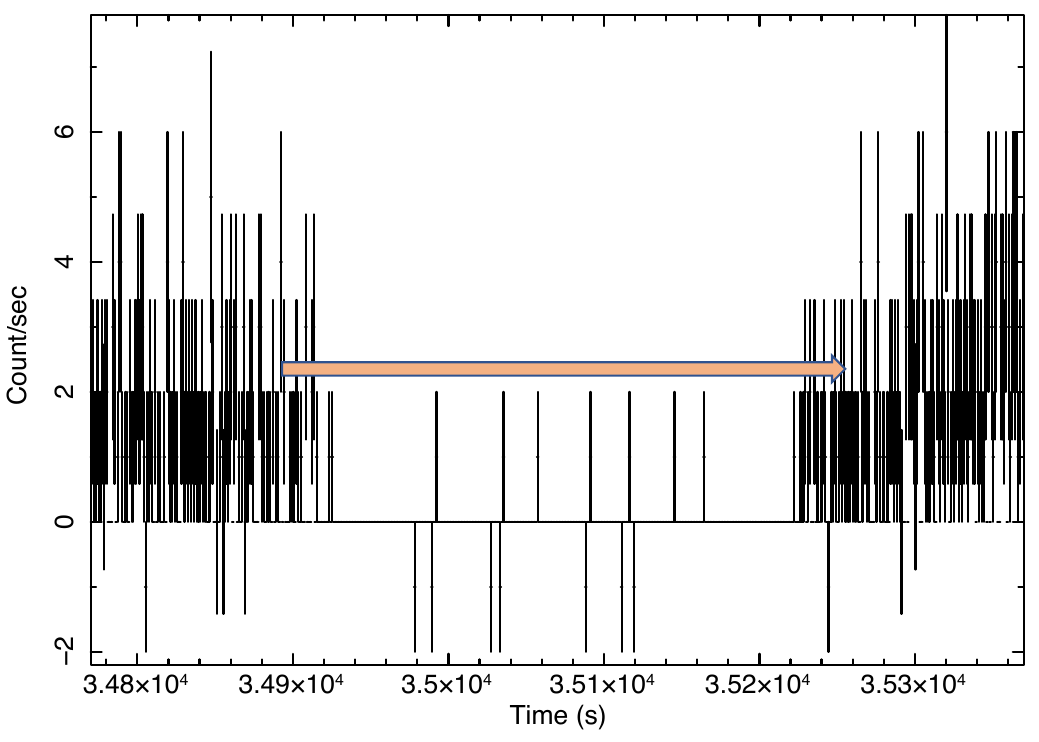}\\
\vspace{-0.1cm}
\includegraphics[height=1.75in,width=3.9in,angle=0]{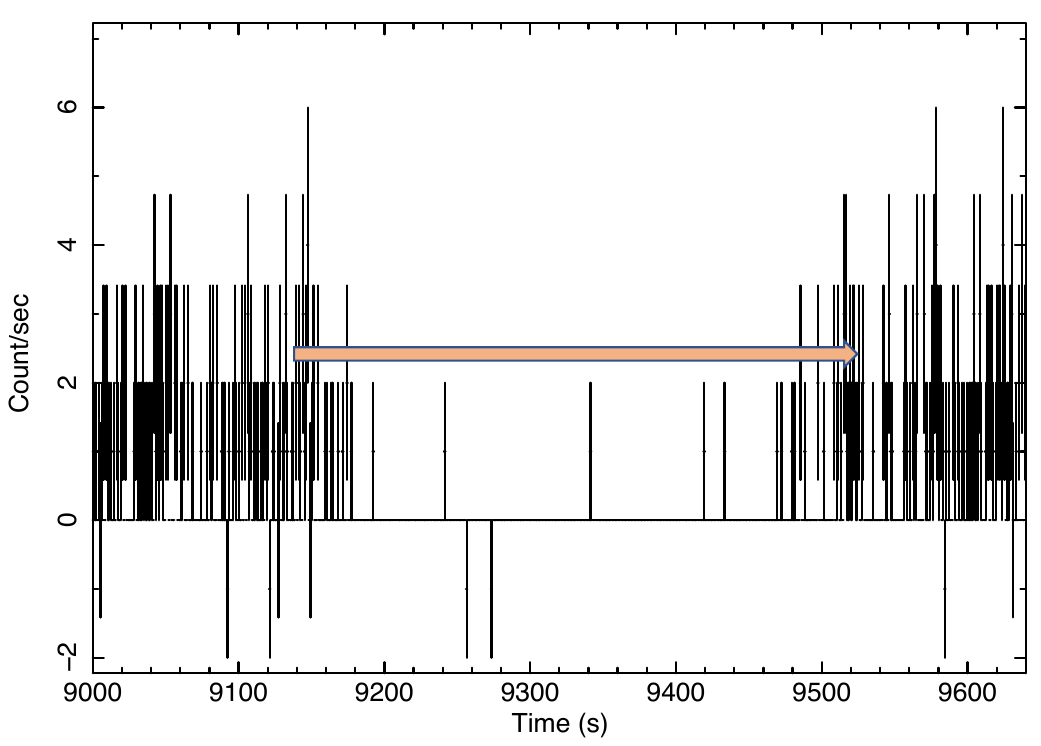}\\
\vspace{-0.1cm}
\hspace{-0.2cm}
\includegraphics[height=1.75in,width=4.0in,angle=0]{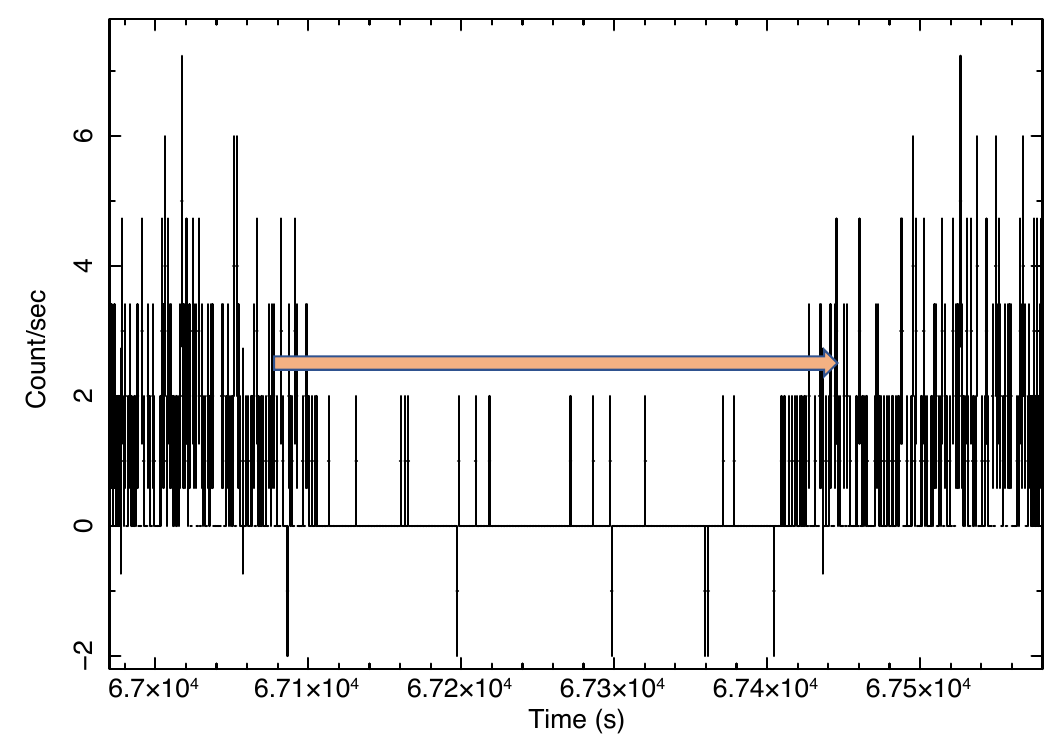}

\vspace{-0.1cm}
\caption{The quiescent X-ray light curve (LC) of Z Cha obtained using the \xmm\ EPIC pn data.
The figure shows the locations of the X-ray/optical eclipses in the X-ray light curve.
A bin size of 1 sec is used in accordance with the source count rate of 1.1 c s$^{-1}$ which yields approximately one photon per bin
in the figure.  A light orange color bar, indicating the X-ray eclipse duration  of 324 s (from mid-ingress to mid-egress) calculated by \citet{2011Nucita}, is approximately overlaid on the plots.\label{fig:eclipse}
}
\end{center}
\end{figure*}

\citet{1984Cook} have performed a detailed analysis of the optical band eclipses of Z Cha and found that the WD eclipse is around phases 0.97-1.03 and
the hot spot eclipse is around 0.99-1.1. We underline that the occultation of the accretion impact zone and the occultation
of the WD overlaps closely in phase for Z Cha which means that the  accretion impact zone is in the line of sight towards the WD at the times
of the eclipses.   
\citet{2011Nucita} find an average EPIC rate of 0.7$\pm$0.1 c s$^{-1}$ and a rate of 0.033$\pm$0.003 c s$^{-1}$ 
during the eclipses. They calculate that the X-ray ingress and egress are about 40 and 54 s, respectively, with an eclipse duration (mid-ingress to mid-egress)
of 324$\pm$9 s. However, the analysis of several consecutive eclipses with the ROSAT HRI (0.2-2.4 keV) yielded eclipse durations of
350-520 s which are longer than \xmm\ \citep{1997Teeseling}. Our reanalysis of the EPIC pn light curve for the source rate during the inferred eclipse times by \citet{2011Nucita}
reveal that in most cases the source is not detected with only few photons in the given time interval except the very beginning and the end of this time interval. 
Figure \ref{fig:eclipse} shows the
quiescent EPIC pn light curve with a high time-resolution of 1 s, at the times of the eclipses. This bin time is particularly chosen so that 
given the 1.1 c s$^{-1}$ EPIC pn count rate, there is one photon per bin in the Figure.  We find that the eclipse times
contain  variable  number of X-ray photons around 4-13 (except for the very beginning and end of the eclipse durations which are likely associated with the ingress and egress).
An inspection of the 13 optical eclipses in the B-band light curve (\xmm\ OM) yields 280-300 s eclipse durations assuming width of the time interval at the half of the brightness that is out of eclipse on the egress side.
We see that the X-ray source is not detected at all during these times.
As a result, we take here that the disappearance of the source for a short time interval given the low count rate of Z Cha may result from
occultation of the X-ray emission region and/or from any absorbers that reside on the disk or the accretion impact zone (given the high inclination). 
We will elaborate on this in our results and the Discussion section.  Therefore, we do not exclude eclipse times from spectral analysis since the source is not detected in the X-rays during these times (see Figure \ref{fig:eclipse}).

\begin{figure*}[ht]
\begin{center}
\includegraphics[height=2.in,width=2.7in,angle=0]{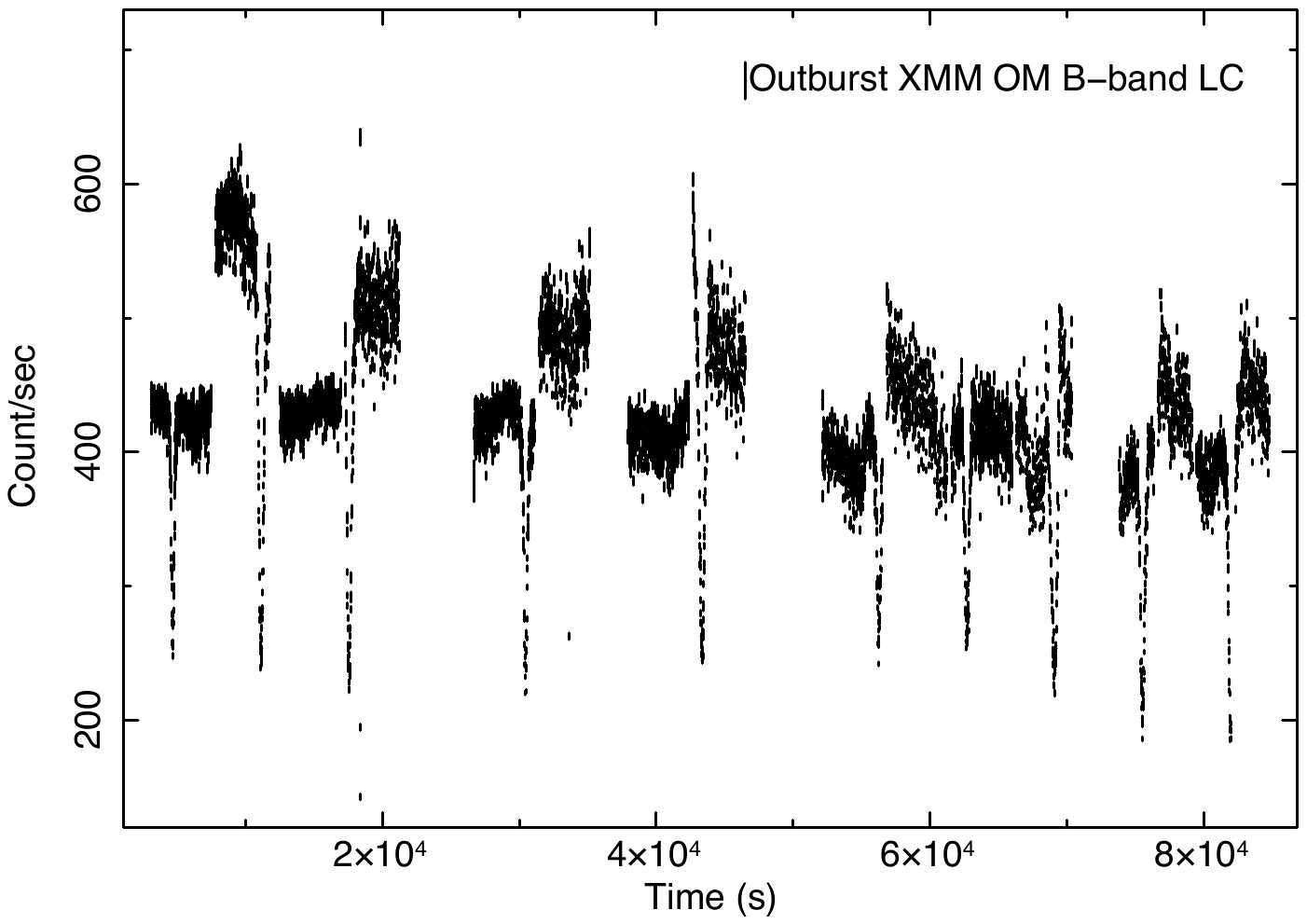}
\vspace{-1cm}
\includegraphics[height=2.in,width=2.7in,angle=0]{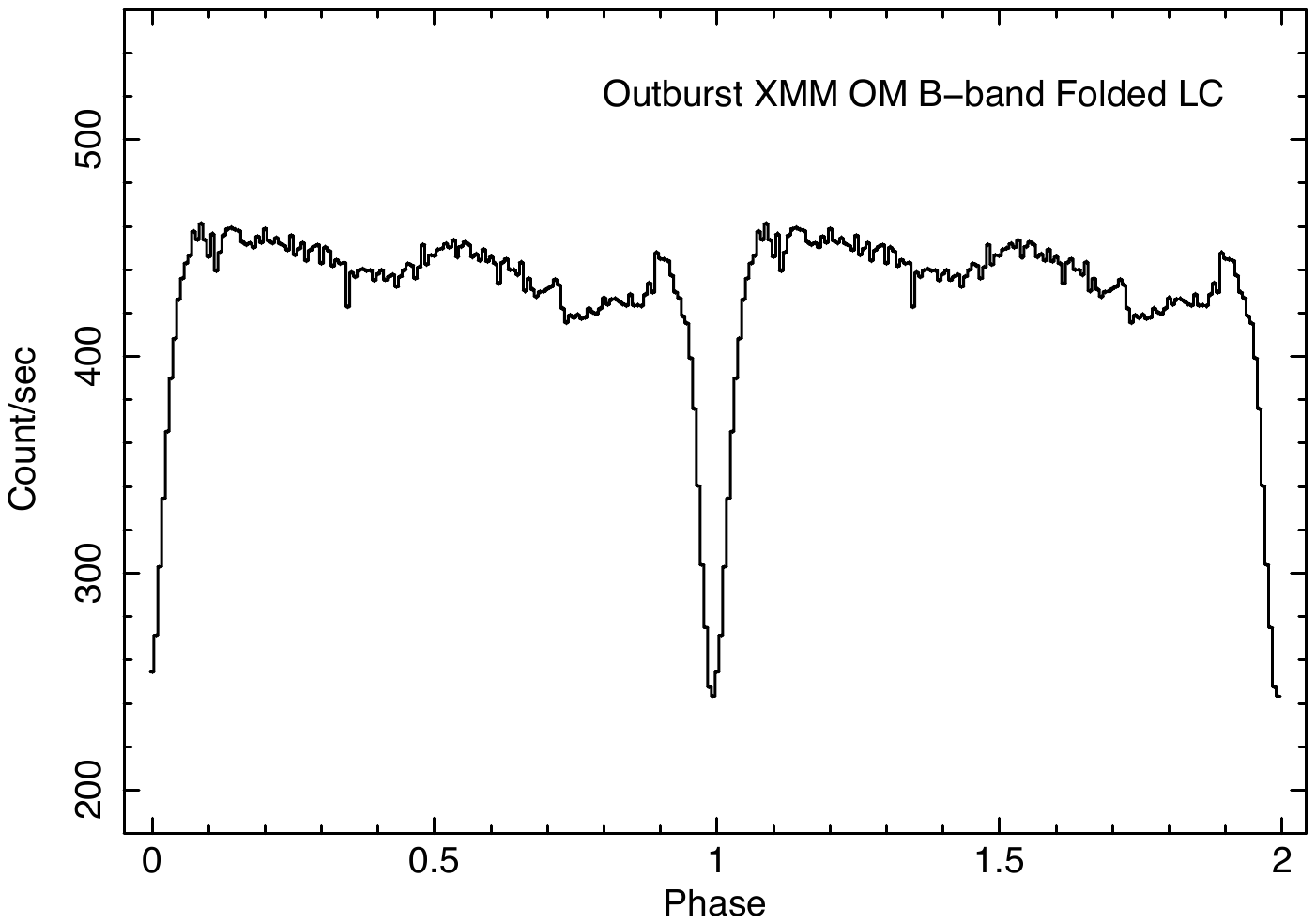}\\
\vspace{-1cm}
\includegraphics[height=2.in,width=2.7in,angle=0]{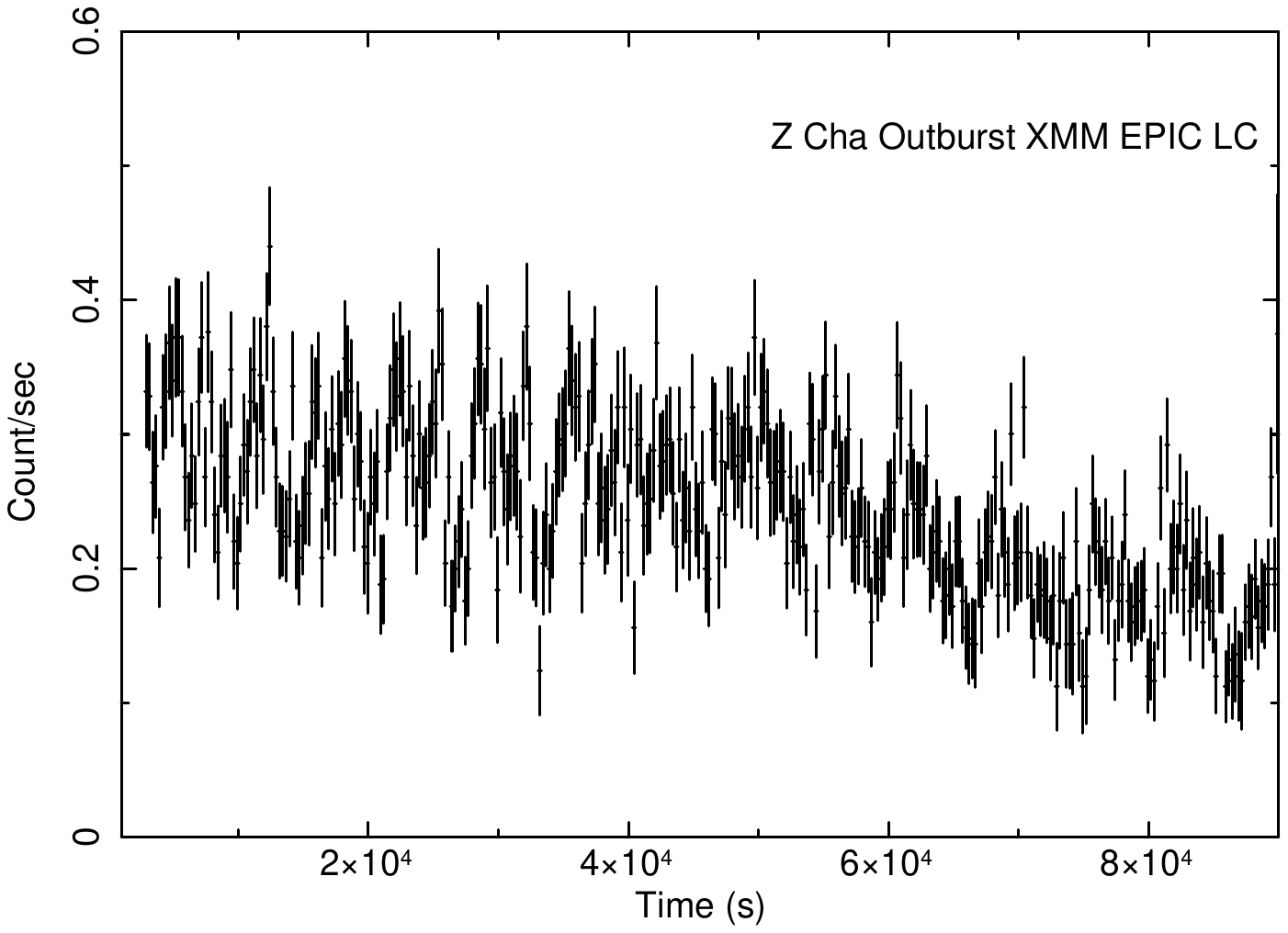}
\includegraphics[height=2.in,width=2.7in,angle=0]{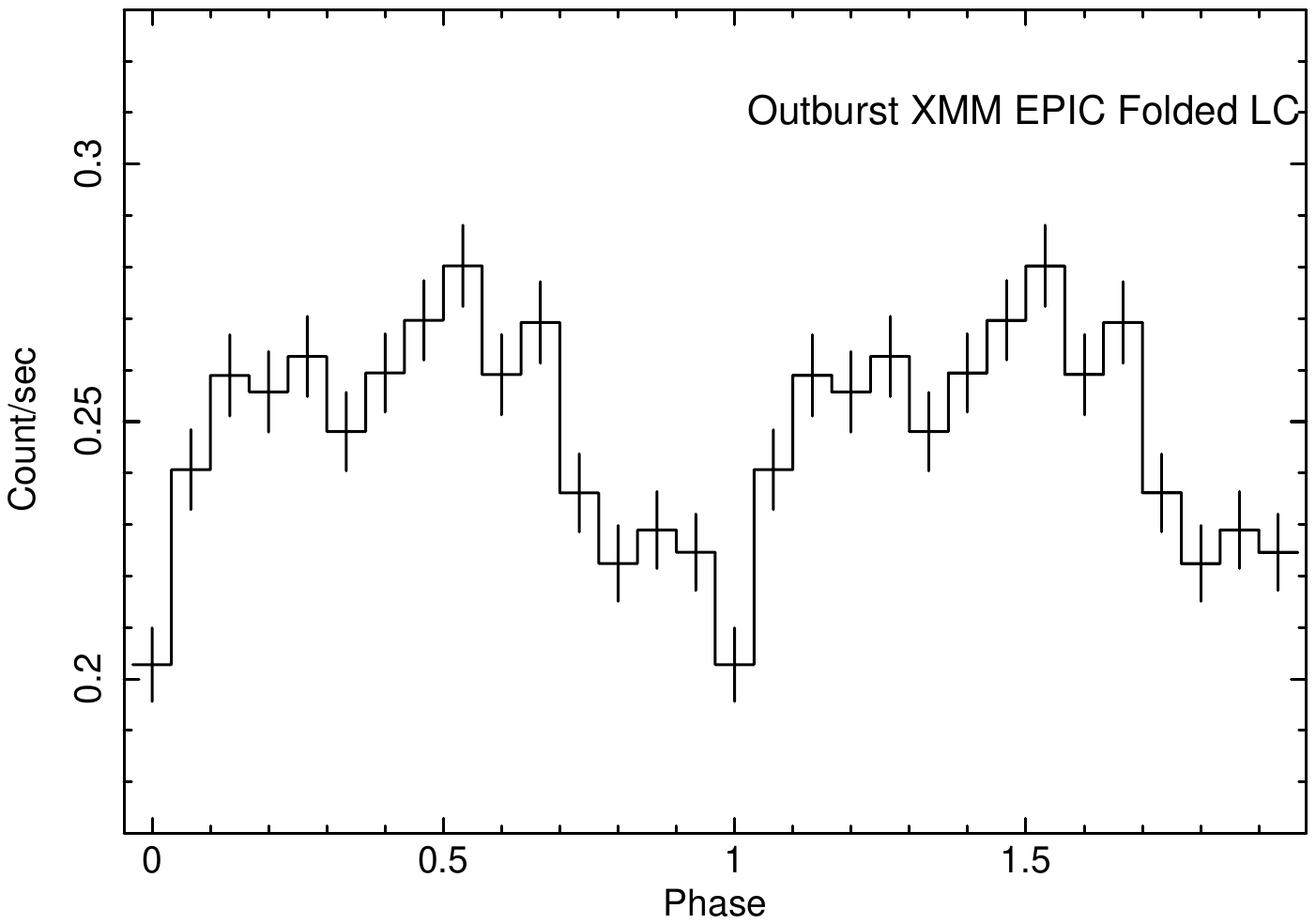}
\vspace{-0.9cm}
\caption{Upper left is the optical OM B-band and lower left is the X-ray light curve (LC)  as observed by EPIC using \xmm. 10 s time binning is used for the B-band LC while 250 s binning is used for the X-ray LC. Right hand panels show the LC in the outburst folded at the orbital period of Z Cha (see text for Ephemerides).  150 and 15 phase bins are used for the B-band and X-ray EPIC mean LC, respectively. \label{fig:lc}
}
\end{center}
\end{figure*}

Figure \ref{fig:lc}, upper left, presents the optical B-band light curve binned at 10 s which shows deep variations at the eclipse times (no full eclipses) during the  
outburst observation. The upper right hand side is the mean B-band light curve folded at the orbital period with the Ephemerides given in the first paragraph of this section. The lower  left  hand panel is the EPIC light curve (250 s bin-time is used) obtained during the outburst and the bottom panel  displays the folded EPIC light curve at the orbital period using 15 phase bins (during the outburst). We find a weak orbital variation with a modulation depth of 13$\%$ ([max-min/max+min]$\times$100). The modulation depth for the B-band data is about 32$\%$ using the mean light curve. However, this value changes between 43\% and 28\% in the B-band light curve. 
We note that combined EPIC light curve was attained
using background-subtracted  pn, and MOS1,2 light curves. We have checked to see if any asynchronicity affected our folding procedure. The individual instrument light curves are slightly different, but our folded EPIC light curve resembles the EPIC pn light curve very closely as expected since the pn has  higher sensitivity (3-4 times more rate) compered with MOS1,2.
The timing analysis (for quiescence) using power density spectra (PDS)  for broadband noise analysis, 
derivation of the break frequencies and cross-correlation analysis are presented in \citet{2020Balman} which we elaborate in the Discussion section.

\section{Spectral Analysis and Results}\label{sec:spec}

\subsection{Analysis of EPIC data in quiescence}\label{sec:eqsp}

The background and source spectra, response and ancillary files were generated for Z Cha as described in  Sec. \ref{sec:obs}.  
The EPIC spectra were grouped with minimum counts of 90, 65, 70  in each spectral bin for pn, MOS1 and 2, respectively, to acquire good \chisq\ statistics.
 The spectral analysis is performed via simultaneous/joint analysis of all three spectra within the XSPEC software (for references and model descriptions see \citealt{1996Arnaud}\footnote{https://heasarc.gsfc.nasa.gov/xanadu/xspec/manual/Models.html}). During the entire spectral analysis, constant multiplicative models (multipticative constants) have been incorporated in the fits to account for the cross-normalization calibration between spectra from different detectors onboard \xmm.
To account for the interstellar and/or intrinsic absorption in the broadband X-ray spectra, the {\it tbabs} model \citep{2000Wilms} was utilized in the fitting procedures (abundances set to abund=wilm \citealt{2000Wilms}).  
However, we used other absorption models in the fitting procedure, {\it pcfabs} and {\it zxipcf}, which are partially covering cold and ionized absorber models, respectively, that yielded better fits when used along with the plasma models.
The joint X-ray spectra (pn, MOS1, MOS2)  are fit with the multi-temperature isobaric cooling flow type of plasma emission model CEVMKL in XSPEC \citep{1995Liedahl,1996Singh}\ with switch=2 where the CEVMKL  interpolates  using the  APEC model and ATOMDB\footnote{http://atomdb.org}\ database.  For comparison and to understand the nature of the flow, we also utilized the VNEI\footnote{https://heasarc.gsfc.nasa.gov/xanadu/xspec/manual/node195.html} (non-equilibrium ionization plasma model in XSPEC) which uses the NEIvers.3.0.9 plasma code with the ATOMDB database.

Quiescent DNe show several  H- and He-like emission lines from
elements N to Fe, and some Fe L-shell lines, revealing
the existence of  hot, optically thin X-ray emitting plasma in these systems \citep[see reviews of][]{2012Balman-mem,2020Balman},
as the accreting material settles onto the WD. This region is found 
to form a structure with continuous temperature distribution in the X-rays (as in a cooling flow plasma).
The emission measure at each temperature is proportional to the
time the cooling gas remains at this temperature.
This emission represents a collisionally-ionized plasma in thermal equilibrium between electrons and ions.

\begin{figure*} 
\includegraphics[height=5.5cm,width=6.5cm,angle=0]{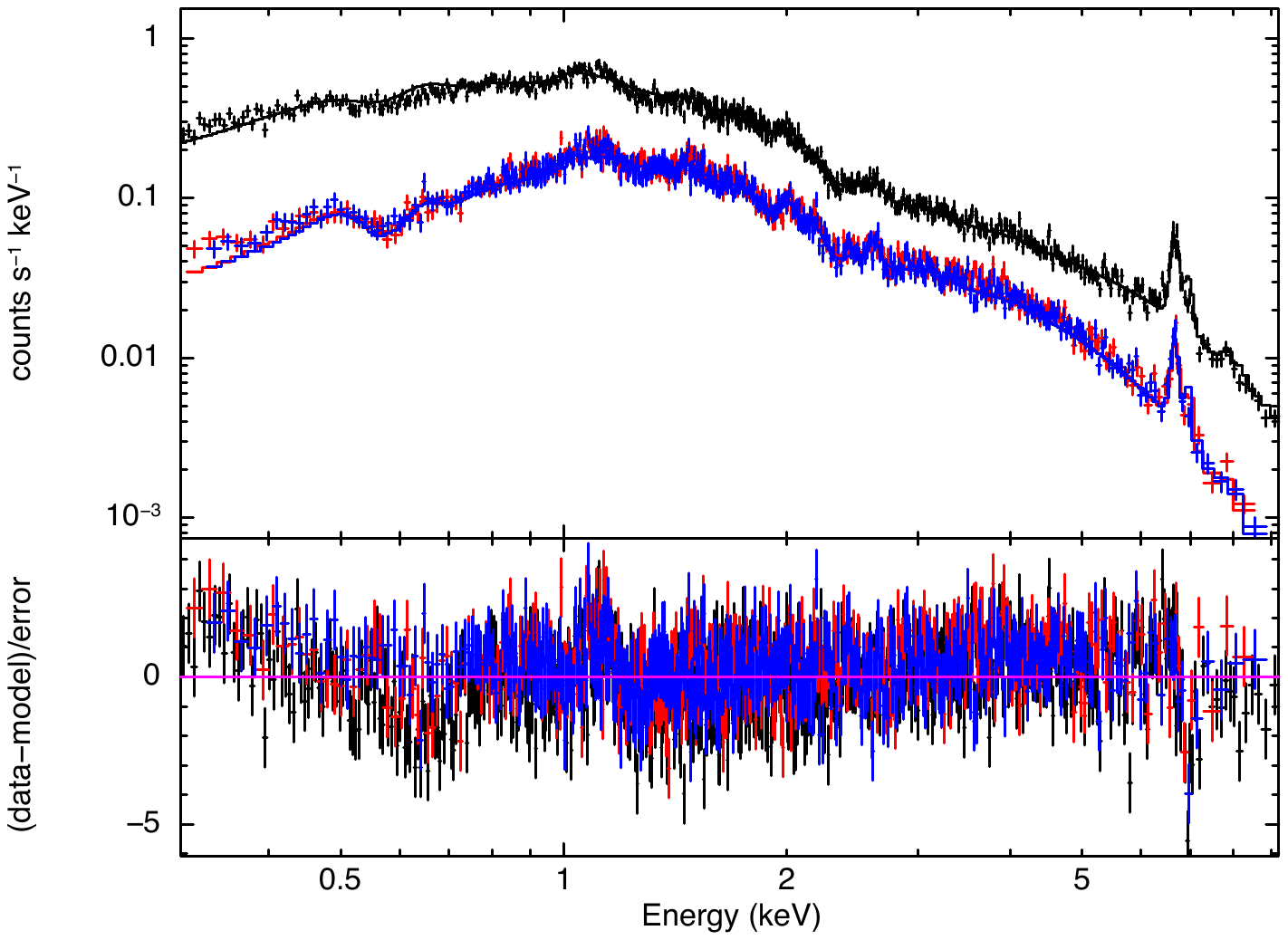}
\hspace{-0.7cm}
\includegraphics[height=5.5cm,width=6.5cm,angle=0]{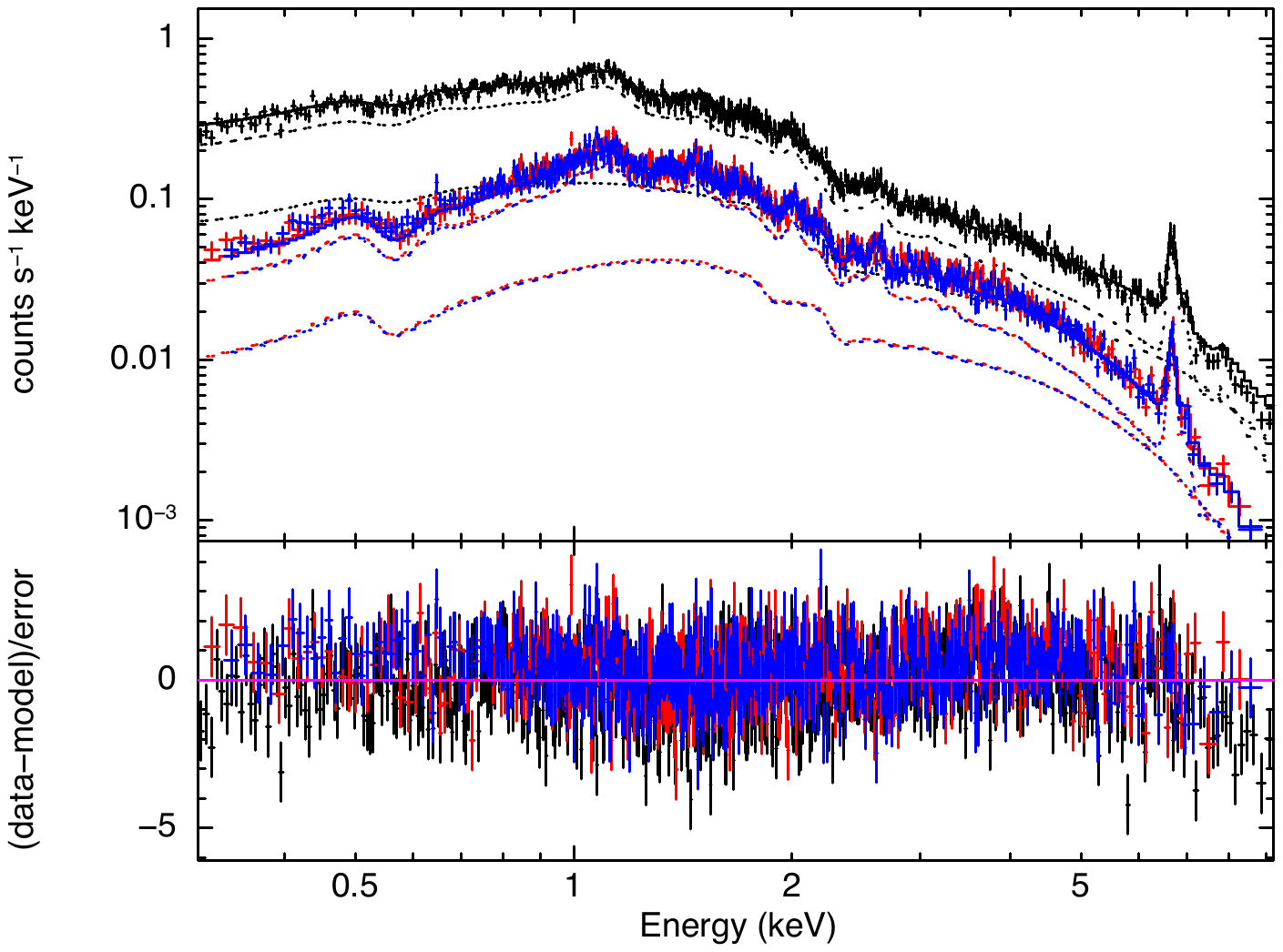}
\hspace{-0.7cm}
\includegraphics[height=5.5cm,width=6.5cm,angle=0]{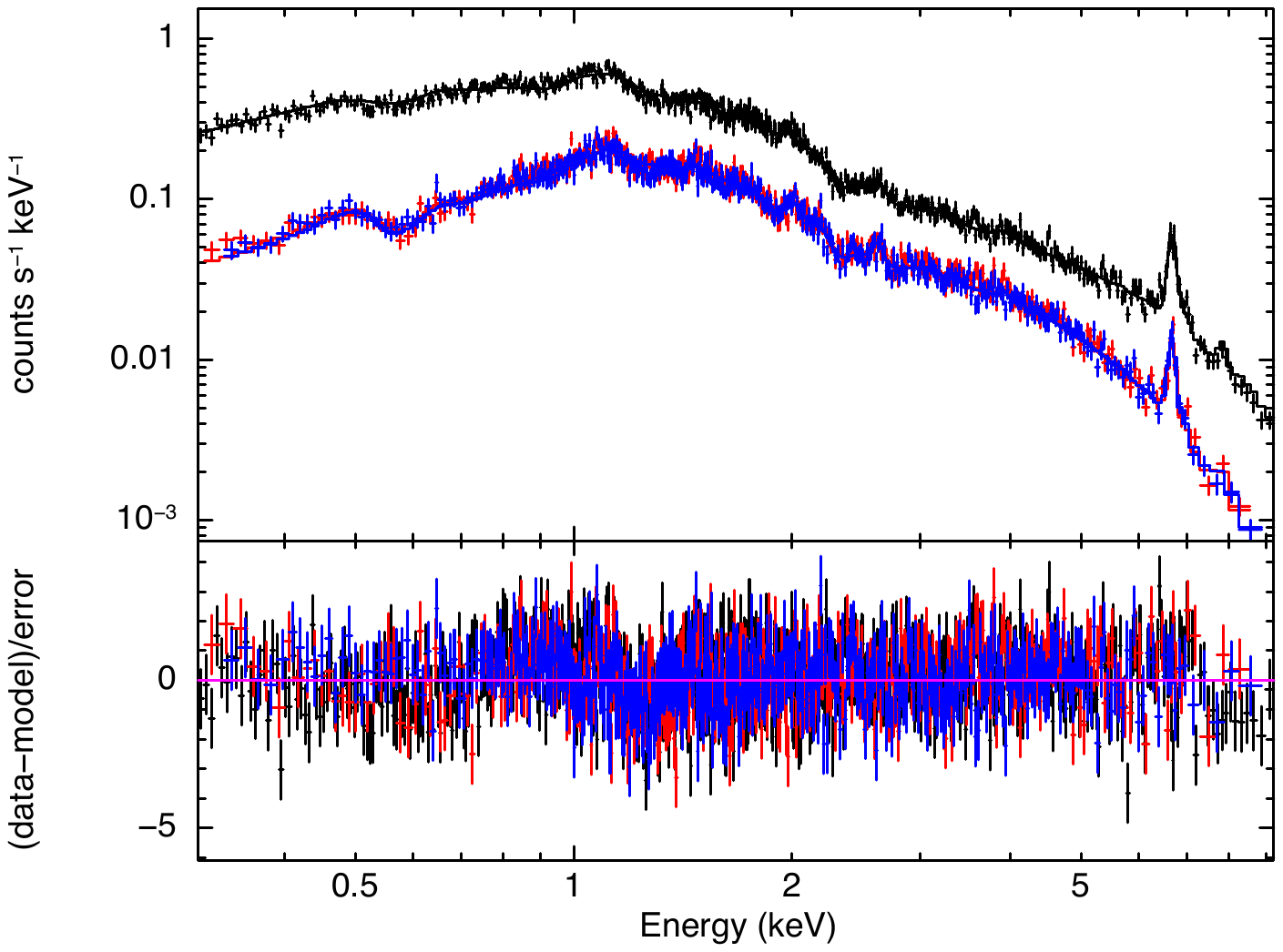}
\caption{
The quiescent EPIC pn and MOS1,2 spectra of Z Cha simultaneously fit with the ($tbabs\times$CEVMKL) model 
(on the left). The middle panel is the EPIC spectra 
simultaneously fit with the composite model of  ($pcfabs\times$(CEVMKL$+ power$)). The right hand panel shows the
spectra fit using the composite model  ($pcfabs\times zxipcf\times$VNEI). In the figure the 
dotted lines show the contribution of different model components.
The lower panels show the residuals  in standard deviations (in sigmas). \label{fig:qsp} } 
\end{figure*}

As a first step, we have fitted the quiescent EPIC pn, and MOS1-2 spectra simultaneously with the simple model
($tbabs\times$CEVMKL) and the $\alpha$ parameter that describes the index of the temperature distribution 
for the CEVMKL model was set free since a \rchisq\ value below 2.0 (i.e., acceptable fit) could not be achieved otherwise.
The spectral parameters are given in column QFit-1 of  Table \ref{tab:sp1} and all the other fit results  QFit-(1-6) are tabulated here, as well.
The spectrum fit with this model is displayed in Figure \ref{fig:qsp} along with some other example fit spectra from the sample QFit-(1-6).
The range of neutral hydrogen column densities derived from QFit-1,
are checked using standard tools that calculate neutral hydrogen column in the line of sight:
(1) COLDEN\footnote{http://cxc.harvard.edu/toolkit/colden.jsp} (using Dickey \& Lockman 1990); 
(2) {\it nhtot}\footnote{http://www.swift.ac.uk/analysis/nhtot/index.php}(using Willingale et al. 2013). 
Willingale et al. (2013) calculate the molecular hydrogen column density, N(H2), in the Milky Way Galaxy
using a function that depends on the product of the atomic hydrogen column density, N(HI), 
and dust extinction, E(B-V), with the aid of the 21-cm radio emission maps and the
\swi\ GRB data. Using these software, we found \nh\ in a range 
1.08-1.36$\times$10$^{21}$ cm$^{-2}$. Our \nh\ value measured with QFit-1 in Table \ref{tab:sp1},  is consistent with interstellar \nh.

Next, we used a different absorption model and  additional components to improve QFit-1 since it had relatively high  \rchisq. 
QFit-2 was constructed with  a partial covering absorber model that could reveal the existence of intrinsic absorption in the system  
since Z Cha is a high-inclination eclipsing system. 
The {\it pcfabs} measured a partial covering of 77.0$^{+5.0}_{-2.0}$\%\ with about twice the interstellar value of 
cold absorption in the line of sight. We calculated that QFit-2 improved over QFit-1  
 using an F-TEST\footnote{https://heasarc.gsfc.nasa.gov/xanadu/xspec/manual/node82.html} within XSPEC that compares the \chisq values and degress of freedom (d.o.f.) of the two different fits.  The F-TEST probability (1--confidence level) was 2.8$\times$10$^{-24}$ that yields an improvement over 10$\sigma$. 
 
 In addition, we tested the possible of existence of a power law 
component which may be due to scattering in the system as was found for some nova-likes and DNe  in outburst \citep{2014Balman,2015Balman,2022Balman,2023Dobrotka}. We used the model ($pcfabs\times$($powerlaw$+CEVMKL)) for QFit-3  which yielded a better  \rchisq\ that strongly suggested existence of a power law component in quiescence. The improvement of QFit-3 over QFit-2 was around  10$\sigma$ significance level, as well (F-TEST probability was 2.4$\times$10$^{-14}$). 
The spectrum fit with this model is displayed in the middle panel of Figure \ref{fig:qsp}.
The QFit-(1-3)  indicate a collisionally-ionized cooling flow type of plasma (that has a temperature distribution) with kT$_{\rm max}$ in a range 9.8-13.0 keV.
The $\alpha$\ parameter (power law index) of the temperature distribution is found to be different than 1.0, between 1.7-2.5 which indicates hard X-ray excess. 

Furthermore, we tested the consistency of the EPIC spectra with nonequilibrium ionization plasma emission using the VNEI model in XSPEC. QFit-4 is similar to QFit-3 with only a replacement  of the CEVMKL model with the VNEI model. The fit result does not improve upon  QFit-3. 
However, we note here that the d.o.f. used in the fits are generally high, as a result of simultaneous fitting of the three EPIC spectra. 
 This yields significant differences in fits even for small variation in \chisq which may not reflect physical differences so that QFit-3 and QFit-4 are equivalent for EPIC analysis with moderate spectral resolution.  
 QFit-5 and QFit-6 are constructed to test the validity of a power law component of emission and existence of complex absorption as in cold partial covering absorbers or warm photoionized absorbers.  
 The fits are performed with a composite model of  ($pcfabs\times zxipcf\times $VNEI), QFit-5 and ($pcfabs\times zxipcf\times$CEVMKL), QFit-6. 
 The spectrum fit with QFit-5 is shown in the right hand panel of Figure \ref{fig:qsp}.
 The \rchisq\ values improve upon the fits with the power laws  (when tested with F-TEST) suggesting that there are complex absorption components on the existing plasma emission of collisional equilibrium or nonequilibrium origin. The power law model may not be physical and may only be  providing modification of the continuum reducing \chisq. We note here that as mentioned before, the power law components are detected in the high states. We find that the N$_{\rm H}$ values of the partial covering of the cold absorber  are consistent with interstellar
 absorption in these fits whereas the ionized warm absorber shows a lower covering fraction (54.0$^{+3.0}_{-8.0}$\% for VNEI  and  36.0$^{+1.0}_{-1.0}$\%  for CEVMKL) with equivalent hydrogen column density of  4.80$^{+1.10}_{-0.06}$$\times$10$^{22}$ cm$^{-2}$ for VNEI and  larger for the fit with the CEVMKL model, 10.3$^{+0.7}_{-0.7}$$\times$10$^{22}$ cm$^{-2}$. These values are a factor of 40-80 times more than the intrinsic and/or line of sight cold absorption.
 QFit-5 and QFit-6 indicate a plasma temperature of 8.4$^{+0.2}_{-0.2}$ keV for the VNEI model and kT$_{\rm max}$=7.30$^{+0.13}_{-0.11}$ keV for the CEVMKL model fits. 
 
\begin{figure}
\begin{center}
\includegraphics[height=2.3in,width=2.35in,angle=0]{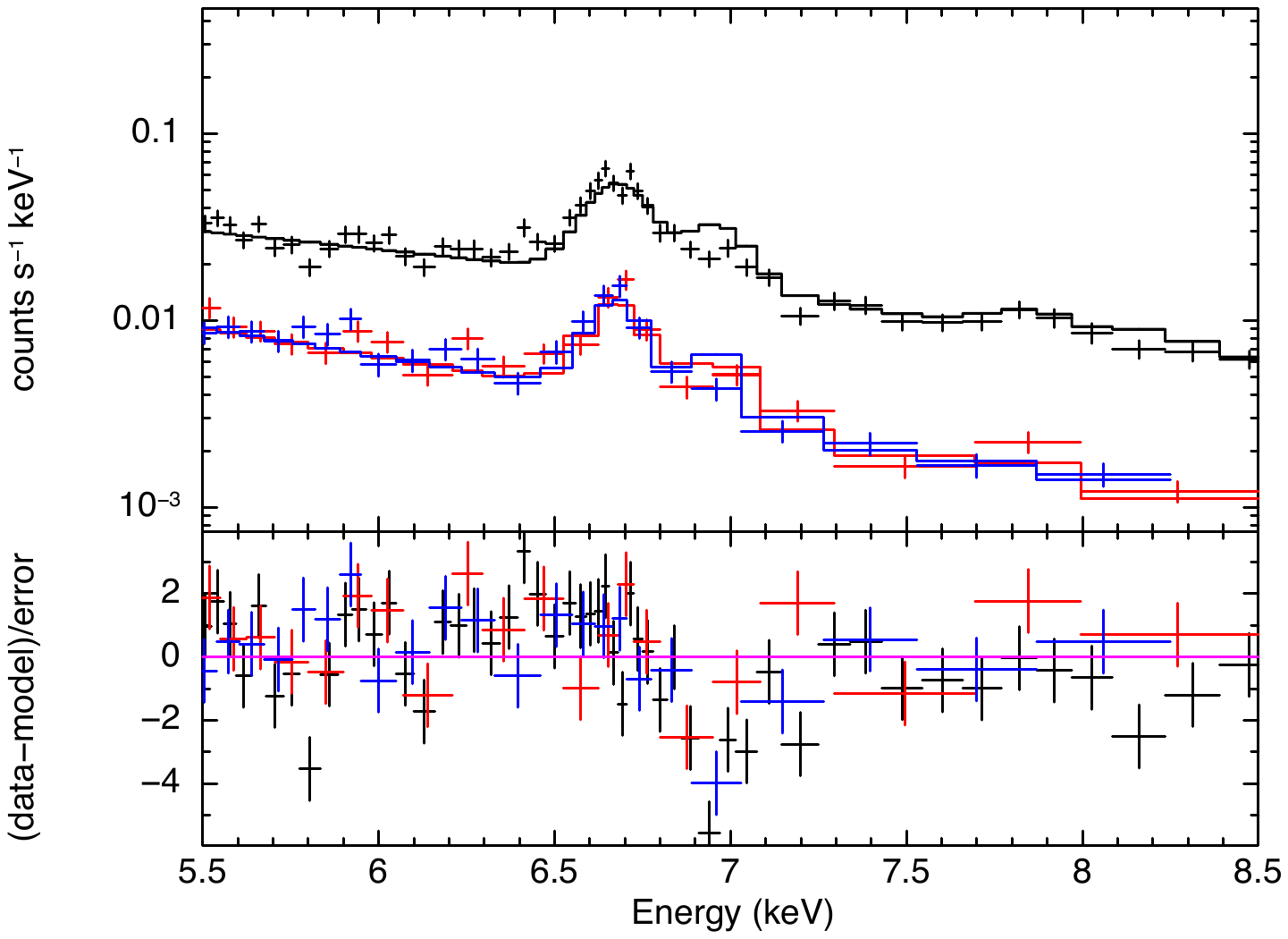}
\hspace{-0.75cm}
\includegraphics[height=2.3in,width=2.35in,angle=0]{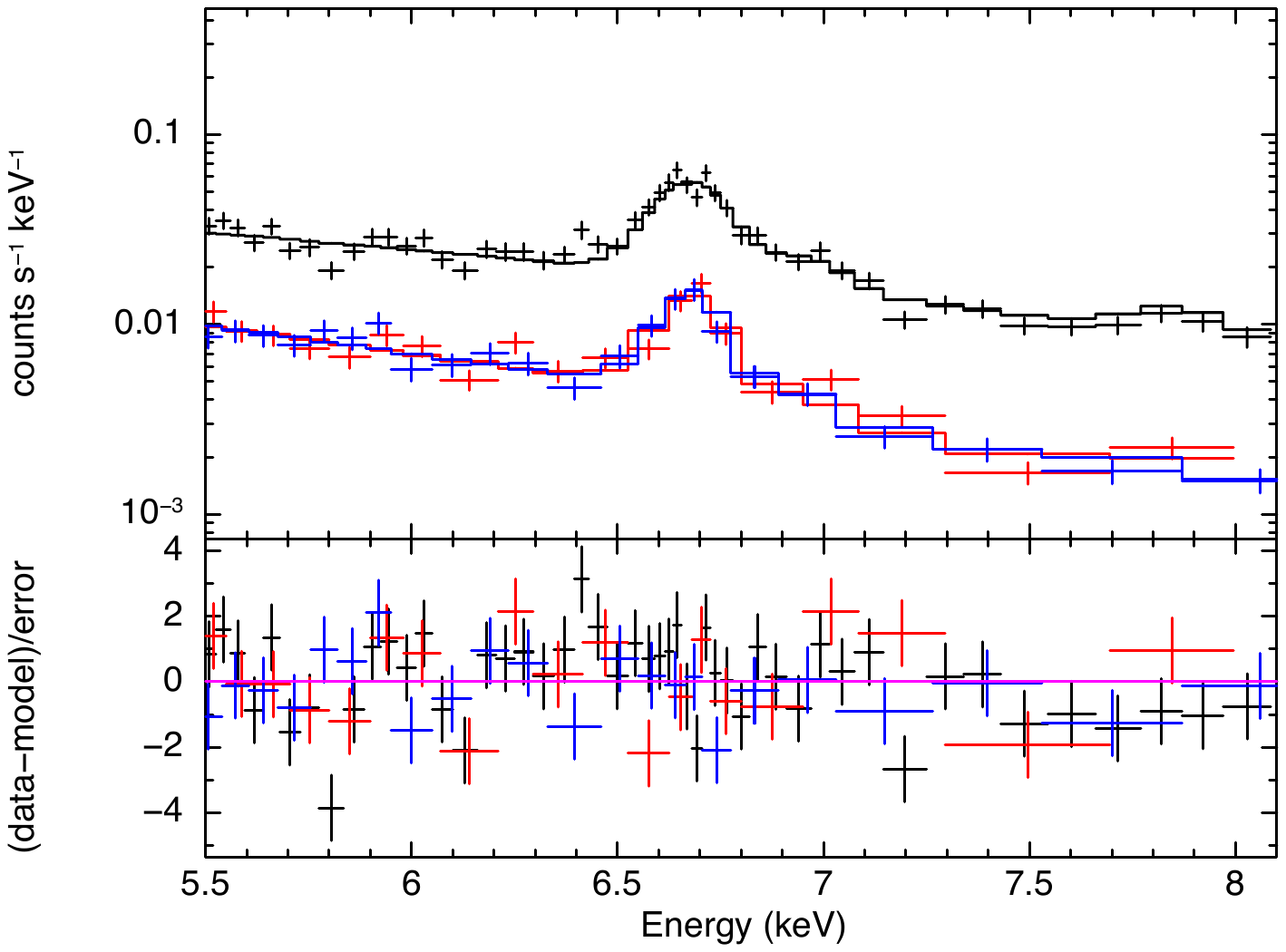}
\caption{Simultaneously-fitted EPIC spectra with the model  ($tbabs\times$CEVMKL). The plots show the fit quality and residuals around the iron complex at 6-7 keV. The residuals on the left reveal an apparent H-like iron absorption line.  The iron abundance is thawed in the fitting process. The right hand panel is a fit using the $warmabs$ model (partially ionized (photoionized)  absorber) where the absorption line is well-accomodated by the absorber model (see Sec. \ref{compare}). \label{fig:feline}
}
\end{center}
\end{figure}

All unabsorbed fluxes for the fits between 0.2-10.0 keV are given in Table \ref{tab:sp1} with the calculated X-ray luminosities using the 97 pc source distance \citep{1986Wood}. 
We find most elemental abundances to be consistent with solar except for some possible over-abundance of  Ne/Ne$_{\odot}$=(1.1-7.4), S/S$_{\odot}$=(1.2-2.7), and under-abundance of
Fe/Fe$_{\odot}$=(0.5-1.0) and perhaps O/O$_{\odot}$=(0.3-1.9). We have checked if  the under-abundance of iron has to do with iron absorption lines within the 6-7 keV band. The left hand panel of Figure \ref{fig:qsp} indicates significant departures in the residuals around the Fe XXV, and Fe XXVI complex where ($tbabs\times$CEVMKL) model was used. Inclusion of a power law or a partially ionized absorber ({\it zxipcf}) model alleviates this problem, but the Fe abundance seem to be problematic. To test the existence of an iron absorption line, we included a gaussian absorption line in the fitting procedure in  the 6.6-7.0 energy range along the  ($tbabs\times$CEVMKL) model which improved the residuals considerably in the 6.0-7.0 keV band (also the \rchisq). This line is found at 6.95$^{+0.05}_{-0.05}$ keV showing that this is a Fe XXVI absorption line (H-like iron absorption line) with a line depth of 0.12$^{+0.04}_{-0.06}$ (unitless) and a line width ($\sigma$) of  0.09$^{+0.01}_{-0.04}$ keV. 
Figure \ref{fig:feline} shows the
region of spectra where the Fe XXV and Fe XXVI emission lines are present and the H-like iron absorption line is detected using the CEVMKL model. The left hand panel shows the fit ($tbabs\times$CEVMKL) in which  iron abundance is thawed yielding  iron under-abundance. An Fe XXVI absorption line is detected with residual deviations at about 4$\sigma$. This fit is irrespective of the plasma emission model used such as  MKCFLOW or APEC in XSPEC. Inclusion of a {\it power law} model diminishes the temperatures and the fitted continuum is modified, reducing the residual variations at the line energy improving the fits, while the {\it zxipcf} model alleviates the absorption feature to a good extent diminishing the residual effects. Thus the iron absorption line is not as apparent
in the middle and right hand panel of  Figure \ref{fig:qsp}. The right hand panel of Figure \ref{fig:feline}  shows the vicinity of the iron lines when the $warmabs$ model is used as a multiplicative partially ionized  (photoionized) absorption model (see Sec. \ref{compare}). On the other hand, the VNEI model which is
a nonequilibrium ionization plasma model, has a weak Fe XXVI emission line and a fit with VNEI yields only a (1.5-2)$\sigma$ dip at the absorption line energy and the detection is then, not very significant. 

Using the same quiescent \xmm\ EPIC data set, \citet{2011Nucita} found a best-fit with \rchisq=1.8 (for 628 d.o.f.) using a composite model of  cold absorption and {\it absori}, an old version of photoionized absorber model and CEMEKL (a multi-temperature plasma emission model similar to CEVMKL with only metal abundances $Z$ to vary).
These fits revealed  \nh=(5.8$\pm$0.4)$\times$10$^{20}$ cm$^{-2}$  for the neutral hydrogen column density, and  \nh=1.7$\pm$0.1$\times$10$^{21}$ cm$^{-2}$ for 
the equivalent ionized hydrogen column density for an ionized absorber  with $\xi$=0.3$\pm$0.1. We find considerably different values for these parameters. 
Nucita et al. find a multi-temperature plasma with $\alpha$=1.6$\pm$0.1 for the power-law index of the temperature distribution with kT$_{max}$=11.4$\pm$0.2 keV. We note that the maximum 
temperature and the $\alpha$ parameter is similar to some of our fits (i.e., QFit-2)  with the partial covering absorber model. However, the partial covering cold and warm absorbers we find are very different. 
The \rchisq\ values we obtain from any of the fits in Table \ref{tab:sp1} are significantly better than the fit in  \citet{2011Nucita}, thus we have improved it. 

 In addition, we have investigated the data for the two orbital phases, 0.30$\pm$0.02 and 0.73$\pm$0.02, noted by \citet{2011Nucita} to show dipping behavior in the mean X-ray light curve. Using the SAS task {\it phasecalc}, we  calculated the phases of the photon arrival times utilizing the Ephemerides of \citealt{2002Baptista}. Then,  we extracted spectra filtering on these calculated phases, and conducted spectral analysis using the QFit5 in Table~\ref{tab:sp1} with the VNEI plasma model. The spectral fit between phases 0.35-0.25 indicates similar parameters with the phase-average fit of the plasma model, and the partial covering cold/neutral plasma. But the partially ionized absorber  though has similar ionization parameter of  log($\xi$)=3.5-3.7, yields an equivalent \nh=(8.2-21.0)$\times$10$^{22}$ cm$^{-2}$ that is 2-4 times larger than the phase-average result with a partial covering of  $>$ 87\%. The spectral fit of the spectrum for the phases 0.7-0.8, give similar plasma model parameters, but with different absorption characteristics. The cold partially covering absorber is 2 times larger and the partially ionized absorber is colder (less ionized) with  log($\xi$)=0.1-0.3 and equivalent \nh=(1.0-1.4)$\times$10$^{22}$ cm$^{-2}$ (partial covering is (70-80)\%). The error ranges are at 90\%\ confidence level and the \rchisq\ of the  two fits are 1.2 (d.o.f. 159) and 1.0 (d.o.f. 97), respectively. The iron abundance in these two fits is found to be between 1.5-2.5 times the solar value which is larger than any fit-result attained in our analysis.

\begin{deluxetable*}{llllllll}
\tablewidth{0pt}
\tablecaption{ Spectral Parameters of the  Fits to the Quiescent EPIC Spectra of Z Cha.  \label{tab:sp1}}
\tablehead{
Model  & Parameter & QFit-1 & QFit-2 & QFit-3 & QFit-4 & QFit-5 & QFit-6            
}
\startdata 
tbabs  & $N_H$ (10$^{22}$) & $0.11^{+0.01}_{-0.01}$ & N/A & N/A & N/A  & N/A & N/A \\     
   &  (ato. cm$^{-2}$)  &  &   &   &   &   &  \\
pcfabs  & $N_H$ (10$^{22}$)& N/A & $0.23^{+0.07}_{-0.06} $ & $0.22^{+0.30}_{-0.20} $ & $0.13^{+0.05}_{-0.06}$   & $0.12^{+0.02}_{-0.01}$   &  $0.18^{+0.01}_{-0.02}$  \\
   &  (ato. cm$^{-2}$)  &  &   &   &   &   &  \\
        & cov. frac. &  N/A  & $0.77^{+0.02}_{-0.02} $ & $0.77^{+0.05}_{-0.02} $  & $0.88^{+0.03}_{-0.03}$  & $0.89^{+0.05}_{-0.03}$  &  $0.88^{+0.01}_{-0.02}$ \\
      zxipcf  & $N_H$ (10$^{22}$) & N/A & N/A &  N/A  & N/A  & $4.8^{+1.1}_{-1.4}$   &  $10.3^{+0.7}_{-0.7}$  \\
   &  (ato. cm$^{-2}$)  &  &   &   &   &   &  \\
          & log($\xi$)  & N/A & N/A &  N/A  & N/A  & $3.56^{+0.10}_{-0.06}$   &  $1.91^{+0.04}_{-0.10}$  \\
        & cov. frac. &  N/A  &  N/A  & N/A   & N/A  & $0.54^{+0.03}_{-0.08}$  & $0.36^{+0.01}_{-0.01}$   \\ 
\hline
CEVMKL       & $\alpha$     & $2.5^{+0.3}_{-0.4} $ & $ 1.7^{+0.1}_{-0.3} $ & $1.95^{+0.13}_{-0.12} $ & N/A    &   N/A & $1.83^{+0.04}_{-0.05} $ \\    
       & $\rm{kT}_{max}$(keV)       &  $10.4^{+0.6}_{-0.6} $   & $11.6^{+1.4}_{-0.3} $    & $ 7.9^{+0.5}_{-0.5} $ &  N/A  &  N/A &  $ 7.30^{+0.13}_{-0.11} $ \\   
 & $K_{cevmkl}$ ($\times$10$^{-3}$) & $9.7^{+2.0}_{-0.1}$ & $7.8^{+0.8}_{-0.9}$ & $6.5^{+0.7}_{-1.5}$  & N/A  & N/A  &  $12.4^{+0.1}_{-0.1}$  \\ 
 \hline
 VNEI      &  $\tau$ (10$^{11}$ s~cm$^{-3}$)  & N/A &  N/A  & N/A  & $5.4^{+0.6}_{-0.3} $    & $2.9^{+0.1}_{-0.1} $   & N/A \\    
       & $\rm{kT}$(keV)     &     N/A          &     N/A        &    N/A     &  $5.5^{+0.1}_{-0.1} $ &  $ 8.4^{+0.2}_{-0.2} $ & N/A \\   
 & $K_{vnei}$ ($\times$10$^{-3}$) & N/A & N/A &  N/A &  $1.70^{+0.02}_{-0.06}$ & $2.08^{+0.02}_{-0.01}$  & N/A  \\ 
 \hline
 & O/O$_{\odot}$   & 1.0 (fixed) & $1.4^{+0.3}_{-0.2} $ & $1.3^{+0.3}_{-0.2} $   & $0.6^{+0.4}_{-0.3} $ & $1.9^{+0.5}_{-0.6} $ & $0.40^{+0.20}_{-0.15} $ \\
 & Ne/Ne$_{\odot}$    & $2.6^{+0.5}_{-0.5}$ & $1.3^{+0.4}_{-0.4} $ & $1.1^{+0.3}_{-0.5} $  & $3.9^{+0.6}_{-0.6}$  &  $7.4^{+1.0}_{-1.0} $ & $0.6^{+0.2}_{-0.3} $ \\
 & Si/Si$_{\odot}$     & $1.9^{+0.3}_{-0.3}$ & $1.3^{+0.2}_{-0.2} $ & $1.6^{+0.2}_{-0.3} $  &$1.4^{+0.2}_{-0.2} $  & $0.7^{+0.2}_{-0.1} $ & $1.13^{+0.14}_{-0.15} $ \\
 & S/S$_{\odot}$      & $2.7^{+0.5}_{-0.4}$ & $2.04^{+0.30}_{-0.30} $ & $2.3^{+0.4}_{-0.3} $   & $1.2^{+0.2}_{-0.2} $ &  $0.8^{+0.1}_{-0.2} $& $1.15^{+0.20}_{-0.19} $\\
 & Fe/Fe$_{\odot}$    & $0.80^{+0.05}_{-0.04}$ & $0.80^{+0.04}_{-0.04} $ & $0.90^{+0.02}_{-0.07} $  &$0.70^{+0.05}_{-0.02} $  &  $0.70^{+0.02}_{-0.05} $& $0.52^{+0.03}_{-0.02} $\\
\hline
Power law & Phot. Ind. &    N/A    &  N/A &  $1.30^{+0.15}_{-0.10}$  &  $1.26^{+0.03}_{-0.02}$ & N/A & N/A \\  
         & $K_{power}$ ($\times$10$^{-4}$)       & N/A  &  N/A &  $1.3^{+1.0}_{-0.2}$  & $1.35^{+0.09}_{-0.08}$ &  N/A & N/A \\  
\hline
 & $\chi^2_{\nu} (\nu)$        &  1.23 (1403)   & 1.14 (1401)     & 1.09 (1399)     &  1.13 (1399) &  1.12 (1398)  &  1.00 (1398)   \\ 
\hline
  Flux & (10$^{-12}$) & 5.2$^{+0.2}_{-0.5}$     &  5.0$^{+1.0}_{-0.2}$ & 5.1$^{+1.4}_{-0.8}$ &  5.1$^{+0.3}_{-0.3}$ & 5.1$^{+0.1}_{-0.1}$ &  6.6$^{+0.6}_{-0.3}$       \\  
  & (erg~cm$^{-2}$s$^{-1}$) &    &   &  &   &  &    \\  
 Luminosity & (10$^{30}$erg~s$^{-1}$) & 5.5$^{+0.3}_{-0.5}$    &  5.3$^{+1.1}_{-0.2}$   & 5.4$^{+1.5}_{-0.8}$   &  5.8$^{+0.3}_{-0.3}$ & 5.8$^{+0.1}_{-0.1}$  & 7.5$^{+0.6}_{-0.3}$    \\
  Flux$_{thermal}$ & (10$^{-12}$) & 5.2$^{+0.2}_{-0.5}$    &  5.0$^{+1.0}_{-0.2}$ & 3.2$^{+0.8}_{-0.5}$  &  3.6$^{+0.2}_{-0.1}$ & 5.1$^{+0.1}_{-0.1}$ &  6.6$^{+0.6}_{-0.3}$   \\  
   & (erg~cm$^{-2}$s$^{-1}$) &   &  &  &  & &    \\  
 L$_{thermal}$ & (10$^{30}$erg~s$^{-1}$) & 5.5$^{+0.3}_{-0.5}$   & 5.3$^{+1.1}_{-0.2}$    & 3.4$^{+0.9}_{-0.6}$ & 4.1$^{+0.2}_{-0.1}$  & 5.8$^{+0.1}_{-0.1}$ &  7.5$^{+0.6}_{-0.3}$    \\
  Flux$_{power}$ & (10$^{-12}$) & N/A    &  N/A    &  2.0$^{+0.8}_{-0.7}$  &  1.5$^{+0.2}_{-0.2}$ & N/A & N/A   \\  
    & (erg~cm$^{-2}$s$^{-1}$) &  &     &    &   &   &     \\  
 L$_{power}$ & (10$^{30}$erg~s$^{-1}$) &  N/A   & N/A   & 2.1$^{+0.9}_{-0.7}$ &  1.7$^{+0.2}_{-0.2}$ & N/A & N/A  \\
\enddata   
\tablecomments{
Fits are performed using EPIC pn and MOS data in the 0.3-10.0 keV range. A constant value model 
is fitted as a free parameter to account for the normalization between different detectors.
$N_H$ is the absorbing column, $\alpha$ is the index
of the power-law emissivity function ($dEM=(T/T_{max})^{\alpha -1} 
dT/T_{max}$), $T_{max}$ is the maximum temperature for the 
CEVMKL model, $K_{CEVMKL}$ is the normalization for the CEVMKL model; 
K=(10$^{-14}$/4$\pi$D$^2$)$\times$EM where EM (emission measure) =${\rm \int n_e\ n_H\ dV}$
(integration is over the emitting volume V).
$K_{powerlaw}$ is the normalization (i.e., photon flux in phot. cm$^{-2}$\ s$^{-1}$) for the power law model. 
 NEIvers.3.0.9 plasma code with ATOMDB database was assumed for VNEI fits.
All errors are calculated at the 90$\%$ confidence limit for a single parameter. 
The unabsorbed X-ray flux and the luminosities are given in the range 0.2-10.0 keV.
The distance of 97 pc is used for luminosity calculations. } 
\end{deluxetable*}

\subsection{Analysis of EPIC data in outburst}\label{sec:eosp}

\begin{deluxetable*}{llllll}
\tablewidth{0pt}
\tablecaption{Spectral Parameters of the
Fit to the Outburst EPIC Spectra of Z Cha. \label{tab:sp2}}
\tablehead{
Model  & Parameter & OFit-1 & OFit-2 & OFit-3 & OFit-4  
}
\startdata
tbabs  & $N_H$ (10$^{22}$ato. cm$^{-2}$) & $0.08^{+0.01}_{-0.01}$ & N/A & N/A & N/A  \\     
pcfabs  & $N_H$ (10$^{22}$ato. cm$^{-2}$)& N/A & $4.1^{+0.7}_{-0.6} $ & $9.6^{+1.1}_{-0.9} $ & $0.56^{+0.09}_{-0.10}$   \\
        & cov. frac. &  N/A  & $0.47^{+0.05}_{-0.03} $ & $0.78^{+0.02}_{-0.02} $  & $0.63^{+0.08}_{-0.06}$   \\
 pcfabs  & $N_H$ (10$^{22}$ato. cm$^{-2}$)& N/A &  N/A & $0.19^{+0.01}_{-0.02} $ & $0.12^{+0.3}_{-0.3}$   \\
        & cov. frac. &  N/A  & N/A & $0.70^{+0.03}_{-0.03} $  & $0.4^{+0.3}_{-0.1}$   \\       
      zxipcf  & $N_H$ (10$^{22}$ato. cm$^{-2}$) & N/A & N/A & $261.1^{+34.0}_{-31.0}$   &  N/A \\
          & log($\xi$)  & N/A & N/A  & $2.81^{+0.02}_{-0.02}$   &  N/A  \\
        & cov. frac. &  N/A   & N/A  & $0.90^{+0.02}_{-0.01}$  & N/A   \\ 
\hline
CEVMKL       & $\alpha$     & $ < 0.1 $ & $< 0.02$ & $ < 0.07 $ & $< 0.09 $ \\    
       & $kT_{max}$(keV)       &  $1.12^{+0.08}_{-0.06} $   & $8.9^{+0.2}_{-0.9} $  & $ 3.0^{+0.2}_{-0.2} $ & $ 1.01^{+0.05}_{-0.04} $ \\   
 & $K_{cevmkl}$ ($\times$10$^{-4}$) & $2.7^{+0.1}_{-0.1}$ & $2.5^{+0.1}_{-0.1}$ & $77.0^{+2.0}_{-2.0}$  & $5.6^{+0.3}_{-0.1}$  \\ 
 \hline
 BBODY &  kT$_{\rm BB}$ (keV) &   $0.027^{+0.008}_{-0.006} $  & N/A      & N/A     &   N/A    \\
               &  K$_{\rm BB}$  ($\times$10$^{-4}$) &  $3.0^{+2.5}_{-0.7}$       &   N/A   &    N/A    &  N/A   \\
 \hline
 & O/O$_{\odot}$    & $0.8^{+0.1}_{-0.1}$ & $0.90^{+0.05}_{-0.04}$ & $0.95^{+0.06}_{-0.05}$   & $0.8^{+0.1}_{-0.1}$  \\
 & Ne/Ne$_{\odot}$     & $1.3^{+0.2}_{-0.2}$ & $0.7^{+0.2}_{-0.2} $ & $2.0^{+0.2}_{-0.2} $  & $1.1^{+0.2}_{-0.2}$   \\
 & Mg/Mg$_{\odot}$     & $1.2^{+0.2}_{-0.1}$ & $2.3^{+0.3}_{-0.3} $ & $1.3^{+0.2}_{-0.2} $  & $0.8^{+0.1}_{-0.1}$   \\
 & Si/Si$_{\odot}$     & $1.3^{+0.2}_{-0.2}$ & $1.8^{+0.3}_{-0.3} $ & $1.50^{+0.02}_{-0.02} $  & $0.8^{+0.1}_{-0.1} $  \\
& S/S$_{\odot}$     & $2.4^{+0.7}_{-0.7}$ & $0.9^{+0.3}_{-0.3} $ & $1.3^{+0.3}_{-0.3} $   & $1.4^{+0.3}_{-0.4} $ \\
& Fe/Fe$_{\odot}$    & $0.6^{+0.1}_{-0.1}$ & $1.14^{+0.04}_{-0.04} $ & $0.60^{+0.02}_{-0.07} $  & $0.6^{+0.1}_{-0.1}$  \\
\hline
Power law & Phot. Ind. &  $1.6^{+0.1}_{-0.1}$  & N/A & N/A & $1.56^{+0.08}_{-0.12}$ \\  
         & $K_{power}$ ($\times$10$^{-5}$)   &  $3.7^{+0.2}_{-0.2}$  & N/A & N/A & $3.4^{+0.1}_{-0.2}$  \\  
 \hline
 GAUSS & Line Energy (keV) &  6.64$^{+0.06}_{-0.05}$  &  N/A & N/A &  6.64$^{+0.06}_{-0.07}$  \\
         & Sigma (keV)  ($\times$10$^{-3}$) & $6.1^{+94.0}_{-\infty}$ &  N/A & N/A &  $6.7^{+90.0}_{-\infty}$ \\
         & $K_{Gauss}$  ($\times$10$^{-7}$)  &  $9.6^{+3.7}_{-3.3}$ & N/A & N/A & $9.6^{+3.6}_{-3.4}$ \\     
\hline
 & $\chi^2_{\nu} (\nu)$        &  1.31 (449)   & 1.46 (454)     & 1.24 (450)     &  1.29 (448)   \\ 
\hline
  Flux & (10$^{-12}$) & 2.5$^{+6.2}_{-1.2}$     &  1.3$^{+0.1}_{-0.1}$ & 28.0$^{+1.0}_{-8.0}$ &  1.7$^{+0.1}_{-0.1}$      \\  
  & (erg~cm$^{-2}$s$^{-1}$) &    &   &  &      \\  
 Luminosity & (10$^{30}$erg~s$^{-1}$) & 2.9$^{+7.0}_{-1.4}$    &  1.50$^{+0.03}_{-0.13}$   & 32.0$^{+1.0}_{-0.9}$   &  1.9$^{+0.2}_{-0.1}$   \\
  Flux$_{thermal}$ & (10$^{-13}$) & 7.7$^{+0.5}_{-0.6}$    &  13.0$^{+1.0}_{-1.0}$ & 280.0$^{+10.0}_{-80.0}$  &  13.8$^{+1.2}_{-0.8}$  \\  
   & (erg~cm$^{-2}$s$^{-1}$) &   &  &  &     \\  
 L$_{thermal}$ & (10$^{29}$erg~s$^{-1}$) & 8.8$^{+0.6}_{-0.7}$  &  15.0$^{+0.3}_{-1.3}$   & 320.0$^{+10.0}_{-9.0}$ & 15.7$^{+1.4}_{-1.0}$     \\
  Flux$_{power}$ & (10$^{-13}$) & 2.9$^{+0.5}_{-0.3}$  & N/A & N/A &  2.8$^{+0.2}_{-0.1}$  \\  
    & (erg~cm$^{-2}$s$^{-1}$) &  &     &    &       \\  
 L$_{power}$ & (10$^{29}$erg~s$^{-1}$) &  3.3$^{+0.6}_{-0.3}$ & N/A & N/A &   3.2$^{+0.2}_{-0.1}$  \\
   Flux$_{BB}$ & (10$^{-11}$) & 1.2$^{+1.7}_{-0.7}$  & N/A & N/A &  N/A \\  
    & (erg~cm$^{-2}$s$^{-1}$) &  &     &    &       \\  
 L$_{BB}$ & (10$^{31}$erg~s$^{-1}$) &  1.3$^{+2.1}_{-0.7}$ & N/A & N/A &  N/A  \\
\enddata  
\tablecomments{
Fits are performed using EPIC pn and MOS data in the 0.3-10.0 keV range.
A constant value is fitted
as a free parameter to account for the normalization between different detectors.
For description of the models, parameters, and normalizations please refer to Table~\ref{tab:sp1}. 
$K_{BB}$ is the normalization of the blackbody model in units of L$_{39}$/D$_{10}$ where L$_{39}$ is the source luminosity in units of 
 10$^{39}$ ergs s$^{-1}$ and D$_{10}$ is the distance to the source in units of 10 kpc.
 $K_{Gauss}$ is the normalization for the Gaussian emission line model  (flux of the line in phot. cm$^{-2}$s$^{-1}$). 
All errors are given
at 90\% confidence limit for a single parameter.
The unabsorbed X-ray flux and the luminosities are given in the range 0.2-10.0 keV
except for  the blackbody model where the ranges are for 0.1-10.0 keV.
For luminosities, the distance of 97 pc is assumed. }
\end{deluxetable*}

The background and source spectra along with the response and ancillary files for the outburst data of Z Cha were generated as described in the Sec. \ref{sec:obs}.  
The EPIC spectra were grouped with a minimum of 70, 45, and 45  counts in each spectral bin for pn, MOS1 and 2, respectively, to achieve a good \chisq\ statistics in the fitting procedure. Spectra are analyzed using the HEASoft and XSPEC softwares as described in Sec. \ref{sec:eqsp} for the analysis of data obtained in quiescence.

\begin{figure*}
\includegraphics[height=5.5cm,width=6.5cm,angle=0]{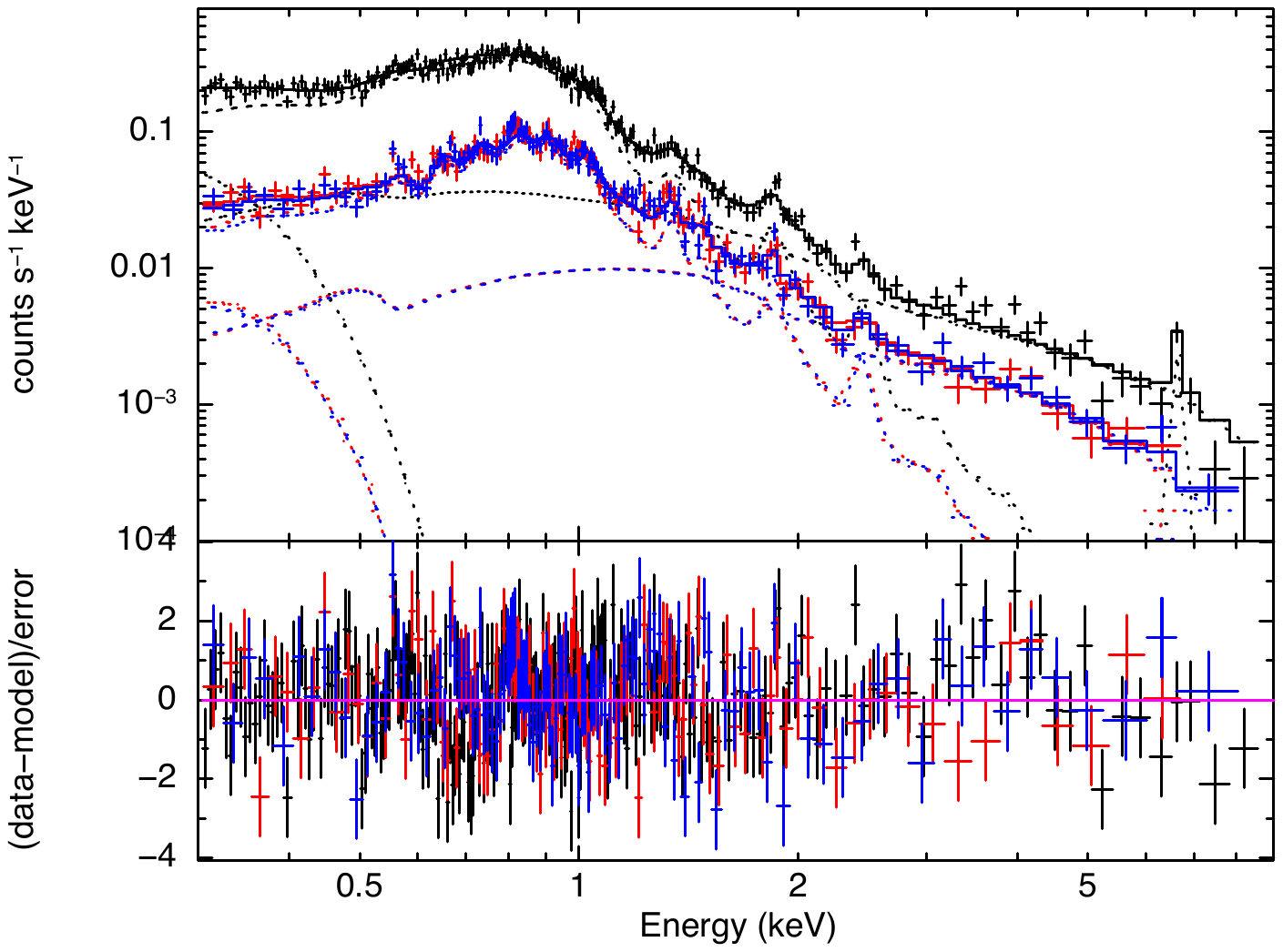}
\hspace{-0.7cm}
\includegraphics[height=5.5cm,width=6.5cm,angle=0]{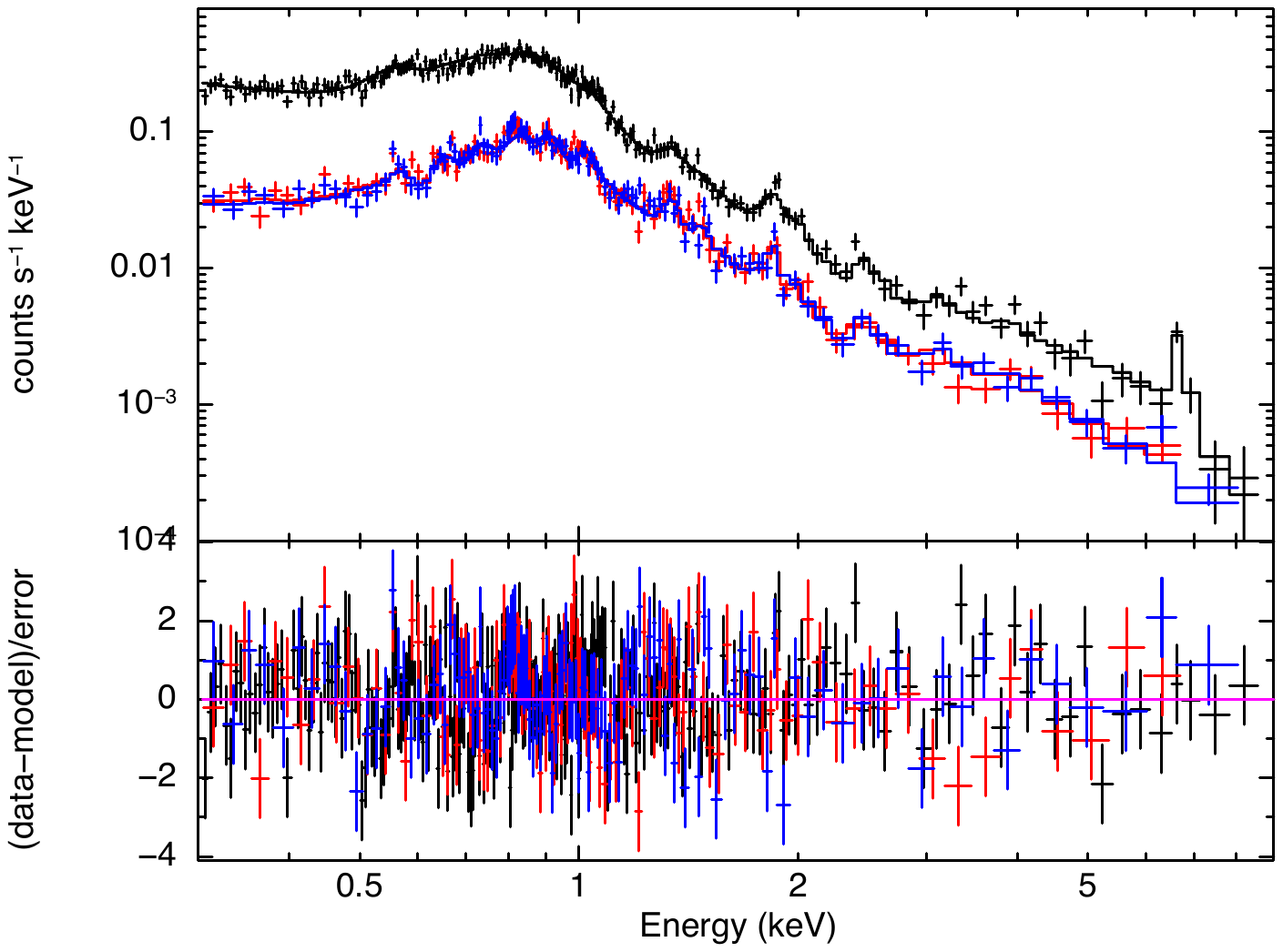}
\hspace{-0.7cm}
\includegraphics[height=5.5cm,width=6.5cm,angle=0]{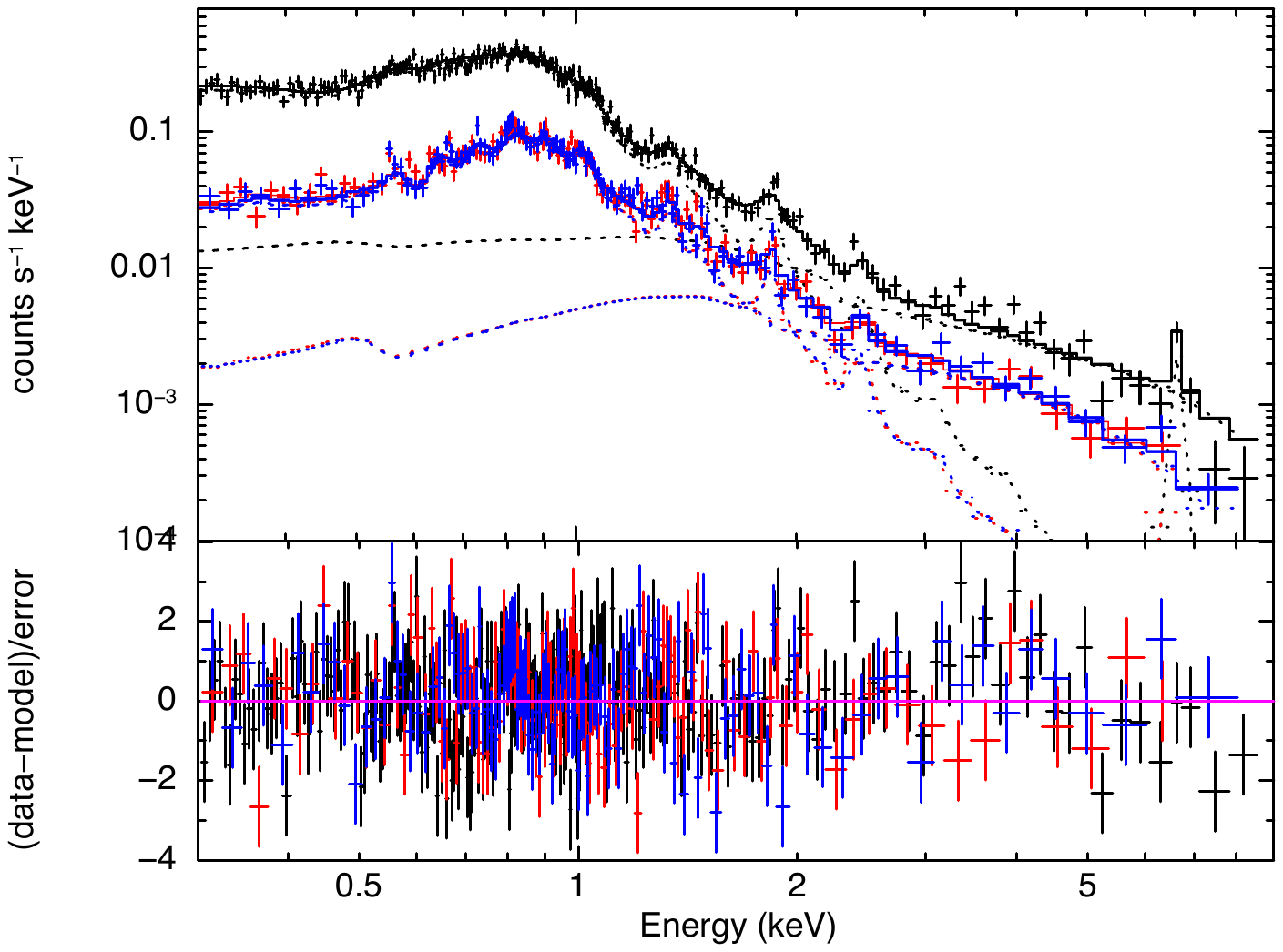}
\caption{
Fits to the outburst EPIC pn and MOS1,2 spectra of Z Cha. All three EPIC spectra are fit simultaneously with the ($tbabs\times$(BBODY$+$GAUSS$+$CEVMKL$+power$)) model 
on the left. The middle panel is the EPIC spectra 
simultaneously fit with the composite model of  ($pcfabs\times pcfabs\times zxipcf\times$CEVMKL). The right hand panel shows the
same fit spectra using the composite model  ($pcfabs\times pcfabs\times$(CEVMKL$+power$)). In the panels, the 
dotted lines show the contribution of different model components.
The lower panels show the residuals  in standard deviations (in sigmas). \label{fig:osp}} 
\end{figure*}

For the spectral analysis of the EPIC  
data obtained in outburst, the same absorption models and plasma emission model components are utilized except for an additional blackbody model BBODY in XSPEC. As in the analysis of data obtained in quiescence, we have  used the multi-temperature isobaric cooling flow type of plasma emission model CEVMKL in XSPEC with switch=2  and the model VNEI, non-equilibrium ionization plasma model in XSPEC when performing simultaneous fits to the EPIC spectra  
(pn, MOS1, MOS2). A constant-value model is also fit along with the composite models to account for cross-normalization calibration between the different detectors. Four composite models are constructed  as  OFit-(1-4) (see Table \ref{tab:sp2}). Resulting parameters of the spectral fits are given in Table \ref{tab:sp2} for these four models. These composite models do not include the plasma model VNEI (non-equilibrium ionization plasma model) because the fits can never reproduce the 6-7 keV single iron line correctly in the fitting process with \rchisq\ larger than 2.0. This may mean that this iron line is produced in a plasma in collisional equilibrium (CIE) or the iron line is not at its rest/expected wavelength for the VNEI model; this will be elaborated in the Discussion section. Thus, all fits involve the CEVMKL model plasma emission for simplicity. However, the $\alpha$ parameter of the temperature distribution is $\sim$ 0.01 similar to the case of NL systems (high state CVs) that bear non-standard advective hot flows. This will be elaborated in the Discussion section, as well.
Example spectral fits with the CEVMKL plasma model are  displayed in Figure \ref{fig:osp}  for OFit-1, OFit-3, and OFit-4. 

The general theoretical expectation of a DN outburst is the existence of a blackbody emission component in the X-ray regime from an optically-thick standard BL  as the accretion rate increases towards the WD 
in the disk.  Thus, we have constructed a composite model fit OFit-1 with a model of BBODY (in XSPEC); ($tbabs\times$(BBODY$+$GAUSS$+$CEVMKL$+power$)). This yields an acceptable \rchisq\  where addition of the 
BBODY model improves the fit at only a 93\% confidence level when an F-TEST is used. 
A  {\it power law} emission component is also necessary to reduce \rchisq\ level of the fit, OFit-1. However, the CEVMKL plasma emission model does not reproduce the iron emission line (in the 6-7 keV band) properly and an additional GAUSS emission line model is added to account for this inadequacy at 6.64$^{+0.06}_{-0.07}$ keV. This line is a narrow line constrained with the  spectral resolution of EPIC that indicates a He-like iron line with a flux of 1.0$^{+0.4}_{-0.3}\times 10^{-14}$ erg\ cm$^{-2}$s$^{-1}$ which is consistent with the low X-ray temperature. The plasma temperatures derived from the CEVMKL models are kT$_{max}$=0.97-3.20 keV given \rchisq $<$ 1.3\  (from all fits).  The composite model OFit-4 is constructed ($pcfabs\times pcfabs\times $($GAUSS$+$CEVMKL$+$power$)) to test if the plausible soft X-ray component (BBODY - blackbody emission component), is a manifestation of the absorption components in the system. This model lacks the BBODY component, but has two partially covering cold absorber components. The \rchisq\ level of this fit is as good as the model with the BBODY component that BBODY model may be accounting for incorrect absorption modeling.  Thus, the existence of a blackbody emitting BL can not be asserted based on the EPIC data in outburst.  

OFit-3 ($pcfabs\times pcfabs\times zxipcf\times $($CEVMKL$))  is constructed to reveal the existence of cold and photoionized warm absorption in the system along with the CEVMKL plasma model. The fit yields the best fitting model (see Table \ref{tab:sp2}).  Therefore, there is complex absorption in the system where one of the cold partial covering absorber components is similar to the one in quiescence, the second one is larger by a factor of 15,  (8.7-10.7)$\times$10$^{22}$ cm$^{-2}$.
The warm absorption modeled by the {\it zxipcf} component shows a very large  equivalent \nh=(2.30-2.95)$\times$10$^{24}$ cm$^{-2}$ with an ionization parameter log($\xi$) in a range 2.79-2.83 (covering fraction is 90\%). This partially-covering warm absorption is about a factor of 25-50 times more than in quiescence. This is the only fit with the CEVMKL model where the He-like weak iron line is adequately fit by the plasma emission model (no need for an additional GAUSS model).  

\subsection{High resolution X-ray spectroscopy with the RGS}\label{sec:rgssp}

The  basic analysis steps of the \xmm\ RGS data of Z Cha have been outlined in Sec. \ref{sec:obs}. The SAS tasks {\it rgsproc} and {\it rgscombine } are used to derive spectra of different orders (1,2) using default extraction regions and to co-add RGS1 and RGS2 spectra (order 1)
to make a total average spectrum with good statistics and high energy resolution ($\lambda/\Delta\lambda \sim$ 200-800\footnote{httprd cfs://xmm-tools.cosmos.esa.int/external/\\xmm$\_$user$\_$support/documentation/uhb/rgs.html}).  
The inferred tasks,  also, create  the necessary background spectra, and response files (for individual or combined spectra).
 The spectral analysis is performed within XSPEC software (for references and model descriptions see \citealt{1996Arnaud}).
The RGS spectra are marginally grouped by 3-6 channels in the range 1-4000 to improve \chisq\ statistics for the fits, mitigating for the best energy resolution. In addition, for the individual line fits using a Gaussian model, binning has been changed to minimum counts of 10-15 in each spectral bin when necessary. 

\begin{figure*}[ht] 
\begin{center}
\includegraphics[height=5.7cm,width=9.2cm,angle=0]{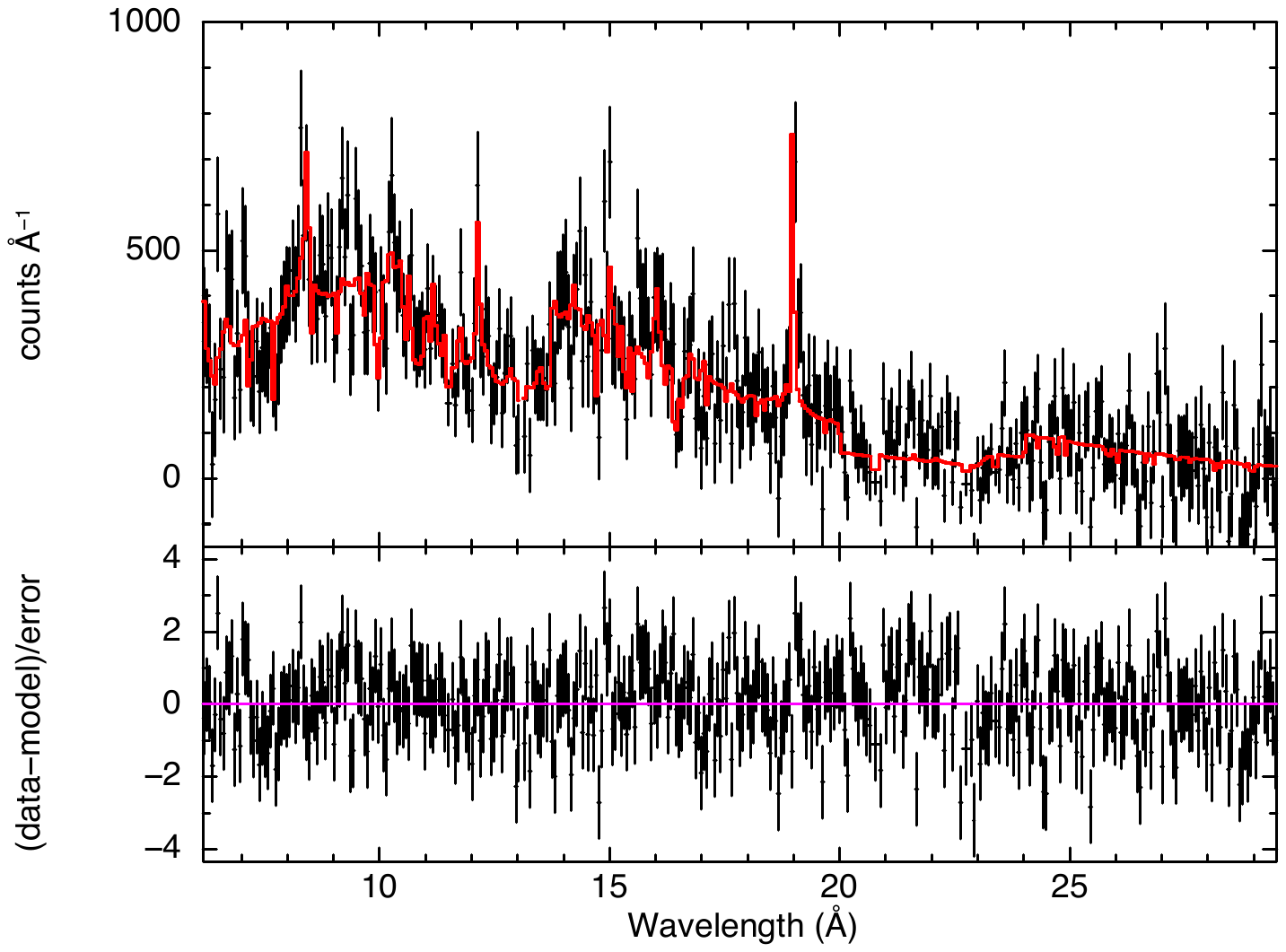}
\hspace{-0.7cm}
\includegraphics[height=5.7cm,width=9.2cm,angle=0]{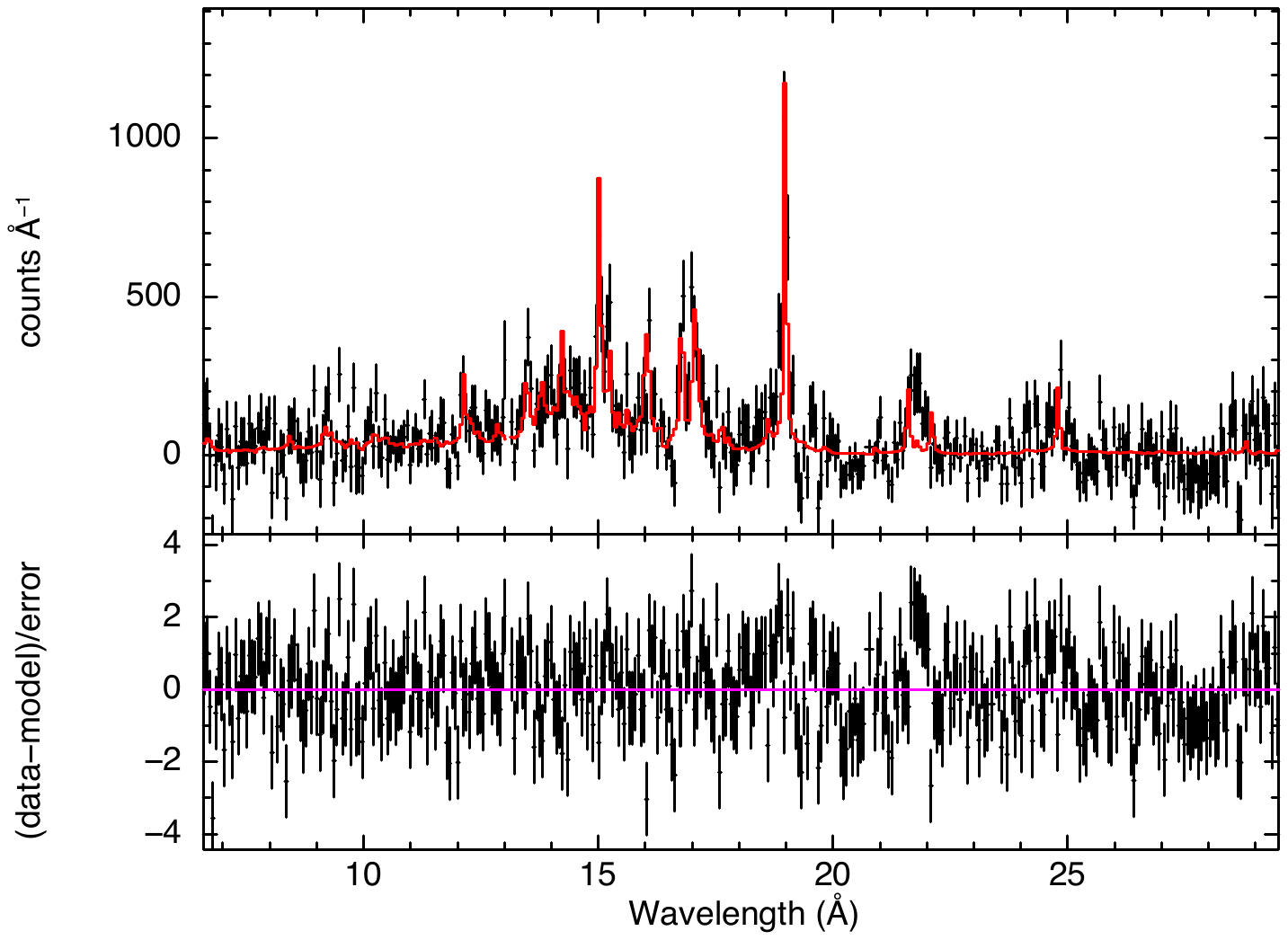}
\caption{
Averaged \xmm\ RGS 1 and RGS 2 spectra of Z Cha. The left hand panel shows  the high resolution X-ray spectrum obtained in quiescence fitted with ($pcfabs\times$CEVMKL) model.
The right hand panel is the averaged RGS 1 and RGS 2 spectra of Z Cha  in outburst fit with a 
($pcfabs\times pcfabs\times$CEVMKL) composite model. In both panels the
solid lines (in red color) show the fitted model. The lower panel shows the residuals  in standard deviations (in sigma).  \label{fig:rgs-sp}}
\end{center}
\end{figure*}

The RGS data analysis is utilized mainly for line diagnostics and to  confirm the spectral results  of the EPIC analysis. We note that the count rates are lower and statistical quality of spectra are adequate but the continuum is detected at a lower level, whereas the line emissions are of high significance compared to EPIC data.  We have fitted the quiescent and outburst RGS spectra with several composite models as in 
Sec. \ref{sec:eqsp} and \ref{sec:eosp}.  Table \ref{tab:rgs-sp} displays the spectral parameters describing the quiescence and outburst RGS spectra of Z Cha and Figure \ref{fig:rgs-sp} shows examples of fitted spectra.  
For quiescence, we present
two composite models $pcfabs\times$CEVMKL, and $pcfabs\times zxipcf\times$VNEI. We used similar models for the outburst as in $pcfabs\times pcfabs\times$CEVMKL and $pcfabs\times zxipcf\times$VNEI. The two plasma models ,
are already described in Sec. \ref{sec:eqsp}.  Table \ref{tab:rgs-sp}  yields a good comparison of both phases of Z Cha as the source makes state transition. We note that the two archival data sets, that we analyze in outburst and quiescence, are not contiguous. 

The neutral absorption towards the source is modeled via the {\it pcfabs} (partial covering neutral/cold absorber) that yields a range of values  (0.11-0.17)$\times$10$^{22}$ cm$^{-2}$ with $\sim$95\% fractional coverage that is similar to the value in Table \ref{tab:sp1} derived from the EPIC results in quiescence. Thus, this value is also detected for the outburst stage as an absorption component, however outburst shows presence of more absorption 
in the form of cold partial covering absorber and/or warm/ionized (photoionized) absorber. We fitted the CEVMKL, collisional equilibrium plasma model, along with the cold/neutral partially covering absorber and the VNEI, nonequilibrium ionization plasma is fitted along with a  partially ionized absorber model {\it zxipcf}. The particular choice of absorption components and plasma models are made relying on the small \rchisq\  values, the goodness of the fits. For example, when a CEVMKL model is used in place of VNEI with the ionized absorber component, the resulting parameters of the ionized absorber are nonphysical for the outburst data. For the quiescence, the \rchisq\  criteria  is used to display the two 
best-fitting composite models. Other composite models do not provide any better fits to the quiescent RGS spectrum. 
The X-ray plasma temperatures found for the RGS spectra are in accordance with Table \ref{tab:sp1} (quiescence) and Table \ref{tab:sp2} (outburst). The X-ray plasma temperature in quiescence is 4.1-5.4 keV using the VNEI model with the {\it zxipcf} absorber and 7.9-10.8 keV for the CEVMKL model using only cold/neutral partial covering absorber. X-ray plasma temperature in outburst is kT$_{max}$= 0.7-1.2 keV (using CEVMKL model with two partial covering cold absorbers) and 0.5-0.7 keV for the VNEI plasma model using two components of absorption (partially ionized and cold)  with partial covering. 

The outburst reveals a cold partially covering absorber  with either $1.7^{+\infty}_{-0.6}$$\times$10$^{22}$ cm$^{-2}$ (Outburst-1) or (0.3-0.8)$\times$10$^{22}$ cm$^{-2}$ (Outburst-2).  Moreover, a partially ionized absorber model with $2.9^{+5.6}_{-2.1}$$\times$10$^{22}$ cm$^{-2}$ (Outburst-2) along with VNEI plasma model is consistent with the co-added RGS spectrum. We note here that the EPIC spectral results in outburst yield considerably higher values for the partially ionized absorber in a range (230.0-295.0)$\times$10$^{22}$ cm$^{-2}$ (see Table \ref{tab:sp2}).  The ionization parameters derived from the RGS spectra for the ionized absorption component are found to be log($\xi$)=$3.6^{+0.4}_{-0.3}$ for quiescence and log($\xi$)=$2.8^{+0.4}_{-0.5}$ for the outburst. 
The $\alpha$\ parameter (power law index for temperature distribution of the CEVMKL model) is found in accordance with EPIC results; 1.7-2.1 for quiescence and 0.1-0.8 for the outburst (similar to NLs). 
The ionization time scale for the VNEI plasma model yields values in a range of $8.7^{+\infty}_{-4.0}\times 10^{11}$  s\ cm$^{-3}$ for quiescence and $1.1^{+0.3}_{-0.4}\times 10^{11}$ s\ cm$^{-3}$ for the outburst revealing near equilibrium plasma, however see the next section on line detections.  EPIC spectral analysis reveal ionization time scales that are different with better statistical errors (better constrained) for the quiescence. The outburst will be elaborated in the Discussion section. 
We have varied the abundances of  N, O, Ne, Mg, Si, and Fe during the fits and found that they are close to solar values or slightly enhanced as the result of global fitting except oxygen seems slightly under-abundant with O/O$_{\odot}$=0.7-1.0 and iron is under-abundant by Fe/Fe$_{\odot}$=0.5-1.1  (see also Sec. \ref{sec:lines}). We find that the statistical errors on the abundances are large.

\begin{deluxetable*}{llllll}
\tablewidth{0pt}
\tablecaption{Spectral Parameters of the
Fits to the Quiescent and Outburst phases of  the RGS Spectra of Z Cha.  \label{tab:rgs-sp}}
\tablehead{
Model  & Parameter & Quiescence-1 & Quiescence-2 & Outburst-1 & Outburst-2  
}
\startdata
pcfabs  & N$_H$  & $0.14^{+0.03}_{-0.02} $ & $0.14^{+0.02}_{-0.02} $ & $0.12^{+0.05}_{-0.05} $ & $0.6^{+0.2}_{-0.3} $ \\
          & (10$^{22}$atoms cm$^{-2}$) &    &    &    &  \\
        & covering fraction & $0.8 < $ & $0.85 < $ & $0.7 < $  & $0.87^{+0.03}_{-0.10} $ \\
pcfabs   & N$_H$ & N/A & N/A & $1.7^{+\infty}_{-0.6}$ & N/A \\
          & (10$^{22}$atoms cm$^{-2}$) &    &    &    &  \\
        & covering fraction & N/A & N/A & $0.78^{+0.10}_{-0.02} $ & NA \\
  zxipcf  & N$_H$  & N/A & $1.1^{+1.3}_{-0.6}$ & N/A   & $2.9^{+5.6}_{-2.1}$ \\
   &  (10$^{22}$atoms cm$^{-2}$) &  &   &   &  \\ 
          & log($\xi$)  & N/A  & $3.6^{+0.4}_{-0.3}$ & N/A  &  $2.8^{+0.4}_{-0.5}$ \\
        & cov. frac. &  N/A   & $0.7 <$ & N/A  &  $0.5^{+0.1}_{-0.2}$  \\         
\hline
VNEI &  kT\ (keV)    &   N/A & $4.7^{+0.7}_{-0.6} $ & N/A & $0.6^{+0.1}_{-0.1}$ \\
      &  $\tau\ (s cm^{-3})$  &  N/A  & $8.7^{+\infty}_{-4.0}\times 10^{11}$ &  N/A & $1.1^{+0.3}_{-0.4}\times 10^{11}$       \\
      & K$_{vnei}$ & N/A &  $1.8^{+0.1}_{-0.1}\times 10^{-3}$ & N/A  & $2.6^{+1.0}_{-1.3}\times 10^{-4}$  \\
\hline
CEVMKL       & $\alpha$   & $1.9^{+0.2}_{-0.2} $ & N/A  & $0.5^{+0.3}_{-0.4} $ & N/A \\
       & kT$_{max}$\ (keV)       & $9.1^{+1.7}_{-1.2}$ & N/A & $0.9^{+0.3}_{-0.2}$ & N/A  \\
 & K$_{cevmkl}$ & $6.5^{+0.3}_{-0.4}\times 10^{-3}$  & N/A & $9.5^{+0.6}_{-1.2}\times 10^{-4}$  & N/A \\
\hline 
 & N/N$_{\odot}$     & 1.0 (fixed)  &                     1.0 (fixed) & $3.6^{+2.3}_{-2.3} $  & $3.1^{+2.3}_{-2.1} $ \\
 & O/O$_{\odot}$      & $1.3^{+0.5}_{-0.4} $ & $2.1^{+0.8}_{-0.6} $  & $1.6^{+0.4}_{-0.4} $ & $0.7^{+0.2}_{-0.2} $ \\
 & Ne/Ne$_{\odot}$      & $2.2^{+1.2}_{-1.1} $  & $3.3^{+1.5}_{-1.5} $ & $1.9^{+1.2}_{-1.1} $ & $1.2^{+0.5}_{-0.5} $ \\ 
 & Mg/Mg$_{\odot}$         & $2.4^{+1.4}_{-1.2} $  &$2.3^{+1.5}_{-1.5} $ &  1.0 (fixed) & 1.0 (fixed) \\ 
 & Si/Si$_{\odot}$   &   $3.6^{+3.0}_{-2.5} $  & $3.6^{+3.5}_{-3.2} $ & 1.0 (fixed) & 1.0 (fixed) \\
 & Fe/Fe$_{\odot}$   &  $1.0^{+0.3}_{-0.3} $ & $0.8^{+0.2}_{-0.3} $ & $1.1^{+0.3}_{-0.2} $ & $0.9^{+0.3}_{-0.2} $  \\
\hline
 & $\chi^2_{\nu} (\nu)$        & 1.18 (372)  & 1.22 (369)  &  1.29 (366) & 1.31 (365)    \\
\hline
 Flux & (10$^{-12}$erg~cm$^{-2}$s$^{-1}$)  & 4.1$^{+0.6}_{-0.6}$ & 4.0$^{+0.4}_{-0.3}$ & 2.2$^{+1.7}_{-0.7}$  & 1.3$^{+0.9}_{-0.7}$    \\
 Luminosity & (10$^{30}$erg~s$^{-1}$)  & 4.7$^{+0.6}_{-0.7}$ &  4.5$^{+0.5}_{-0.3}$ & 2.5$^{+1.9}_{-0.8}$   &  1.5$^{+1.0}_{-0.8}$ \\
\enddata  
\tablecomments{
Fits are performed using RGS1+2 data in the 0.4-2.0 keV range. 
$N_H$ is the absorbing column, $\alpha$ is the index
of the power-law emissivity function of CEVMKL model with ($dEM=(T/T_{max})^{\alpha -1} 
dT/T_{max}$), $T_{max}$ is the maximum temperature for the
CEVMKL model,
$K_{CEVMKL}$ is the normalization for the CEVMKL model. 
 NEIvers.3.0.9 plasma code with ATOMDB database was assumed for VNEI fits and $K_{VNEI}$ is the normalization for the VNEI model.
All errors are calculated at 90\% confidence limit for a single parameter. 
The unabsorbed X-ray flux and the luminosities are given in the range 0.2-10.0 keV. To calculate luminosities, a the distance of 97 pc is assumed.}
\end{deluxetable*}

\subsection{Line emissions in quiescence and outburst revealed by RGS}\label{sec:lines}

\begin{figure*}[ht] 
\begin{center}
\includegraphics[height=6cm,width=8cm,angle=0]{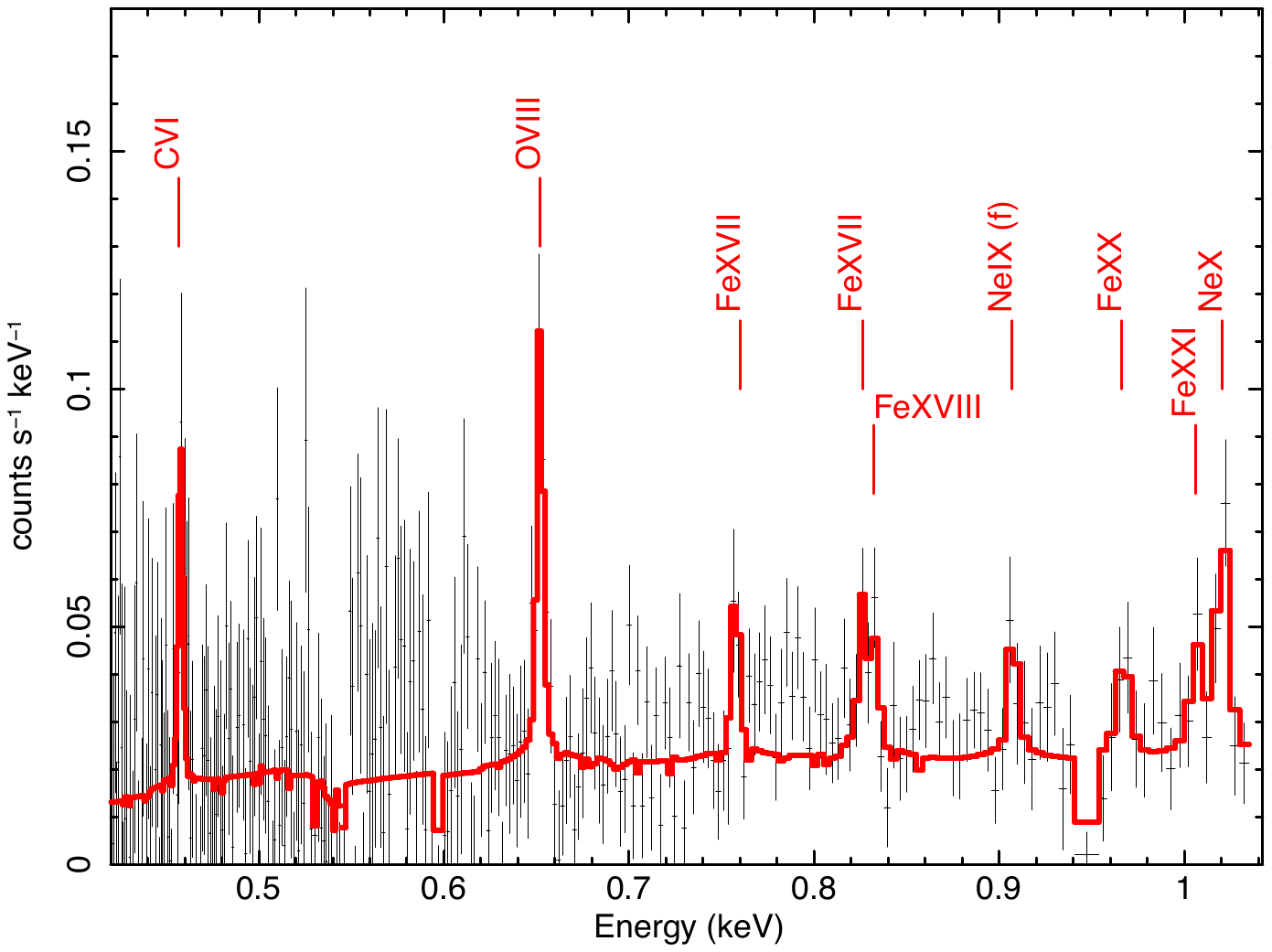}
\includegraphics[height=6cm,width=8cm,angle=0]{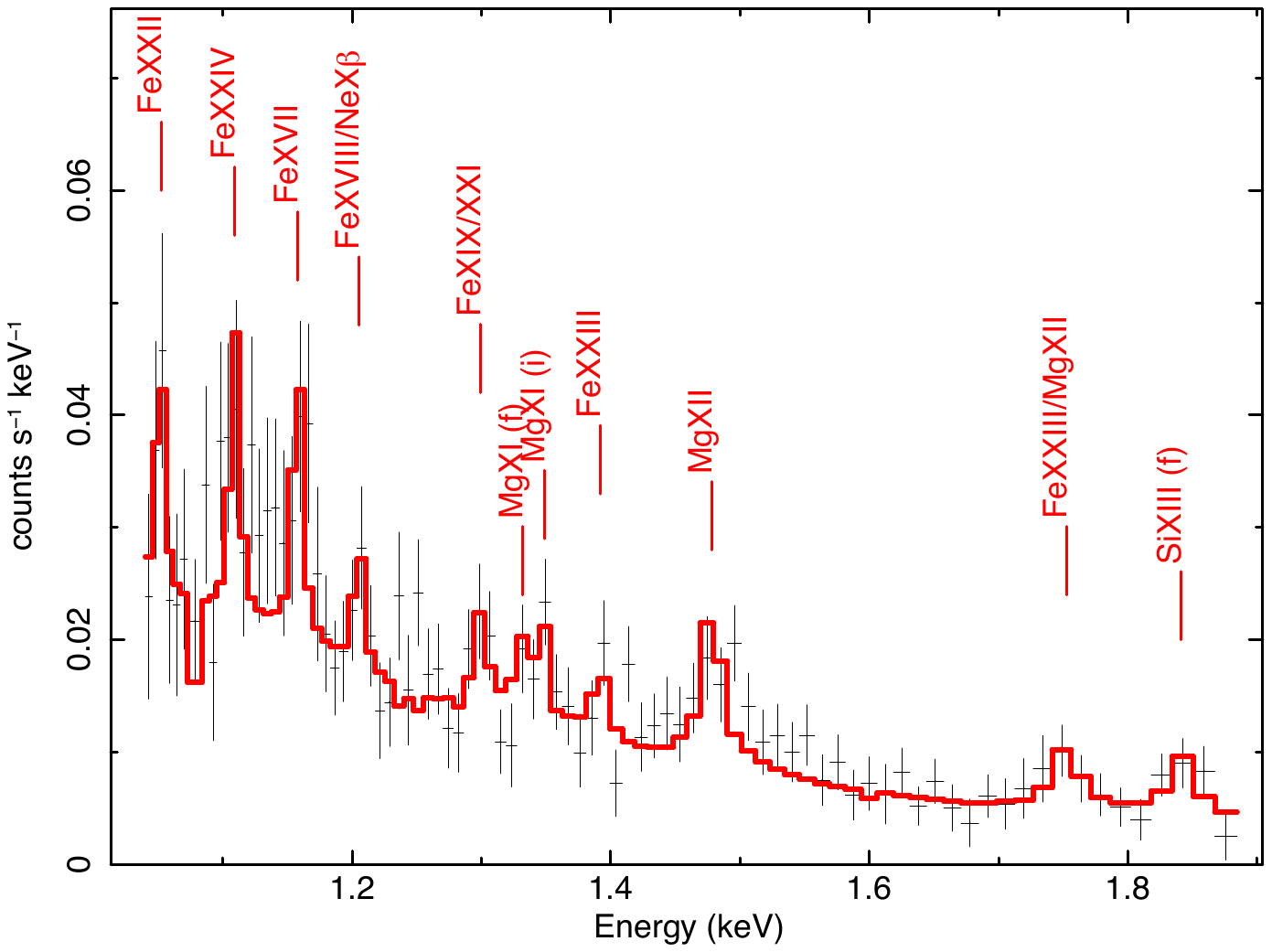}\\
\vspace{-0.5cm}
\includegraphics[height=4.5cm,width=5.5cm]{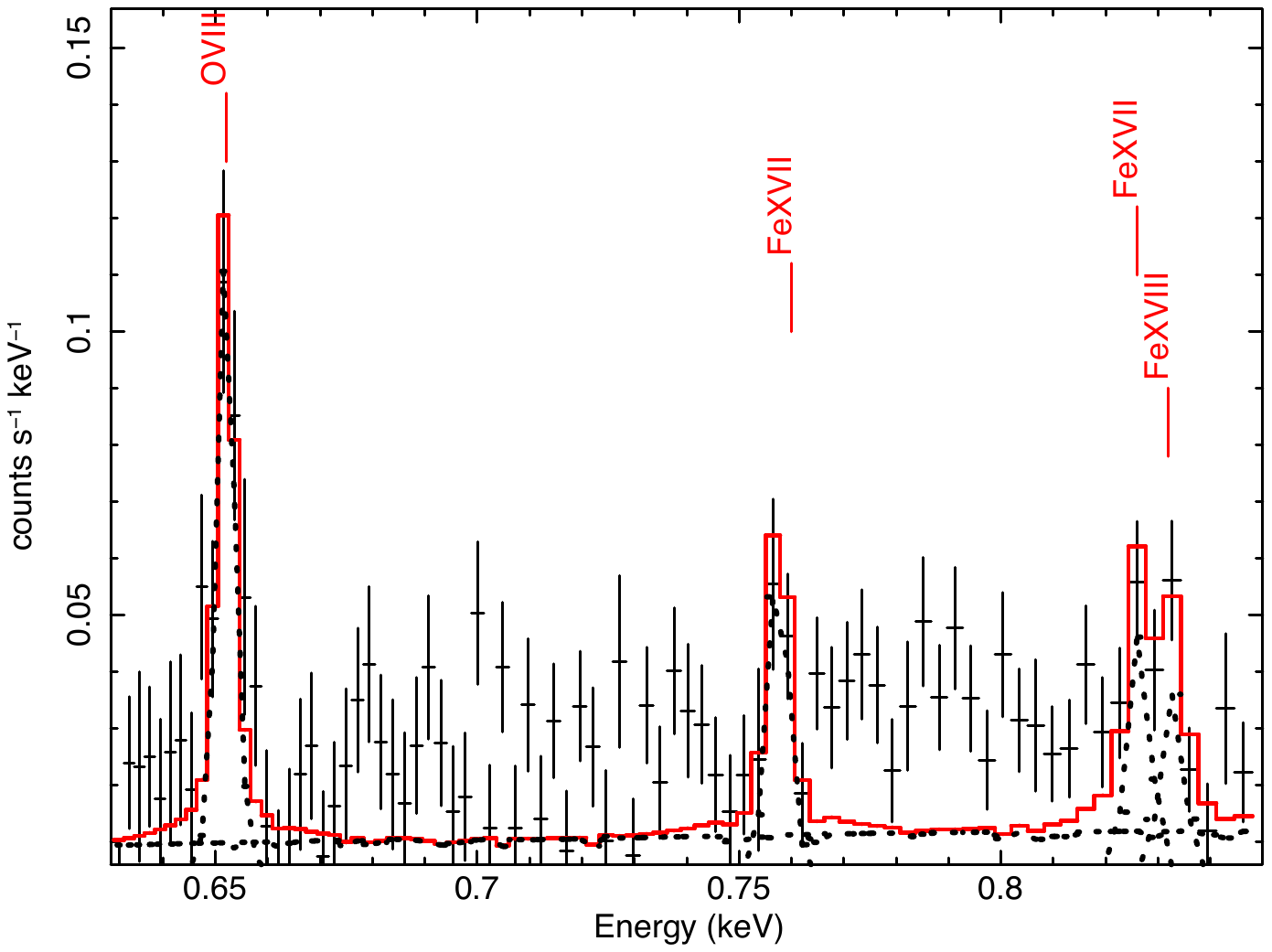}
\includegraphics[height=4.5cm,width=5.5cm]{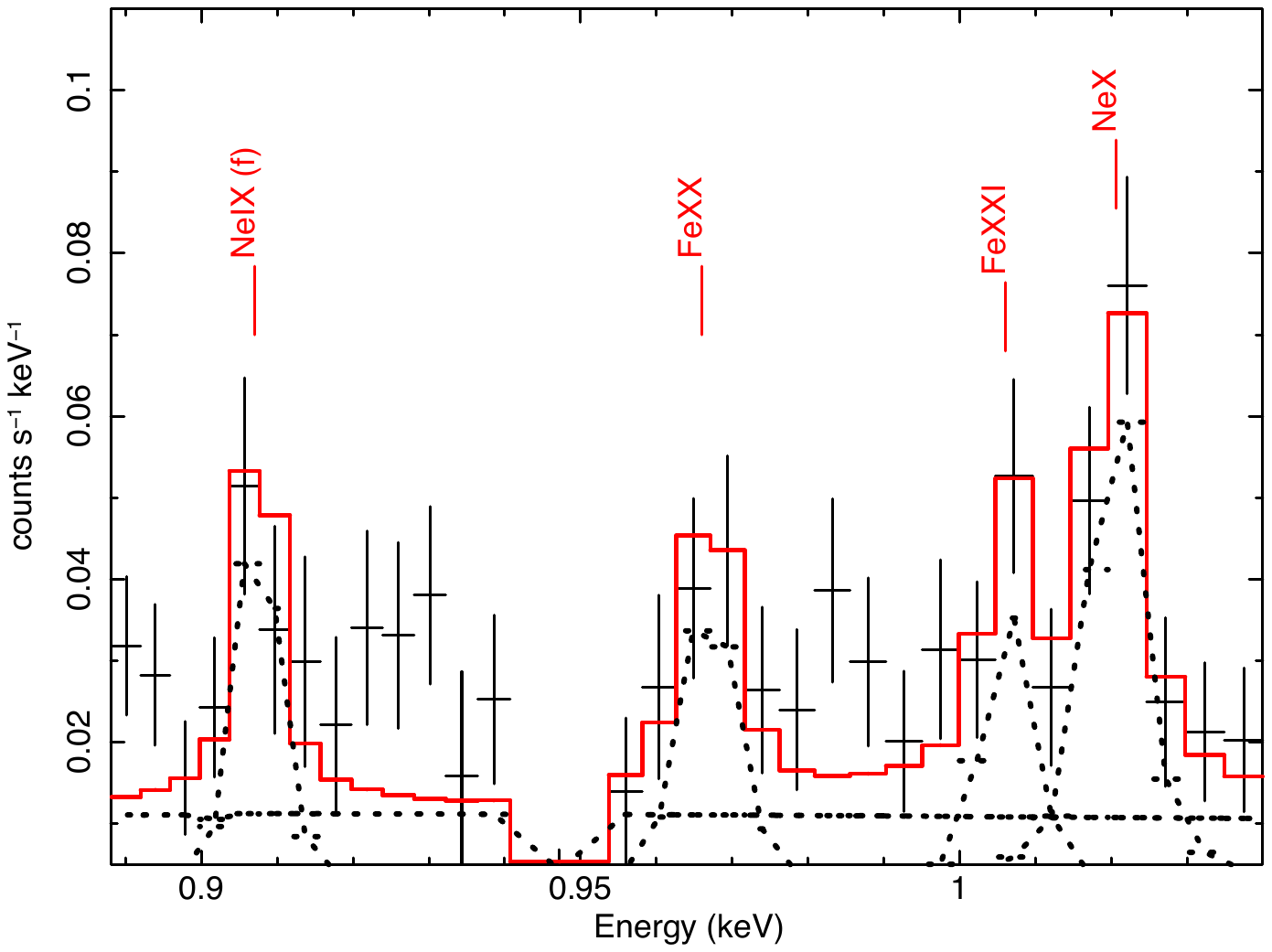}
\includegraphics[height=4.5cm,width=5.5cm]{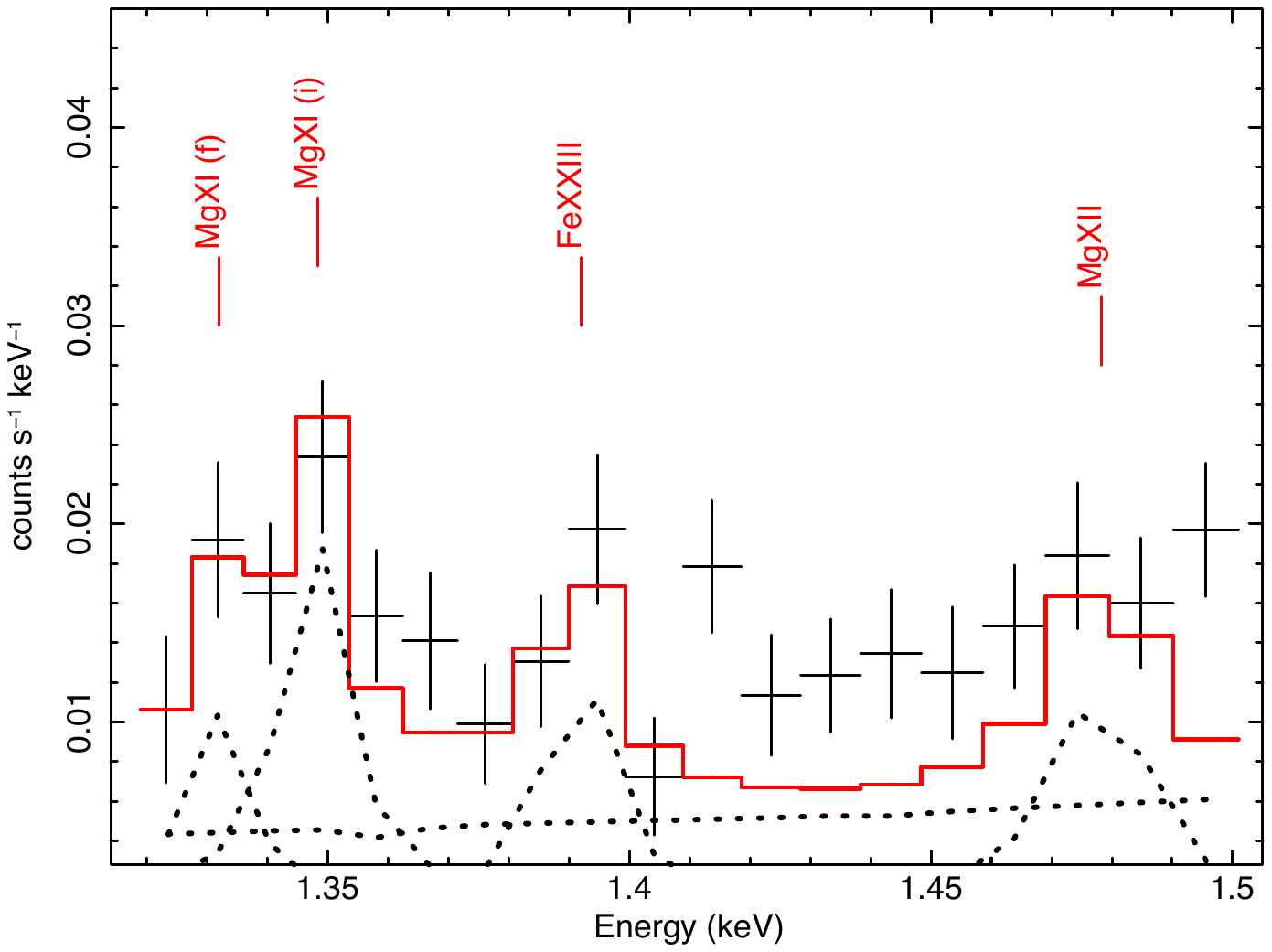}
\caption{
Averaged RGS 1$+$RGS 2 spectra of Z Cha in quiescence. The top panel shows identified lines labeled individually.
The fitted high resolution RGS spectrum is in the left hand panel of Figure \ref{fig:rgs-sp}. 
The three bottom panels show selected examples of fitted lines with a GAUSS emission model as displayed on Table \ref{tab:qlin}. \label{fig:qlin}
}
\end{center}
\end{figure*}

\begin{figure*}[ht] 
\begin{center}
\includegraphics[height=6cm,width=8cm]{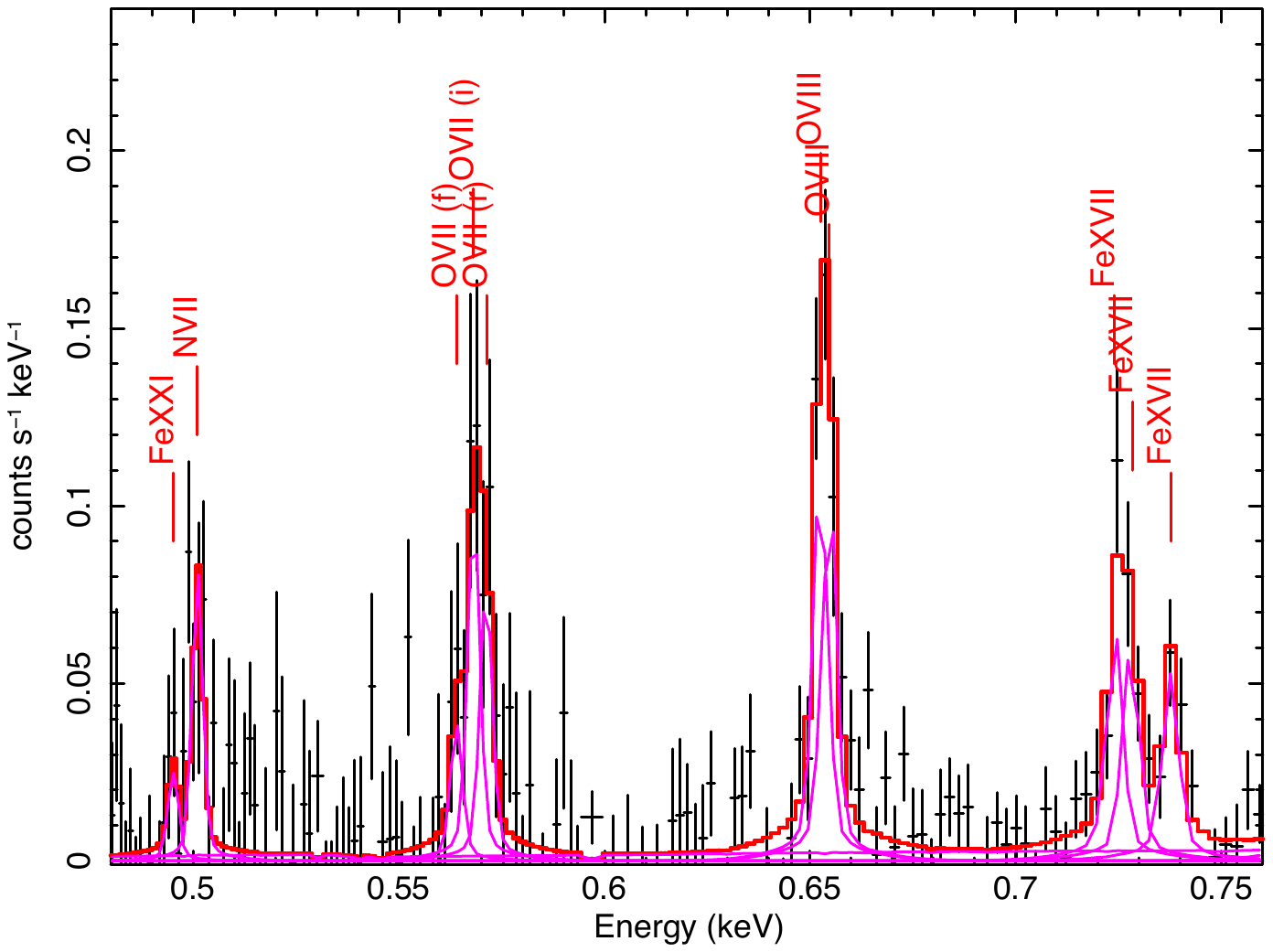}
\includegraphics[height=6cm,width=8cm]{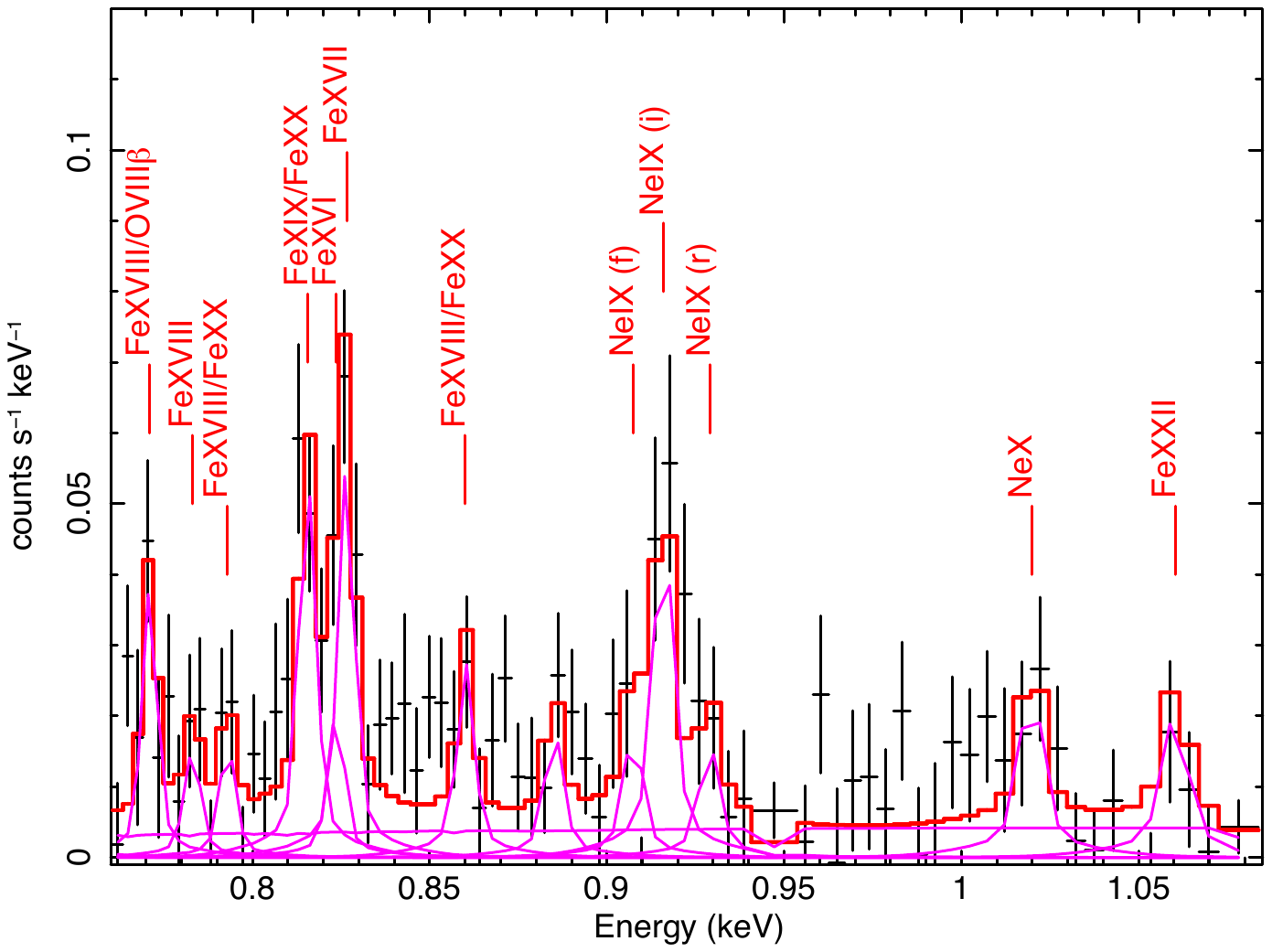}\\
\vspace{-0.5cm}
\includegraphics[height=4.5cm,width=5.5cm]{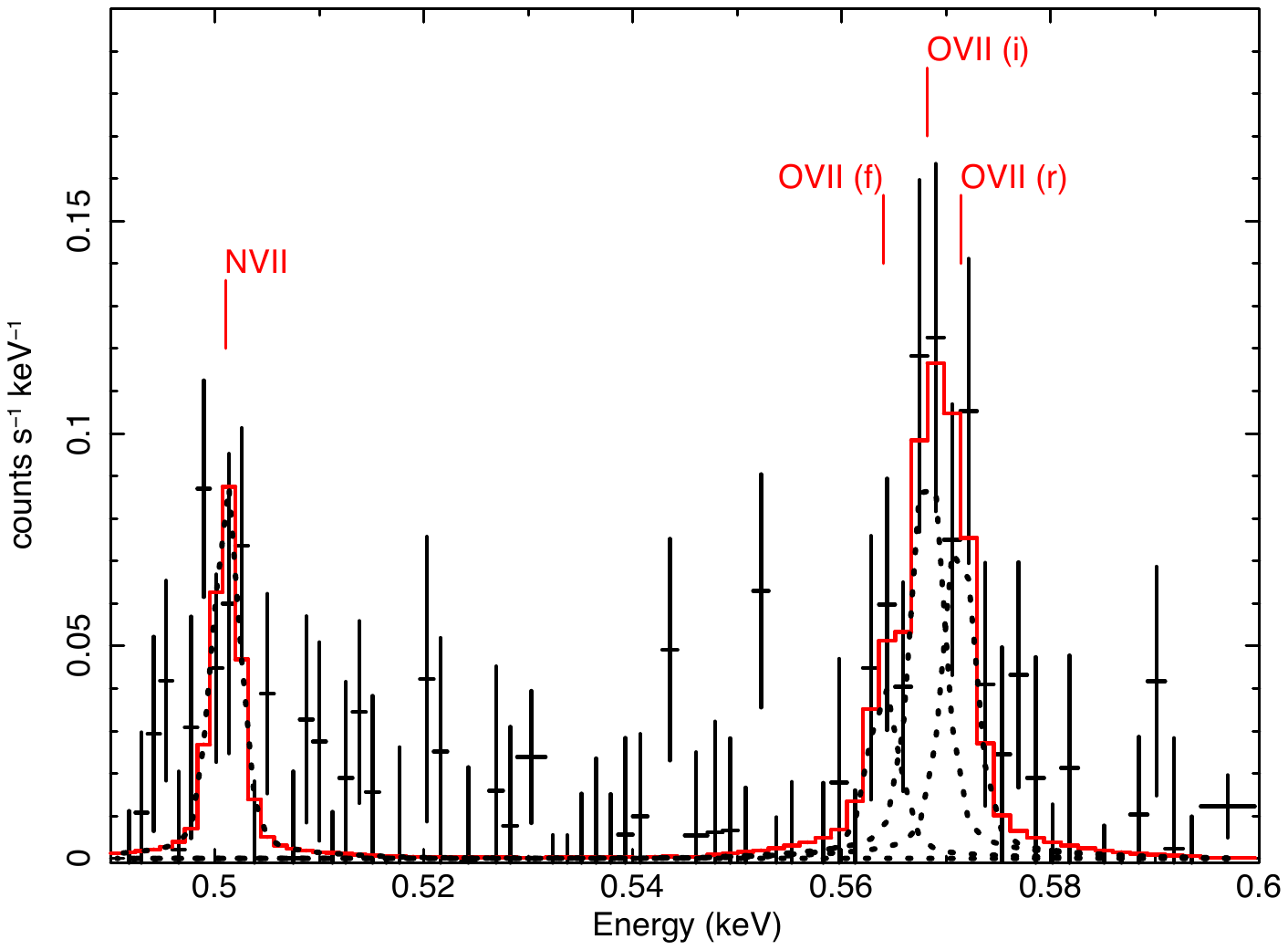}
\includegraphics[height=4.5cm,width=5.5cm]{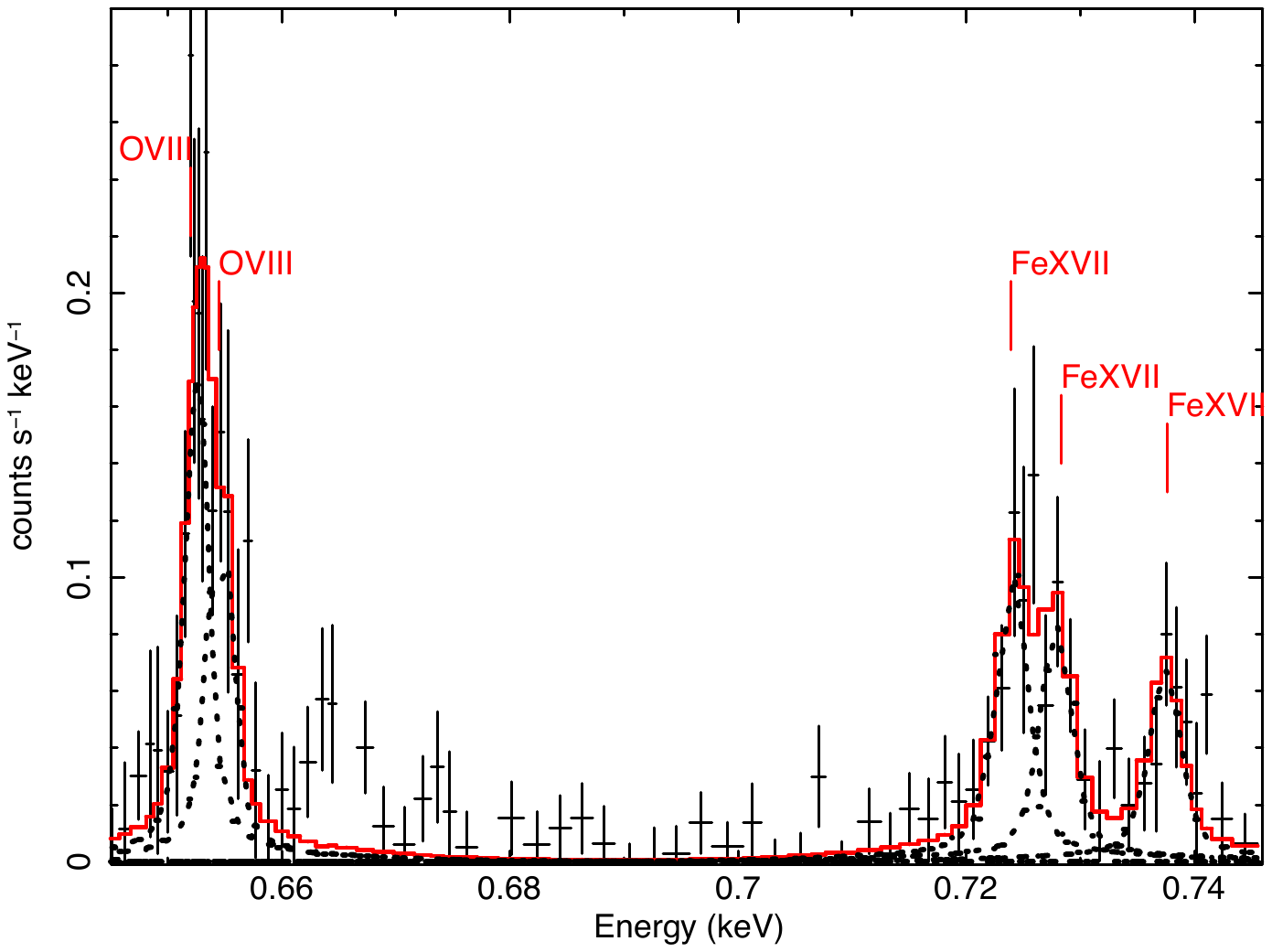}
\includegraphics[height=4.5cm,width=5.5cm]{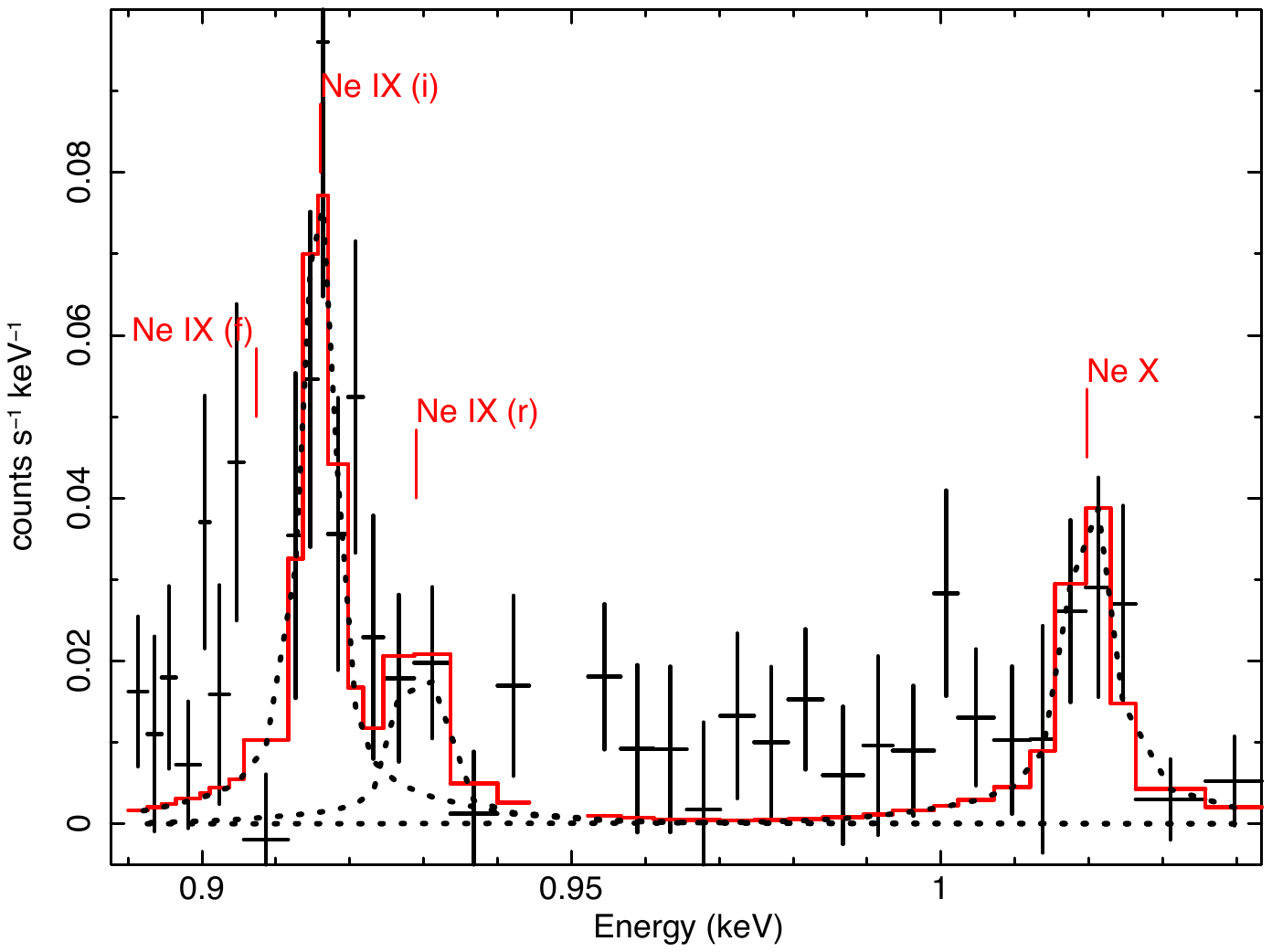}
\caption{
Averaged RGS 1$+$RGS 2 spectra of Z Cha in outburst. Top left and right hand panels show identified lines labeled individually.
The fitted high resolution RGS spectrum is in the right hand panel of Figure \ref{fig:rgs-sp}. The bottom three panels show selected examples of fitted lines with a GAUSS emission model as displayed on Table \ref{tab:olin}. \label{fig:olin}}
\end{center}
\end{figure*}

For a better  understanding of  the characteristics of X-ray emission, we exploit the RGS spectra to detect lines that would reveal the state of the plasma independent of the
continuum characteristics as would be dominant in the moderate spectral resolution EPIC data.  We note here that \xmm\ RGS energy range is between 0.4-2.2 keV which is smaller than the band for EPIC data (0.3-10.0 keV). Table \ref{tab:qlin} and  Table  \ref{tab:olin} list the
detected lines for the RGS spectra, in quiescence and outburst, respectively. For line identifications XSPEC task {\it identify} and ATOMDB database (see Sec. \ref{sec:eqsp}) have been used.
The calculations are made via fits with a Gaussian model to individual or collections of 2-3 lines if there is blending along with a power law for the base/continuum around the line feature. We have tabulated lines with photon fluxes
F$_{line} > $3$\times$10$^{-6}$phot.~cm$^{-2}$s$^{-1}$ ($>$4$\times$10$^{-15}$erg~cm$^{-2}$s$^{-1}$).  Tables denote line centers in keV and \AA\ together with the photon fluxes of the lines as derived from the normalization of the fits and the corresponding energy flux. All errors 
are given at the 90\% confidence limit for a single parameter.  The detected lines and example fits to individual lines are displayed in Figures \ref{fig:qlin} and \ref{fig:olin} for the quiescence and outburst, respectively.

In the quiescent state, we detect all the L$\alpha$ (H-like) lines of C (C VI), O (O VIII), Ne (Ne X), and Mg (Mg XII) that are in the energy range of the RGS. The strongest line is the H-like O VIII with  a flux of (2.7-4.6)$\times$10$^{-14}$erg~cm$^{-2}$s$^{-1}$.  In general,  the He-like lines of metals (Ne IX,  Mg XI, Si XIII), except for C which is not available, are recovered.  None of the O VII lines (He-like oxygen) have been detected. RGS resolves the three components of the He-like lines that are forbidden, intercombination, and resonance line emissions. Among these, no resonance line of any of the He-like triplets have been recovered along with no intercombination lines except for Mg XI. All the forbidden emission lines of the triplets of Ne, Mg, and Si have been detected with a range of energy fluxes (0.6-3.9)$\times$10$^{-14}$erg~cm$^{-2}$s$^{-1}$.  Oxygen reveals no He-like emission lines (of the triplet).  The rest of the detected emission lines are L-shell lines of iron from  Fe XVII to Fe XXIII. The statistical quality does not allow disentangling some of the Fe lines. There may be blending or the spectral resolution is not high enough to tell some of the line emissions apart. 

In the outburst state, we detect the H-like (L$\alpha$) lines, N VII, O VIII, and Ne X along with the He-like emission lines of  O (O VII) and Ne (Ne IX). The L$\alpha$ emissions show a range of energy fluxes (0.1-4.9)$\times$10$^{-14}$erg~cm$^{-2}$s$^{-1}$.  The data quality (and the X-ray temperature) do not allow for detections above 1.1 keV. We do not detect Mg or Si lines.  The detected H-like line fluxes (energy flux) of O, and Ne are consistent with quiescent values with some increase in O VIII and some decrease in Ne X. More importantly, this time the resonant emission lines of the He-like triplets are detected. They are weak, but relatively stronger than the forbidden lines. The highest flux is in the intercombination lines of the triplets of O and Ne.  The forbidden emission lines of O and Ne are detected in outburst where  the flux of Ne IX  in  outburst  is half the flux in quiescence. No comparisons can be made for  O VII. The energy flux of the forbidden line of oxygen (outburst) is about half the intercombination emission line of the triplet. O VIII (H-like emission line) have similar fluxes in quiescence and outburst only slightly higher in outburst. Several iron L-shell lines are detected between Fe XVII--Fe XXIII with energy fluxes in a range (0.3-1.8)$\times$10$^{-14}$erg~cm$^{-2}$s$^{-1}$ where the strongest emission is in Fe XVII. 

In general, total emission line flux is similar in both quiescence and outburst with $\sim$ 3$\times$10$^{-13}$erg~cm$^{-2}$s$^{-1}$.

\begin{deluxetable*}{llll}
\tablewidth{0pt}
\tablecaption{The list of recovered emission lines in the RGS1$+$2 spectrum obtained in quiescence. \label{tab:qlin}}
\tablehead{
Ion  &  Line Center $E_c$ & $K_{Gaussian}$  & Flux 
}
\startdata
 &   (keV)--(\AA) & 10$^{-5}$ & 10$^{-13}$ \\
 &    & Phot. cm$^{-2}$\ s$^{-1}$  & erg~cm$^{-2}$\ s$^{-1}$ \\
 \hline
C VI   &   $0.4577^{+0.0023}_{-0.0007}$--$27.088^{+0.04}_{-0.13}$ & $1.9^{+1.3}_{-1.1}$  &  0.14   \\
O VIII &    $0.6520^{+0.0013}_{-0.0005}$--$19.015^{+0.015}_{-0.038}$  & $3.6^{+0.8}_{-1.0}$ & 0.39  \\
Fe XVII   &    $0.7574^{+0.0014}_{-0.0017}$--$ 16.370^{+0.036}_{-0.030}$  & $ 0.9^{+0.6}_{-0.4}$ & 0.12  \\
Fe XVII &    $0.8261^{+0.0013}_{-0.0010}$--$ 15.008^{+0.009}_{-0.023}$  & $0.50^{+0.30}_{-0.12}$ & 0.07  \\
Fe XVIII/Fe XX  &  $0.8326^{+0.0009}_{-0.0014}$--$ 14.891^{+0.025}_{-0.016}$  & $0.6^{+0.2}_{-0.3}$ & 0.08  \\
Ne IX (f) &  $0.9074^{+0.0019}_{-0.0035}$--$13.663^{+0.053}_{-0.028}$ & $0.7^{+0.7}_{-0.4}$  & 0.10   \\
Fe XX  &    $0.9694^{+0.0026}_{-0.0059}$--$12.789^{+0.079}_{-0.034}$  & $0.6^{+0.2}_{-0.3}$ & 0.08  \\
Fe XXI &    $1.0065^{+0.0030}_{-0.0033}$--$12.318^{+0.037}_{-0.041}$  & $0.6^{+0.2}_{-0.3}$ & 0.08  \\
Ne X  &     $1.0206^{+0.0023}_{-0.0020}$--$12.148^{+0.024}_{-0.027}$ &  $1.0^{+0.6}_{-0.6}$ &  0.17 \\
Fe XXII &   $1.0516^{+0.0035}_{-0.0041}$--$11.790^{+0.046}_{-0.039}$ & $0.6^{+0.5}_{-0.4}$  & 0.11 \\
Fe XXIV &   $1.1094^{+0.0031}_{-0.0095}$--$11.175^{+0.096}_{-0.031}$ & $1.1^{+0.5}_{-0.5}$  & 0.20 \\
Fe XIX/Ni XXIII  &   $1.1579^{+0.0107}_{-0.0064}$--$10.708^{+0.059}_{-0.097}$ & $0.7^{+0.5}_{-0.5}$  & 0.13 \\
Fe XVIII/Ne X$\beta$ &  $1.2053^{+0.0085}_{-0.0079}$--$10.287^{+0.067}_{-0.072}$ & $0.40^{+0.30}_{-0.35}$ & 0.06 \\
Fe XXI  &    $1.2993^{+0.0073}_{-0.0070}$--$9.542^{+0.051}_{-0.053}$ & $0.5^{+0.3}_{-0.3}$  & 0.07 \\
Mg XI (f) &    $1.3319^{+0.0141}_{-0.0055}$--$9.308^{+0.038}_{-0.097}$ & $0.40^{+0.25}_{-0.20}$ & 0.06 \\
Mg XI (i) &  $1.3484^{+0.0057}_{-0.0035}$--$9.195^{+0.024}_{-0.038}$ & $0.9^{+0.3}_{-0.3}$ & 0.18 \\ 
Fe XXIII &  $1.3923^{+0.0004}_{-0.004}$--$8.907^{+0.022}_{-0.032}$ & $0.6^{+0.3}_{-0.3}$ & 0.13 \\ 
Mg XII  &  $1.4783^{+0.0084}_{-0.0111}$--$8.387^{+0.063}_{-0.047}$  & $0.7^{+0.5}_{-0.4}$ & 0.16 \\
Mg XII/Fe XXIII  &   $1.7525^{+0.0090}_{-0.0143}$--$7.074^{+0.058}_{-0.036}$  & $1.2^{+0.6}_{-0.5}$  & 0.36 \\
Si XIII (f)  &   $1.8410^{+0.0174}_{-0.0080}$--$6.735^{+0.029}_{-0.063}$ & $1.3^{+0.6}_{-0.6}$  & 0.39 \\
\hline
 &    &  total line flux = &  3.08  \\
\enddata   
\tablecomments{
Gaussian line fits are performed using co-added RGS1 and RGS2 data in the 0.4-2.2 keV range.
A constant Hydrogen column density of $0.18\times10^{22}$atoms cm$^{-2}$ is used 
in the fits in accordance with Table \ref{tab:sp1}. 
$K_{Gauss}$ is the normalization for the Gaussian Line model. 
All lines are assumed to
be narrow as derived from the fits where the line widths $\sigma$ are constrained with the
spectral resolution of RGS and taken to be fixed at 0.001 keV.
All errors are calculated
at 90\% confidence limit for a single parameter. Any upperlimit is given at 2$\sigma$
significance level.}
\end{deluxetable*}

\begin{deluxetable*}{llll}
\tablewidth{0pt}
\tablecaption{The list of recovered emission lines in the RGS1+2 spectrum obtained in outburst.  \label{tab:olin}}
\tablehead{
Ion  &  Line Center $E_c$ & $K_{Gaussian}$   & Flux
}
\startdata
 &   (keV)--(\AA) & 10$^{-5}$ & 10$^{-13}$  \\
 &    & Phot. cm$^{-2}$\ s$^{-1}$ &  \fluxcgs\  \\
\hline
Fe XXI/Ar XVI & $0.4952^{+\infty}_{-0.004}$--$25.037^{+0.023}_{-\infty}$ & $0.55^{+0.62}_{-\infty}$  &  0.04  \\
N VII &   $0.5010^{+0.0005}_{-0.0010}$--$24.747^{+0.053}_{-0.021}$ & $1.9^{+0.8}_{-0.8}$  &  0.15  \\
O VII (f) & $0.5640^{+0.0011}_{-0.0025}$--$21.983^{+0.097}_{-0.042}$ & $1.4^{+1.2}_{-1.1}$  &  0.12  \\
O VII (i) & $0.5682^{+0.0003}_{-0.0006}$--$21.821^{+0.023}_{-0.011}$ & $3.2^{+1.5}_{-1.5}$  &  0.27  \\
O VII (r) & $0.5712^{+0.0016}_{-0.0017}$--$21.706^{+0.064}_{-0.060}$ & $2.6^{+1.6}_{-1.6}$  &  0.23  \\
O VIII &    $0.6525^{+0.0005}_{-0.0004}$--$19.001^{+0.011}_{-0.014}$  & $2.5^{+0.7}_{-0.6}$  & 0.28 \\
O VIII  &    $0.6545^{+0.0013}_{-0.0005}$--$18.943^{+0.014}_{-0.037}$  & $2.0^{+0.7}_{-0.9}$  & 0.21 \\
Fe XVII &    $0.7239^{+0.0009}_{-0.0006}$--$17.127^{+0.014}_{-0.021}$  & $1.6^{+0.5}_{-0.6}$  & 0.18 \\
Fe XVII &    $0.7279^{+0.0008}_{-0.0010}$--$17.033^{+0.023}_{-0.019}$  & $1.3^{+0.5}_{-0.5}$  & 0.15 \\
Fe XVII & $0.7376^{+0.0017}_{-0.0004}$--$16.809^{+0.009}_{-0.038}$  & $1.1^{+0.4}_{-0.4}$  & 0.13 \\
Fe XVIII/O VIII$\beta$ & $0.7710^{+0.0006}_{-0.0030}$--$16.081^{+0.062}_{-0.012}$  & $0.55^{+0.30}_{-0.30}$  & 0.07 \\
Fe XVIII &    $0.7830^{+0.0043}_{-0.0043}$--$15.834^{+0.087}_{-0.086}$  & $0.44^{+0.30}_{-\infty}$  & 0.06 \\
Fe XVIII/Fe XX & $0.7930^{+0.0017}_{-0.0004}$--$15.635^{+0.008}_{-0.033}$  & $0.33^{+0.30}_{-\infty}$  & 0.04 \\
Fe XX/ Fe XIX & $0.8155^{+0.0004}_{-0.0012}$--$15.203^{+0.022}_{-0.007}$  & $1.3^{+0.4}_{-0.4}$  & 0.16 \\
Fe XVI/Fe XIX/Fe XX & $0.8237^{+0.0010}_{-0.0031}$--$15.052^{+0.056}_{-0.018}$  & $0.75^{+0.40}_{-0.40}$  & 0.09 \\
Fe XVII & $0.8266^{+0.0018}_{-0.0006}$--$14.999^{+0.023}_{-0.02}$  & $0.8^{+0.4}_{-0.4}$  & 0.10 \\
Fe XX/ Ne K$_{\alpha}$/Fe XVIII & $0.8600^{+0.0008}_{-0.0090}$--$14.417^{+0.15}_{-0.013}$  & $0.5^{+0.3}_{-0.2}$  & 0.07 \\
Fe XXI  & $0.8860^{+0.0057}_{-0.0039}$--$ 13.994^{+0.06}_{-0.09}$  & $0.4^{+0.3}_{-0.3}$  & 0.06 \\
Ne IX (f)  &  $0.9074-13.663$ (fixed) & $< 0.37 $  & $< 0.05$  \\
Ne IX (i)  &  $0.9160^{+0.0009}_{-0.0010}$--$13.535^{+0.014}_{-0.013}$ & $1.0^{+0.6}_{-0.4}$  & 0.15  \\
Ne IX (r)  &  $0.9290^{+0.0049}_{-0.0028}$--$13.335^{+0.040}_{-0.039}$ & $0.5^{+0.4}_{-0.4}$  & 0.08  \\
Ne X  & $1.0198^{+0.0040}_{-0.0040}$--$12.157^{+0.048}_{-0.047}$ &  $0.6^{+0.4}_{-0.4}$  &  0.10 \\
Fe XXII &   $1.0603^{+0.02}_{-\infty}$--$11.693^{+0.224}_{-\infty}$ & $0.2^{+0.4}_{-\infty}$  & 0.03 \\
\hline
 &    &  total line flux = &  2.82  \\
\enddata   
\tablecomments{
Gaussian line fits and line fluxes performed using co-added RGS1 and RGS2 data in the 0.4-2.2 keV range.
A constant Hydrogen column density of $0.12\times10^{22}$atoms cm$^{-2}$ is used
in the fits in accordance with Table \ref{tab:sp2}.
$K_{Gauss}$ is the normalization for the Gaussian Line model. All lines are assumed to
be narrow as derived from the fits where the line widths $\sigma$ are constrained with the 
spectral resolution of the RGS and taken to be fixed at 0.001 keV.
All errors are calculated
at 90$\%$ confidence limit for a single parameter. All upper limits are given at 2$\sigma$
significance level.}
\end{deluxetable*}

\section{Discussion}\label{sec:discuss}

\subsection{Line diagnosis and the nature of the accretion flows in Z Cha}\label{linedia}

Collisional coronal plasma diagnosis utilize  line ratios that are sensitive to electron density and electronic temperature. Several processes are involved. Collisional excitation form lines from the ground level with or without recombination. H-like radiative recombination can be important at high temperatures or depending on temperature and densities, He-like recombinations may even dominate. In general at high temperatures (in the X-ray emission regime), the plasma can be dominated with collisional  excitation or with radiative recombinations revealing photoionized plasma. Hybrid plasmas show both characteristics where recombination and collisional processes occur. The ratios of H-like and He-like ions can be used for collisional plasma diagnostics.  The He-like triplet comprises  the resonance line ($r$: 1s$^2$ $^1$S$_0$ -- 1s2p $^1$P$_{1}$), the intercombination lines ($i$: 1s$^2$ $^1$S$_0$ -- 1s2p $^3$P$_{2,1}$) and the forbidden line ($f$: 1s$^2$ $^1$S$_0$ -- 1s2p $^3$S$_{1}$) emissions corresponding to transitions between n=2 and the n=1 ground state shell.  There are two particular ratios between these triplet line emissions: 1)  $R= f / i$, $R(n_e)$ ratio yielding electron density diagnostics, and 2) 
$G= (f+i) / r$ which is a function of electron temperature T$_e$. Large $G$ ratios and a low resonance line indicates photoionization (and radiative recombination) dominated plasmas where $G<$ 4 along with an intense resonance line is a signature of hybrid plasmas where both collisional processes and photoionization is effective. At low $G$ ratios collisional excitation (processes) dominates.  G ratios can yield a good measure for electron temperature however this ratio can vary by two orders of magnitude  when column density increases/changes (from 10$^{17}$ to 10$^{24}$ cm$^{-2}$; \citealt{2007Porter}). The $R$ ratio is a good density probe.  This ratio tends from high values at high  $f$ intensity, to low values where intercombination lines are dominant. The low R ratios are indicative of high densities, so prominent forbidden lines signal low electron densities.  

We have derived R and G ratios using the O, Ne, Mg triplets that have been detected for Z Cha. The R and G ratios using these triplets have also been calculated in \citet{2014Schlegel} for other accreting CVs.  In quiescence, the triplets of  Ne, Si show only the forbidden line emission, so  no R or G ratios can be derived. Only Mg shows forbidden and the intercombination lines. We calculate that R$\simeq$0.44 .  No resonance line is detected for any He-like triplet in quiescence. Therefore no ratio of the H-like line to the resonance {\it r} line of the He-like triplet can be calculated.  However, the same ratio is 1.4-1.7 using the forbidden lines, {\it f}, of neon and magnesium. In outburst, the He-like triplets of O and Ne are detected with the {\it f, i} and {\it r} emission lines. For oxygen (O VII) R$\simeq$0.44 and G$\simeq$1.8 . For neon R$\simeq$0.4 and G$\simeq$2.7 .  The ratio of the H-like line to the resonance, {\it r}, line of the He-like triplet is $\sim$ 1 for both oxygen and neon.  The same ratio is 1.7 using the forbidden lines (similar for both O and Ne), and the intercombination lines yield a ratio of  0.6-0.8 ({\it i} line intensities are brighter).  As a result, the low R ratios in outburst indicate higher densities compared to quiescence with n$_e \simeq$ 7$\times$10$^{11}$ cm$^{-3}$ for oxygen and n$_e \simeq$ 9$\times$10$^{12}$ cm$^{-3}$ for neon following from \citet{2000Silver}. The G ratios of oxygen and neon are indicative of regions with T$_e \simeq$ (2-3)$\times$10$^6$ K in accordance with \citet{2000Porquet}.  In quiescence, the non-detection of {\it r}, and {\it i} lines and  detection of {\it f}  lines (except for Mg IX) indicate  that the X-ray emitting plasma has lower density compared to the outburst and  the plasma is not collisional, most likely not in equilibrium, and  is not photoionized much, either. The comparable forbidden line emission to the H-like species also supports this. This condition improves in outburst and most regions have higher densities and more collisional (i.e., collisional excitation) owing to the {\it r} lines.  However,  prominent  {\it i} lines compared with even the H-like species indicate photoionization is appreciable (i.e., recombination) and thus, the X-ray emitting plasma may be  "hybrid". However, the relatively low flux in the resonance lines compared to intercombination and H-like lines indicate that a full collisional ionization equilibrium is not set in the plasma even in outburst. 

The iron L-shell lines can also be used as density and electron temperature probes.  We have detected some of these lines securely in the outburst RGS data. The ratio of  Fe XVII lines at 0.81-0.82 keV (15.01-15.26 \AA)  is correlated with electron temperature \citep[][for CVs]{2001Brown,2014Schlegel}. The lines at 15.01 \AA\ and 15.26 \AA\ are also known as 3C, resonance and 3D intercombination lines, respectively.  The entire set of Fe XVII 3s/3d (0.72-0.82 keV) 
defines  a temperature measure with intensity ratio I(16.77+17.05+17.10 \AA)/I(15.01+15.26+15.45 \AA) \citep{2005Chen}. This ratio is about 1.4 (assuming we have not recovered the 15.45 \AA\ line). This yields an electron temperature of T$\sim$ 6$\times$10$^{6}$ K consistent with the RGS spectral results. The ratio of Fe XVII lines which are 17.01/16.77 \AA\ and 17.10/17.05 \AA\  are calculated in a range (1.2-1.5) yielding electron densities $<$ 10$^{13}$ cm$^{-3}$ ; \citet{2014Schlegel} show the dependence of the ratio on densities (for CVs) and related references. The densities are in accordance with those calculated using triplet lines.
 
A more fundamental characteristic of a plasma is the ionization balance. The line spectrum is sensitive to deviations from ionization equilibrium.  Bremsstrahlung continuum immediately follows the changing plasma conditions so that the line-to-continuum ratio indicates deviations from the equilibrium (note that significant abundance changes can also effect line-to-continuum ratio).  Typically a plasma will reach CIE conditions when n$_e$t $\sim$ several$\times$10$^{13}$ s\ cm$^{-3}$ (i.e., the ionization parameter). For most atoms equilibrium may be achieved by a factor 10 earlier depending on electron density and plasma temperature where light atoms (e.g., C, O) may achieve equilibrium even earlier.
At low T ($<$10$^6$ K) L-shell ions dominate, and at high T, the ion population is dominated by He-like, H-like or bare ions \citep[see,][]{2010Smith}. For iron to reach ionization equilibrium, an ionization parameter of  several $\times$10$^{12}$ s\ cm$^{-3}$ may be sufficient. Given these characteristics and the derived values of ionization parameters for Z Cha, spectral results reveal plasma that is near ionization equilibrium, but not in equilibrium (see the discussion in the next section).

\subsection{Comparison of the quiescence and outburst in the X-rays}\label{compare}

An overall elaboration of the quiescence and outburst spectral analyses is necessary to understand the accretion flow characteristics in the Z Cha system.  Comparative studies of dwarf nova in quiescence and outburst is an essential tool to understand accretion physics in cataclysmic variables and related objects and the DIM phenomena. The outburst EPIC pn  light curve yields only a 13\% pulsed fraction at the orbital period of Z Cha which is 
1.788 hrs (see Sec.~\ref{sec:temp}) and lies below the period gap.  In quiescence the source shows deep eclipses in the optical and the X-ray light curve indicates non-detections in a similar time scale to eclipse durations of 319-329 s given by  \citet{2011Nucita} which indicates very large modulation depths. These modulation discrepancies (outburst and quiescence) show that the X-ray-emitting region is quite extended during the outburst which is a characteristic of hot accretion flows in the X-rays. The nondetection in quiescence can be due to occultation or absorption in the system of full/partial covering cold or ionized absorbers intrinsic to the system.  This is also revealed in the mean light curves in the B-band and 
the X-rays during outburst where the former shows very narrow deep dips and the latter shows a wider depression of the pulse profile over 0.3 phases which is about 1/3 of the orbit (at the least).

For the analysis in quiescence, we used CEVMKL or nonequilibrium ionization plasma model, VNEI,  in XSPEC for the continuum and line modeling (see the descriptions in Sec.~\ref{sec:eqsp}). Table~\ref{tab:sp1} shows six different fits using these plasma models along with different absorber models. The \rchisq\ of the fits indicate not much difference between the CEVMKL model that indicates collisional equilibrium or the nonequilibrium ionization model VNEI.  The plasma temperatures are a kT of 5.4-8.6 keV with the VNEI model and kT$_{max}$=7.2-13.0 keV for the CEVMKL  model fits with a source luminosity of (5.0-6.0)$\times$10$^{30}$\lumcgs using both plasma models. The lower limits on the temperatures represent values when a power law is added to the composite spectral model which yields a photon index of  index of 1.2-1.5 at a nonthermal luminosity of  (1.4-3.0)$\times$10$^{30}$\lumcgs. The power law component  statistically improves the fits, but are not any better than when a complex absorber model is applied to the spectrum. The spectral fits yield better \rchisq\ values when using partial covering absorbers as opposed to {\it tbabs} absorber model.  A partially-ionized absorber model, {\it zxipcf} gives an equivalent \nh=(3.4-5.9)$\times$10$^{22}$ cm$^{-2}$ and a log($\xi$)=3.5-3.7 with (50-60)\% covering fraction when VNEI model is used.  A CEVMKL plasma model yields an equivalent \nh=(9.4-11.0)$\times$10$^{22}$ cm$^{-2}$ and a log($\xi$)=1.81-1.95 with (35-37)\% covering fraction. These values are 40-80 times more than the hydrogen column density of cold partially covering absorber with  \nh=(0.11-0.19)$\times$10$^{22}$ cm$^{-2}$ that is consistent with all plasma models and the interstellar absorption towards the system. Finally,
the quiescent spectrum also shows Fe XXVI absorption line at 6.95$\pm$0.05 keV detected for the first time in a dwarf nova with line $\sigma$=0.05-0.10 and a depth in a range 0.04-0.16.  The optical depth at the line center follows from the line depth parameter as 0.32-0.65 which indicates an  optically thin absorbing media. Such iron absorption lines are indicators of warm absorbers in X-ray binaries \citep{2009Balman,2016Diaz-Trigo}.   We find that the partially covering ionized absorber that we use with the VNEI plasma model is consistent with this absorption line detection. In general, CEVMKL model finds a colder ionized  absorber with  higher equivalent hydrogen absorption by a factor of 3-5. 

The line diagnosis of quiescent RGS data shows no resonance lines of the He-like O, Ne, Mg, Si, and N. Only forbidden lines of  Ne, Mg, Si exist together with the H-like C, O, Ne, and Mg. The strongest line is O VIII with (2.7-4.6)$\times$10$^{-14}$\fluxcgs. The forbidden line emitting He-like lines of  Ne, Mg have  fluxes of (0.6-3.9)$\times$10$^{-14}$erg~cm$^{-2}$s$^{-1}$ which is only a  factor of 1.4-1.7 times less than the H-like line fluxes (of the same species). The quiescent X-ray emitting plasma, therefore, is not collisional lacking the resonance lines of all the He-like species and not in ionization equilibrium as line diagnosis indicates (see Sec.~\ref{linedia}). Note that EPIC reveals ionization timescale of $(2.8-6.0)$$\times$10$^{11}$ s~cm$^{-3}$ (consistent with the RGS result) indicating that plasma is near to but not in ionization equilibrium \citep[see also,][]{1999Liedahl,2010Smith}. This nonequlibrium ionization condition in quiescence is consistent with the nature of the ADAF-like hot advective accretion flows. Moreover, the broadband noise structure and the band-limited noise (f$_{break}$=1.0$\pm$0.4 mHz) detected in this system together with the X-ray lags (i.e. lagging UV) of 112-209 s\  \citep{2020Balman}; all support the ADAF-like, hot advective accretion flows in the inner disk during quiescence.  

During the outburst, EPIC data yield a plasma temperature of kT$_{max}$= 0.97-3.20 keV at a luminosity of (1.4-2.5)$\times$10$^{30}$\lumcgs\ where the temperatures are much lower compared to quiescence along with a diminished X-ray luminosity. A CEVMKL model of plasma is used in all four fits displayed on Table~\ref{tab:sp2} since a VNEI plasma model did not yield an acceptable \rchisq\ below 2 because it did not fit the iron line between 6-7 keV properly. The $\alpha$\ parameter which is the power law index of the distribution of plasma temperatures for the CEVMKL model is found to be $<$ 0.1 with best fit results $\sim$0.01 which is the value recovered for the NL systems \citep{2022Balman}, consistent with the high state nature of Z Cha in outburst.  There is more complex absorption in the system during the outburst where one of the cold partial covering absorber components is similar to the one in quiescence,  and the second one is larger by a factor of 15,  (8.7-10.7)$\times$10$^{22}$ cm$^{-2}$. On the other hand, the outburst EPIC spectra is also consistent with a partially ionized absorber model, {\it zxipcf}, that shows a very large  equivalent \nh=(2.30-2.95)$\times$10$^{24}$ cm$^{-2}$ with an ionization parameter log($\xi$) in a range 2.79-2.83 (covering fraction is 90\%), which is  about a factor of 25-50 times more than quiescence-value. This is the only fit with the CEVMKL model where the He-like weak iron line is correctly attributed by the plasma emission model (no need for an additional GAUSS model). However, this amount of partially ionized absorption is physically not consistent with observations of DN \citep{2017Mukai,2020Balman}. Outburst spectral modeling allows for a blackbody model or a power law model assumed together with CEVMKL plasma emission. This gives a plasma temperature $\sim$ 1 keV and an additional iron line at 6.64$^{+0.06}_{-0.07}$ keV is necessary, to yield acceptable fits. The blackbody model at 0.021-0.035 keV is significant at only 93\%  confidence level (slightly less than 2$\sigma$). The blackbody  luminosity is 3.4-0.7$\times$10$^{31}$\lumcgs\ (unabsorbed, 0.1-10 keV) which is several orders of magnitude less than standard disk model expectations at high states (also, below expectations of DIM). This value of luminosity is about only $\sim$ 20 times less than the bolometric blackbody luminosity  at this temperature.  A power law model along with CEVMKL  without  a blackbody, models the EPIC spectra well with a similar \rchisq\  along with the additional He-like iron line. The associated power law photon index is 1.5-1.7 at a luminosity of (3.0-3.9)$\times$10$^{29}$\lumcgs similar to NLs , but with a lower luminosity \citep{2022Balman} and also see
\citet{2015Balman,2023Dobrotka}\ .  

The outburst line-diagnosis show He-like O, and Ne with intercombination lines being the strongest along with  weaker resonance lines.  The R-ratios in outburst shows electron densities (7-90)$\times$10$^{11}$ cm$^{-3}$ calculated from O and Ne lines and the G-ratios yields electron temperatures of (2-3)$\times$10$^{6}$ K.  These indicate that the plasma is more collisional (resonance lines exist) and denser than quiescence, but yet not in a full collisional equilibrium as the RGS data indicate ionization timescales of (0.97-1.40)$\times$10$^{11}$ s\ cm$^{-3}$ (less than quiescence by 2-6 times). 
All detected lines are narrow limited with the resolution of RGS yielding Keplerian rotational velocities $<$1000 km/s which is low for any boundary layer and consistent with the nature of extended advective hot flows at 
sub-Keplerian speeds. The flow is most likely not homogeneous, but patchy with higher and lower density regions.  In the outburst the accretion flow is denser by a factor of $\sim$ 60 compared to quiescence calculated using ratio of ionization parameters (i.e., under the assumption of similar ionizing X-ray luminosity and similar location; $\xi$=L/(nr$^2$). Overall, X-ray emission from Z Cha shows ADAF-like advective hot flow characteristics in quiescence and outburst.
 
\subsubsection{A closer look into the comparison of EPIC and RGS spectra}

There are two handicaps in the spectral analysis of Z Cha in quiescence and outburst. The first is the very high equivalent \nh\ of partially ionized absorption attained in the outburst using EPIC data (only, but not by RGS).  
For this matter, in place of the {\it zxipcf} (an old version for the {\it warmabs} model for partially ionized absorption),
we utilized the XSTAR warm absorber model {\it warmabs}\footnote{https://heasarc.gsfc.nasa.gov/xstar/docs/html/node102.html; latest vers. 2.47 is used that involves better atomic database, level populations, line, rrc opacities and line emissivities}, which is a multiplicative local model that can be incorporated in any XSPEC distribution for spectral analysis. The model is a collection of warm absorption and photoionized emission models derived from the XSTAR program that calculates physical conditions and spectra of photoionized gasses. We used  {\it warmabs}  as a substitute of {\it zxipcf}  in QFit-5  that uses the VNEI model (see Table~\ref{tab:sp1}) for quiescence and OFit-3 in outburst.  The fit for quiescence yields the same \rchisq\  for  d.o.f. of 1399 (there is no covering fraction parameter, it is 1.0). The equivalent \nh=(0.5-0.7)$\times$10$^{22}$ cm$^{-2}$ with log($\xi$)=3.09-3.17. The ionization parameter is similar but the equivalent \nh\ is about 10 times less than predicted by {\it zxipcf}. We applied the same replacement in OFit-3 for the outburst 
using a CEVMKL model. We obtained \nh=(0.020-0.009)$\times$10$^{22}$ cm$^{-2}$ with log($\xi$)=1.33$^{+0.80}_{-0.80}$. The \rchisq = 1.38 for  d.o.f. of 459. This is not as good as OFit-3 where \rchisq = 1.24 for  d.o.f. of 450, however, it is quite acceptable. The discrepancy in the equivalent \nh\ parameter  between outburst and quiescence is about 30-50, but with more physically relevant equivalent \nh\ values in the outburst . Therefore, the equivalent \nh\ for the partially ionized absorption model in Table~\ref{tab:sp2} should be taken with caution.   Moreover, the ionization parameter for the warm absorber in outburst is different and lower (determined by the {\it warmabs} fit), indicating a colder absorber. Note here that the line diagnosis with RGS finds nonequilibrium plasma in quiescence, but the EPIC spectra can be modeled well by both NEI and CEI models.

 The RGS finds similar results to EPIC in quiescence, but a factor of 5 times less equivalent \nh\ for the  {\it zxipcf} model; all RGS results are similar to the results of  the {\it warmabs} fit for EPIC. In the outburst, the RGS fits have similar  ionization parameter to the  {\it zxipcf} absorber model for EPIC, but have a factor of 90  times less equivalent \nh\ than the {\it zxipcf} model. Compared with the  {\it warmabs} model-fit of EPIC, the equivalent \nh\ determined by RGS is only about 10 times more, but the ionization parameter is still, different and lower, indicating a colder absorber in the {\it warmabs} fit for EPIC. We note here that for a successful fit with RGS using
a {\it zxipcf} model, only VNEI plasma code gives fits with acceptable  \rchisq\ values under 2. This is consistent with the line diagnosis but not with EPIC fits.
 
The second handicap follows from the fits in the above paragraph and the fact that for the outburst EPIC spectra, VNEI model fits do not yield any acceptable \rchisq\ because the He-like iron line can not be fitted by this plasma emission model.  For spectral fits with EPIC, the CEVMKL plasma model is assumed for the outburst.  Note here that line diagnosis by RGS, as noted above, show that the X-ray emitting plasma is not in CIE state in both phases of the source. The outburst RGS spectra can be reasonably well-modeled, with the nonequilibrium ionization plasma model, VNEI.  To account for  this dichotomy, we have applied a composite model of a complex absorber and two plasma emission models, CEVMKL and VNEI, simultaneously-fitted to  the EPIC spectra.  The resulting fit had a \rchisq=1.29 (dof=444) which is not the best fitting model, but is the best fitting physical model. The complex absorber, had two components, 1)  partial covering  cold absorber,  2) partial covering ionized absorber (warm absorber). The cold absorber were similar to the fitted interstellar absorption in quiescence or outburst. The warm absorption was \nh=1.6$\times$10$^{23}$ cm$^{-2}$ with log($\xi$)=1.08\ and a covering fraction of  80\% .  This is less than the value in Table~\ref{tab:sp2} by factor of 16. This also corrects the over-predicted ionized absorption in the EPIC data during outburst. 
The ionization parameter is much lower depicting a significantly colder warm absorption in outburst (consistent with the {\it warmabs} fit in the first paragraph of this subsection). The plasma model parameters in Table~\ref{tab:sp2} (for CEVMKL) and in Table~\ref{tab:rgs-sp} (for VNEI) are very similar to this two-plasma model fit. Particularly, we were able recover the same temperatures as displayed in the different tables (OFit-3 and Outburst-2) which makes the EPIC and RGS results consistent. The normalizations and ionization timescales  hold within a factor of 2 (OFit-3 and Outburst-2). Therefore, we suggest that the X-ray emission region in the outburst indicates local NEI and CIE conditions which would maintain the EPIC and RGS consistency in outburst.

\subsection{The DN state transition of Z Cha and the disk outburst phenomena}\label{DOP}

\subsubsection{The state transition in the optical and UV}
In order to fully understand the X-ray evolution of the dwarf nova  Z Cha  through state transitions, we need to peek at the UV and the optical wavelength characteristics of  these transitions. During peak outburst the accretion disk dominates the UV light curve of Z Cha,  with a mass accretion rate of the order $1 \times 10^{-9}M_\odot$/yr.  This value of accretion rate amounts to a disk luminosity $\sim$ 5$\times$10$^{33}$ \lumcgs\ given a WD mass of 0.7\msun\ and a radius of  7.7$\times$10$^{8}$ cm.
In the UV and optical the disk seems to remain optically thick for two weeks after outburst.  The WD temperature after outburst is about 20,000 K and declines to 15,700 K in quiescence, with the bright spot (the region where the stream impacts the edge of the disk) being hotter around 16,700 K. By the third week, the disk is optically thin in the UV  \citep[see][]{1995Robinson}. The bright spot appears to be a non-negligible emitting component in the system, and it is a source contributing to the flickering light of Z Cha during quiescence \citep{1996Bruch} indicating it is a rather active region. 

UV and Optical observations of Z Cha during quiescence reveal a continuum slope of -1.7 (in log units) over an orbital period/phase  which is much flatter than  the expected -3.5 (or so) for an accretion disk \citep{1981Rayne}. The main reason for this discrepancy is the high inclination of the system where the disk is seen nearly edge on. Furthermore, it has been shown \citep{2012Smak} that in quiescence, the stream material at L1 overflows the edge of the disk in a manner similar to the eclipsing  system OY Car, which is known to be heavily veiled in the UV \citep{1994Horne}.  This explains the need for an ionized warm absorber (in addition to the ISM) in our modeling of the X-ray data during quiescence and could be responsible for the more complex nature of the absorption components. In fact, in a number of CV systems, the WD appears to be masked by the L1 stream material overflowing the disk edge  severely affecting the continuum flux level \citep{2019Godon}, especially for higher inclination systems near orbital phase $0.75\pm0.10$ and even near $0.25\pm0.10$.  Stream overflow is expected to occur more easily during quiescent dwarf novae \citep{1999Hessman}, however, in outburst the disk in Z Cha becomes flared \citep{1999Robinson} and it is likely that the disk edge occults to some extend the inner part of the disk and the WD.  For example, EM Cyg (with an inclination of $67^\circ$) in outburst exhibits a  disk UV flux and slope decreasing dramatically near orbital phase 0.75 where the stream overflows the disk  \citep[][Figs. 1 \& 2 bottom left]{2019Godon}, due to both an inflated disk edge and stream material flowing over the disk.  For a high inclination eclipsing system like Z Cha, the flared disk in outburst  is likely veiling the inner disk and WD at all orbital phases. This could explain the warm absorber being 30-50 times larger in outburst than in quiescence and the need for a cold absorber, too. 

The spectrum of the secondary in Z Cha is contaminated by the accretion disk  making proper identification of its spectral type uncertain. Assuming an M5 classification \citep{2011Hamilton} find the CO features (in the IR) much weaker than they should be, indicating the possibility that the secondary (and therefore accretion disk) might be deficient in C (CNO - processed?).  While here we find some agreement with solar abundances, the under-abundance of oxygen we find is consistent with weaker CO features. 

\subsubsection{The state transition in the context of compact X-ray binaries}

Black hole (BH)  and Neutron star (NS) Low-mass X-ray binaries (LMXBs) spend most of their time in a quiescent state and exhibit disk outbursts in the theoretical framework of DIM where they increase in emissivity, in luminosity and go through distinct  spectral states during outbursts \citep{2006Remillard,2011Belloni} where timescales of occurrence and durations are not very similar to DN (i.e., longer). In addition, some turn out to be failed outbursts \citep{2016delSanto}. Timing properties (e.g., noise characteristics) also change dramatically between states. Some  QPOs, some time lags  and band limited noise (BLN)  resemble to the ones detected for CVs and DNe  \citep[see][for a review and comparisons]{2019Balman} where broad-band noise structure in the optically thick disk decreases in strength in the outbursts, the noise moves to higher frequencies and gets more coherent (e.g., appreance of QPOs, \citealt{2022Balman}). LMXB transients are believed to move in a counter clockwise fashion trough a q-diagram (in hard and soft color space or luminosity versus spectral-color space)  as they change from quiescent  and hard state (hot flow \& jet dominated state) to a soft high state (with disk winds) which is disk dominated and finally back to quiescence. Few DNs do show this behavior where a soft X-ray blackbody emitting BL (e.g., SS Cyg \citealt{2008Kording}) is observed with a detection of nonthermal synchrotron emission interpreted as jet formation (SS Cyg). Not all DN show such behavior (see Sec.~\ref{DNX}) where mostly the hard X-ray emission in quiescence  diminish in flux and temperature getting softer in the outburst and recovering back to the quiescent higher fluxes and temperatures. This requires a clockwise rotation as opposed to the counterclockwise rotation on a q-diagram.  The physical explanation of this is beyond the scope of this paper, but we believe it resides in the ADAF-like structures that form in the inner disks that persist throughout a DN outburst as detected and studied for Z Cha in this paper. This is different than the standard LMXB picture. DNe and CVs in general, aside from super soft X-ray sources, are not very bright in accretion luminosity. 

NSs, as opposed to BHs, have a hard surface encountered by the accretion flow where  shocks and BLs are observed. NS LMXBs also show cooler coronae than BH LMXBs, cooled perhaps by the extra source of soft photons from the NS or the inner disk.  BH and NS LMXBs indicate a strong power law component that evolves throughout  the outburst from flatter photon indices in hard and intermediate states to steeper ones ($\Gamma > 2.1$) in the soft states. This component is connected to the hot coronal flows and jet emissions in these systems (unless it is from the NS itself; e.g., Milli Second Pulsars). High state CVs, DNe and as in this work, Z Cha show power law component in the outburst state at photon indices similar to LMXB transients provided that soft high states in the X-rays (blackbody emitting disks)  are rarely encountered thus, the photon indices remain flatter.  Note that some of our spectral fits for Z Cha with equally acceptable \rchisq\ do not require a power law model in quiescence or outburst. For  LMXB transients, jets, winds, and accretion states are well-studied where jets dominate hard states and winds are stronger in the soft states with 200-3000 km s$^{-1}$ where such winds  exhibit warm absorber effects on the emission internal to the absorbing regions or the central source \citep{2004Fender,2008Miller,2009Neilsen,2012Ponti,2016Diaz-Trigo}. For high state CVs and DNe in the outburst, we do observe similar effects of warm absorbers  (i.e., derived from spectral fits), most likely  a result of disk wind formation in the outburst phases as described for Z Cha.  Furthermore, we detect warm absorber effects, also in quiescence similar to e.g., LMXB-dippers \citep[][and references therein]{2009Balman}. Jet formation (during  outbursts) is suggested for DNe with radio detections of a handfull of systems at the onset of the outbursts \citep{2020Coppejans}, but weak detections hinder any further study.  

General characteristics of ADAF-like flows and disk geometry are described, also in the context of CVs, in \citet{2022Balman,2020Balman} and will not be repeated here. In general, the accretion flow and the disk structure can be described as an interplay between an outer standard disk (SAD) and an inner advection dominated flow which is responsible for ejection of the jets (JED: jet emitting disk) along with an accretion disk coronae found to co-exist  with this dichotomic structure  \citep{1994Narayan,1997Esin,2014Yuan}.  This JED-SAD hybrid disk provide successful explanation that describes the state transitions, and also matches evolution of QPO structures detected in LMXB transients \citep{1997Ferreria,2019Marcel,2020Marcel}.  This type of hybrid structure is also described to form a torus (e.g., nonself-gravitating axisymmetric thick tori in NS-LMXBs) in the central regions of these transients where the radial and vertical  oscillations of this torus structure is used to describe the QPOs (e.g., kHz QPOs) in LMXBs \citep{2003Rezzolla,2017Parthas,2018Avellar}.  Such a torus or the inner advective (ADAF-like)  hot flow is believed to be large in hard states where optically thick disk is more distant and the torus or inner advective hot flow can expand. There is a small/er torus or inner advective/coronal structure in soft states when inner disk is compressed as a result of increased accretion rate. This prescription is in accordance with spectroscopic results for Z Cha since we find that the X-ray emitting region is denser and cooler in outburst, thus smaller than the quiescence where the X-ray temperature is about 10 times more with low densities as the flow is not very collisional and forbidden lines of He-like elements persists (with no intercombination and resonance lines) the geometry is larger than the outburst. Note here that the break frequency for Z Cha (1.0$\pm$0.4 mHz) in quiescence indicates transition (e.g., into an advective hot flow) around 1.5$\times$10$^{10}$ cm (using $\nu_0 = 1/2\pi (GM_{WD}/R_{in}^3)^{1/2}$). The brightness temperature measurements in quiescence (optical)  imply deviation from standard structure around 0.5R$_{L1}$ where R$_{L1}$ (distance between inner Lagrangian point and  the WD) is calculated to be 3.0$\times$10$^{10}$ cm \citep{1990Wood} which makes our disk structure model for Z Cha consistent across wavelengths.
  
 \section{Conclusions}\label{conc}
  
Comparative studies of dwarf nova in quiescence and outburst is an essential tool to understand accretion physics in cataclysmic variables and related objects. Z Cha being a close-by eclipsing SU UMa-type dwarf nova below the period gap, has been studied over most of the electromagnetic spectrum in quiescence and outburst.  In this paper, we presented X-ray spectroscopy of Z Cha using the EPIC and RGS instruments of  \xmm\ Observatory (i.e., an archival study). Such data is scarce and worthwhile for an in-depth and  detailed study to draw information on state transitions of DN and CVs though disk outbursts.  Our main aim was to do rigorous spectral modeling to understand the conditions in the X-ray emitting plasma during quiescence and how that changed in the outburst. 

Our results are consistent with the general observational picture where the X-ray luminosities and temperatures are higher in quiescence where a hard X-ray emitting plasma always remains. A soft X-ray, blackbody emission is found as part of a composite spectral fit to the outburst data, but  not significant enough to be the best fitted model and the soft X-ray luminosity was about 1000 times below the standard disk and DIM theory expectations
for a mass accretion rate of $10^{-9}M_\odot$/yr.  The quiescent hard X-ray emission can be modeled by collisional equilibrium (CEVMKL in XSPEC) or nonequilibrium plasma models (VNEI in XSPEC), yielding a kT of 5.4-8.6 keV with the VNEI model and kT$_{max}$=7.2-13.0 keV (with CEVMKL) where the lower limits come from composite-fits including a power law model of emission . The spectra yield better \rchisq\ values using partial covering absorbers of cold and photoionized nature. This is in accordance with the observations in the optical and UV where stream-overflow (and also hotspot region) and winds in the outburst are detected. The ionized absorber has an equivalent \nh=(0.5-0.7)$\times$10$^{22}$ cm$^{-2}$ with log($\xi$)=3.09-3.17  when the {\it warmabs}  model is used (along with VNEI which has $\tau$=5.4$\times$10$^{11}$ s\ cm$^{-3}$). Moreover, we find that in quiescence at two different phases over the orbit (0.3 and 0.75), the warm absorber characteristics change and either the equivalent \nh\ is larger (less pronounced dip in the light curve) or the warm absorber is colder and less ionized (a relatively deeper dip) similar to LMXB-Dipper characteristics.
The first dip near phase 0.75 is due to the L1-stream material overflowing the disk edge, near phase 0.9, and directly veiling the inner disk and the WD near phase 0.75 as it moves inwards. The overflowing material is the most likely origin of the varying warm and cold absorption in the X-rays. In its ballistic trajectory, the stream material continues and hits the disk surface near phase 0.50 \citep{1986Marsh,1989Lubow} where it bounces off the disk surface outwards and veils the inner disk and the WD a second time near phase 0.30, thereby causing a second dip \citep{2019Godon} as we view the warm absorption by this material in the X-rays. In outburst, the flaring of the disk viewed at an angle above 80\degmark\ is likely responsible for much of the complex phase-average X-ray absorption; particularly around phase 0.9 the L1-stream impacts the disk edge thereby increasing its vertical extend, and we detect orbital modulation of the X-rays dipping between phases 0.7-1.0 as a result of  the inferred absorption characteristics.

The line diagnosis in quiescence shows no resonance lines of the He-like O, Ne, Mg, Si, and N. Only forbidden lines of  Ne, Mg, Si are detected together with the H-like C, O, Ne, and Mg. The strongest line is O VIII. The quiescent X-ray emitting plasma is not collisional and not in ionization equilibrium as line diagnosis indicates which is consistent with hot advective accretion flows  in nature.  The quiescent spectrum also indicates an Fe XXVI absorption line detected at 6.95$\pm$0.05 keV for the  first time in a dwarf nova using collisional equilibrium plasma models and when a nonequilibrium ionization model is used the line detection is only 1.5-2$\sigma$ due to the weakness of the H-like Fe. The line diagnosis of the outburst RGS spectrum shows He-like O, and Ne with intercombination lines being the strongest along with relatively weaker resonance lines indicating electron densities (7-90)$\times$10$^{11}$ cm$^{-3}$  (using R-ratios). The fits yield electron temperatures 0.5-0.7 keV using VNEI plasma model. The X-ray plasma as seen by RGS is more collisional and denser in the outburst, but not in ionization equilibrium with ionization timescales of (0.97-1.4)$\times$10$^{11}$ s\ cm$^{-3}$. The EPIC data indicates that there is dichotomy in the X-ray plasma in outburst. The best-fitting physical model consistent with both RGS and EPIC spectra is a combination of CEVMKL$+$VNEI models where the lower temperatures are consistent with the NEI characteristics and thus, the RGS spectrum. The higher energies, the higher plasma temperatures, are consistent with a collisional equilibrium model around $\sim$ 3 keV. In general, X-ray emitting region is extended and  inhomogeneous with perhaps higher and lower density regions.  The radiative efficiency of the X-ray  region is about 0.0004 comparing the disk and X-ray luminosities of the source in outburst.  All detected lines are narrow with Keplerian rotational velocities $<$1000 km s$^{-1}$ consistent with the advective hot flow characteristics. 

We can list the basic observational characteristics of advective hot flows in the X-rays for DNe (and CVs) as low emission efficiencies 0.01-0.0001, sub-Keplerian flows of nonionization equilibrium plasma with extended non-optically thick disk structure (some power law emission in high-states), a BLN noise with break frequencies indicating transition radius in a hybrid disk and propagation lags (as in X-rays lagging the UV/optical). Overall, X-ray emission from Z Cha shows ADAF-like advective hot flow characteristics in quiescence and outburst.  Our results have implications on all state transitions and disk outbursts together with accretion physics of high-state CVs (NLs) and accreting WDs. This work will help to improve theoretical studies of disk modeling, MHD formalism and shock formation and outflows. 
Our study also adds to the general understanding of radiatively inefficient hot  flows in accretion disks in X-ray binaries and AGNs as it comprises a non-relativistic limit at lower temperatures 
with hard surface on the compact accretor (with significantly low or no magnetic field).  

\begin{acknowledgments}
Authors thank M. Mendez and J. P. Lasota for comments on the manuscript.
\c{S}B acknowledges grant by the Scientific Research Projects Coordination Unit of Istanbul University through the BAP Project No: 40017. This research is based on observations obtained with \xmm, an ESA science mission with instruments and contributions directly funded by ESA Member States and NASA. This work has made use of data and software provided by the High Energy Astrophysics Science Archive Research Center (HEASARC) maintained by NASA/GSFC.
\end{acknowledgments}

\end{document}